\def\prd{Phys. Rev. D}

\def\jcap{JCAP}

\newcommand{\ra}{\;\raise1.0pt\hbox{$'$}\hskip-6pt\partial\;}
\newcommand{\lo}{\;\overline{\raise1.0pt\hbox{$'$}\hskip-6pt\partial}\;}

\def \<{\langle}
\def \>{\rangle}

\RequirePackage{snapshot}

\documentclass[a4paper,11pt]{article}
\usepackage{jcappub}

\usepackage{graphicx,epsfig,natbib,color,times,bm,amsmath,multirow}
\usepackage[ddmmyyyy,hhmmss]{datetime}
\begin{document}
\title{On the time lags of the LIGO signals}

\author[a]{{James Creswell},}
\author[a]{{Sebastian {von Hausegger}},}
\author[b]{\\{Andrew D. Jackson},}
\author[a,c]{{Hao Liu},}
\author[a]{and {Pavel Naselsky}}
\affiliation[a]{The Niels Bohr Institute \& Discovery Center, Blegdamsvej 17, DK-2100 Copenhagen, Denmark}
\affiliation[b]{Niels Bohr International Academy, Blegdamsvej 17, DK-2100 Copenhagen, Denmark}
\affiliation[c]{Key laboratory of Particle and Astrophysics, Institute of High Energy Physics, CAS, 19B YuQuan Road, Beijing, China}
\emailAdd{dgz764@alumni.ku.dk, s.vonhausegger@nbi.dk, jackson@nbi.dk, liuhao@nbi.dk, naselsky@nbi.dk}

\abstract{
To date, the LIGO collaboration has detected three gravitational wave (GW) events appearing in both its Hanford and Livingston detectors.  In this article we reexamine the LIGO data with regard to correlations between the two detectors.  With special focus on GW150914, we report correlations in the detector noise which, at the time of the event, happen to be maximized for the same time lag as that found for the event itself.  Specifically, we analyze correlations in the calibration lines in the vicinity of 35\,Hz as well as the residual noise in the data after subtraction of the best-fit theoretical templates. The residual noise for the other two events, GW151226 and GW170104, exhibits similar behavior. A clear distinction between signal and noise therefore remains to be established in order to determine the contribution of gravitational waves to the detected signals.
}

\maketitle

\section{Introduction}
\label{sec:introduction}

With its initial publication by the LIGO collaboration, the event GW150914 became the prime candidate for the first direct detection of gravitational waves~\citep{Ligo1}.  Detectors in Hanford, WA (H) and in Livingston, LA, (L) separated by approximately 3000\,km, measured a well-pronounced signal of a shape consistent with the merger of two massive black holes and with a time delay of 6.9\,ms.   Independent detection by each of the two detectors within a time window of $\pm 10$\,ms is clearly essential for validation of detection; the observed time delay additionally provides significant constraints on the direction from which the gravitational waves originated.  Following this announcement, considerable effort has been invested in discussing the consequences of this discovery in cosmology and astrophysics.  However, substantially less attention has been paid to independent reexamination of the data and to the reliability of its physical interpretation \citep{Naselsky16,Liu16}.  There is little doubt that the signal is unique regarding both its amplitude and shape.  It stands out clearly in the time record in a manner that is largely independent of the method adopted for its analysis, i.e. with or without the use of templates (\citep{Ligo1,Ligo2,Ligo3} and \citep{Naselsky16}, respectively).  Both methods reach the conclusion that the significance of the event is in excess of $5\sigma$.\footnote{It should be stated, however, that these estimates and their meaning depend on the properties of the statistical ensemble to which the event is compared, e.g. instrumental noise~\citep{Ligo1} or an ensemble of cross-correlations between detectors~\citep{Liu16}.}  Of these, only the method based on the use of binary-merger templates is capable of determining masses and the distance of these spiraling objects.\\

Given its evident significance, it is important to understand the characteristics of GW150914 as well as relevant features of the noise associated with the LIGO detectors.  The primary focus in this paper will be the difference in the times at which the event is registered in the two detectors.  The fundamental assumption of the LIGO experiment is that the measured time difference of \mbox{$\tau_{\rm GW}\approx6.9\,$ms} is uniquely related to the event itself and thus provides conclusive confirmation of its existence.  It is assumed that such correlated detection is independent of detector noise before or after the event as well as residual noise during the event.  We have previously suggested \citep{Liu16} that this is not the case by demonstrating that the noise before and after the event is maximally correlated (on average) for precisely this time difference.  Here, we will extend the analysis of this peculiarity to specific components of the noise, i.e. to calibration lines in the detectors at $\approx$35\,Hz and the residual noise at the time of the event.  We will present evidence suggesting that the observed time delay is not specific to the event itself.  In particular, we will show that there exist similar features in the time-ordered data (TOD) due to the interference of two calibration lines for the Hanford detector (at $35.9$ and $36.7$\,Hz) and for the Livingston detector (at $34.7$ and $35.3$\,Hz).  Their relevance is due to the fact that instrumental lines (such as the calibration lines) are capable of producing signals morphologically similar to those expected from GW events~\citep{LIGO warning}.  Due to the different frequencies of these calibration lines, the interference patterns for the two detectors are shifted in the TOD.  Within the $200$\,ms window containing the GW150914 event, the calibration lines in the H and L data are found to have a time lag of $\tau_{\rm GW}$.  Moreover, the residual noise in the H and L data also have the same time lag.  Clearly, the more general presence of correlations with time delay $\tau_{\rm GW}$ would significantly reduce the ability to identify and measure gravitational wave signals with a similar time delay.  In fact, after subtracting the best-fit GW template from the cleaned data, the remaining signals in the two detectors are also maximally correlated for the same time lag.  Since similar effects are also found for the GW151226 and GW170104 events, this suggests the presence of unwanted systematic effects. Following a suggestion from the referee of this paper, we have added an appendix in which we extend our analysis for the time domains from 16.2 to 16.7 s, including the GW150914 event which lies between 16.25 and 16.45 records. This tells us more about the properties of the noise and residuals of cleaning. Using an $\approx$20 ms running window, we have detected a "precursor" and an "echo" of GW150914 that have the same time delay as GW150914 itself.

The organization of this paper is as follows. In Section~\ref{sec:data} we consider both the amplitude and the phase of the Fourier transforms of the Hanford and Livingston GW150914 time records.  Although it is conventional to focus on the Fourier amplitudes and the closely related power spectrum, we place greater emphasis on the Fourier phases that are known to play an important role in determining the morphology of signals.  This analysis will reveal surprisingly large phase correlations.  Section~\ref{sec:correlations} will consider the nature and possible origin of the 7\,ms time delay that has been offered as the evidence for the discovery of this gravitational wave event.  In Section~\ref{sec:otherevents} the time lag in the other two GW events will be discussed, and the final section will contain observations and concluding remarks.  In the interest of relevance, we will use LIGO raw and cleaned data as well as LIGO cleaning procedures except as noted.

\section{Remarks on the data for GW150914 and its Fourier amplitudes and phases}
\label{sec:data}

The noise present in LIGO's detectors is neither Gaussian nor stationary as discussed in~\citep{LigoTransientNoise}.  Nevertheless, estimations of e.g.\,detection rates and noise background inherently assume the opposite.  Attempts to remove non-stationary noise using methods intended for the treatment of stationary noise can have the unintended consequence of introducing spurious residual effects.  It is thus important to identify sources of non-Gaussianity and non-stationarity and to eliminate them in order to ensure reliable GW detection.  (See e.g.~\citep{Bose} and~\citep{Bose1}.)

Stationary random processes are conveniently characterized by their Fourier amplitudes and phases.  Despite the above concerns, we will, in this section, perform a Fourier analysis of the TOD streams for the GW150914 32\,s records with particular focus on phase information.  The transition from the TOD, $x(t)$, to the Fourier amplitudes, $X(\omega)$, is governed by the following equations:
\begin{align}
\begin{split}
x(t) &= \int_{-\infty}^{\infty}d\omega\, e^{i\omega t}X(\omega)\hspace{0.2cm} \\
X(\omega) &= \frac{1}{2\pi} \int_{-\infty}^{\infty}dt\, e^{-i\omega t}x(t)
\end{split}
\label{eq1111}
\end{align}
where $\omega=2\pi f$.  The complex Fourier coefficients, $X(\omega)$, can be represented in terms of their amplitudes and phases as $X(\omega)=|X(\omega)|\exp(i\varphi(\omega))$, where $|X(\omega)|$ is related to the power spectrum of the signal through $P(\omega ) \propto |X(\omega)|^2$.\\

Given that the strains observed in the raw data are some three orders of magnitude larger than the GW150914 signal, various cleaning operations must be performed.  It is our intention to follow closely LIGO's prescription for cleaning the data.  We will therefore employ the method described in~\citep{LIGO soft}, which will be described briefly below.  However, to provide an intuitive understanding of the processes involved, we will first describe a simpler method used previously~\citep{Naselsky16} that gives results in excellent agreement with those obtained with LIGO's cleaning. The first step is to perform a Fast Fourier Transform (FFT) of the full 32\,s record.\footnote{Keeping possible effects of the treatment of non-stationary data in mind, the steps described here can be applied to the 4096\,s data sets as well (or any sufficiently long part thereof).  If desired, a 32\,s time segment centered on GW150914 can be extracted from the resulting TOD.  This 32\,s TOD can then be Fourier transformed to determine the amplitudes $|X(f)|$ and the phases $\varphi(f)$ to be compared with the results shown here.}  There are a variety of sources of noise including quantum noise, seismic noise, suspension thermal noise and Brownian noise.  The effects of these noise sources can be reduced dramatically by applying a simple band-pass filter to select only frequencies with $35 \lesssim f \lesssim 350$\,Hz and thus to restrict our attention to the domain of lowest detector noise and highest sensitivity.  In addition, the detectors have a myriad of narrow resonances including known mechanical resonances, mains power harmonics, and signals injected for purposes of calibration and alignment.  It is generally acknowledged that these narrow resonances have nothing to do with gravitational waves.  Since their effect is several times larger than the GW150914 signal, they can and should be removed by ``clipping''.  The resonances that will be removed by clipping are found at frequencies 14.00, 34.70, 35.30, 35.90, 36.70, 37.30, 40.95, 60.00, 120.00, 179.99, 304.99, 331.49, 510.02, and 1009.99\,Hz (See~\citep{LIGO soft}).\footnote{Remarkably, this list contains only 5 lines in the range of greatest detector sensitivity between 50 and 350\,Hz.}  Once the data has been cleaned, an inverse FFT is performed to obtain a band-passed (and possibly clipped) stream of TOD, and more details and discussions can be found in the appendix.

LIGO's filtering method~\citep{LIGO soft} follows essentially the same approach.  However, instead of working in the Fourier domain, they make use of a linear infinite impulse response filter that acts on the data in the time domain.  The band-pass is thereby performed by application of a Butterworth filter in effectively the same range as quoted above, while the narrow resonances are removed via notch filters.  As already noted, we will use this method in the following in order to stay as close as possible to the LIGO analysis. By comparing our cleaning result with the available LIGO cleaning result in the 0.2 s time range for GW150914, we see that the two differ relatively by only $3\times10^{-4}$ (see appendix for more details).

Fig.~\ref{fig:straindata} shows the 32\,s strain record for the LIGO detectors.  The top and bottom curves show the uncleaned data \citep{Ligo1}.  The middle two curves show the same data band-passed and clipped and magnified by a factor of 100 for visibility.  The dashed vertical line marks the onset of GW150914, and its small bumps can be seen in the cleaned data.
\begin{figure}
\centering
\includegraphics[width=0.98\textwidth]{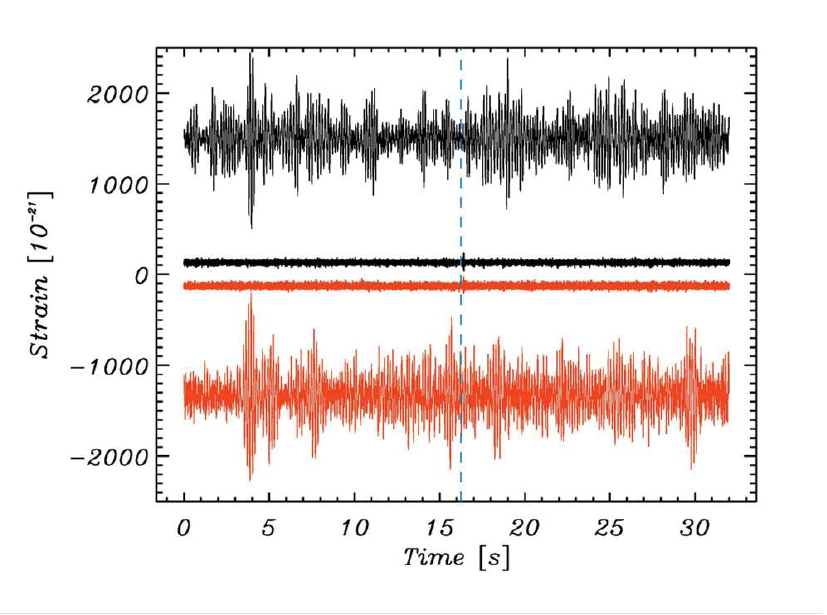}
\caption{Comparison of the LIGO 32\,s data (black for Hanford and red for Livingston).  The top and bottom records are raw data, the middle records have been band-passed and cleaned (as described in the text), then amplified by a factor of 100 for visibility.  All four curves have been manually shifted vertically for ease of comparison.}
\label{fig:straindata}
\end{figure}

\begin{figure}[h!]
\centering
\includegraphics[width=0.32\textwidth]{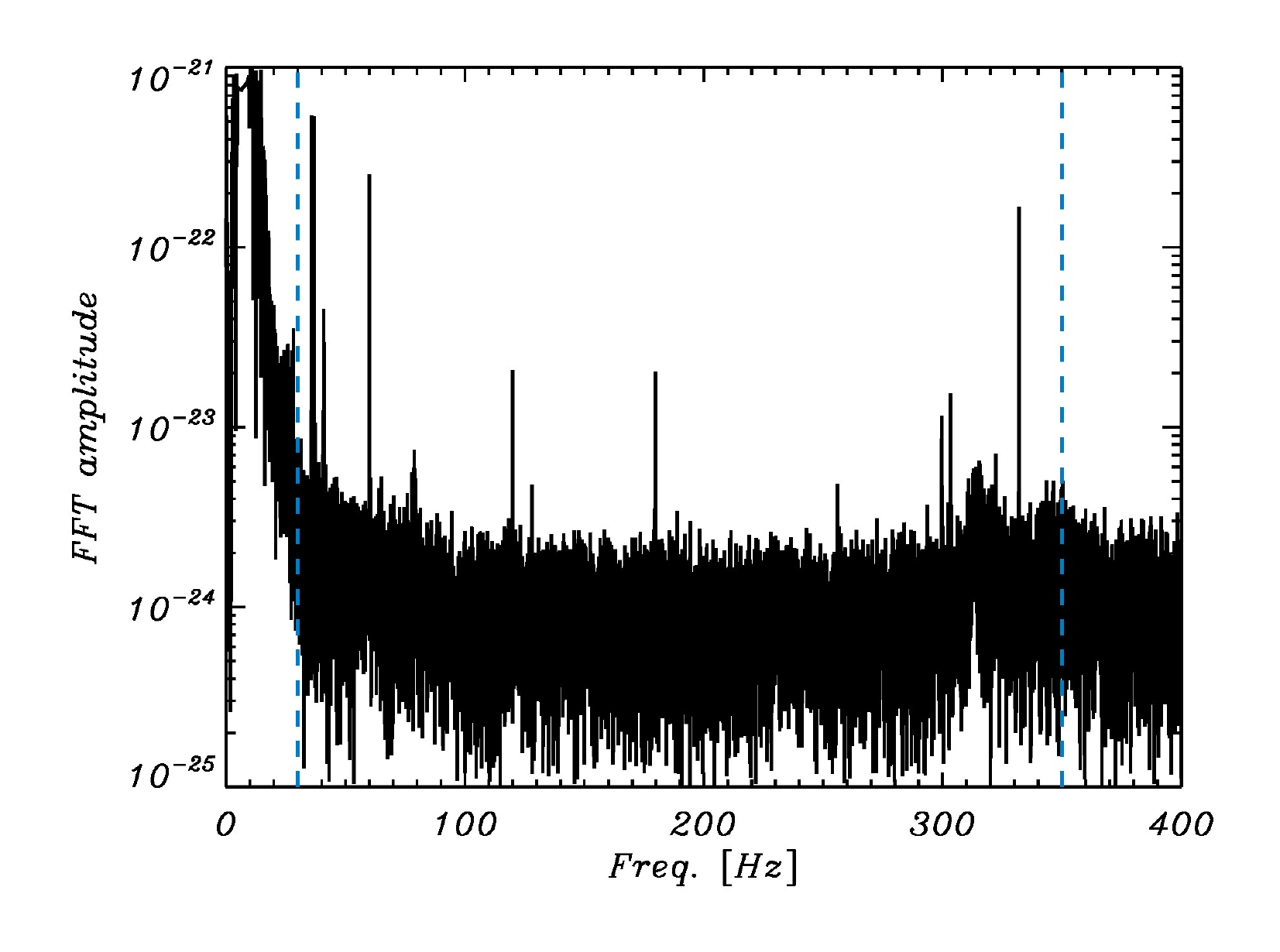}
\includegraphics[width=0.32\textwidth]{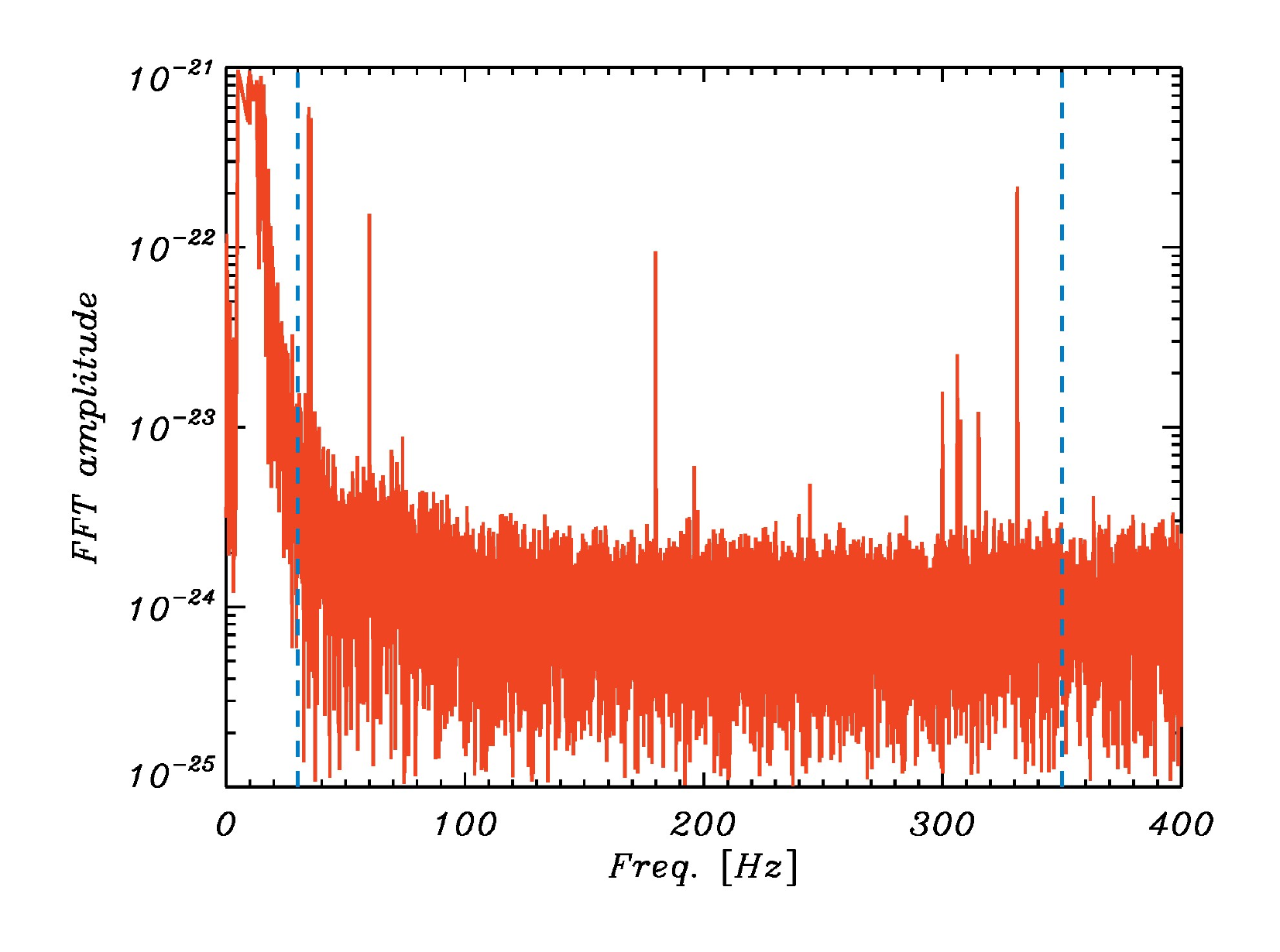}
\includegraphics[width=0.32\textwidth]{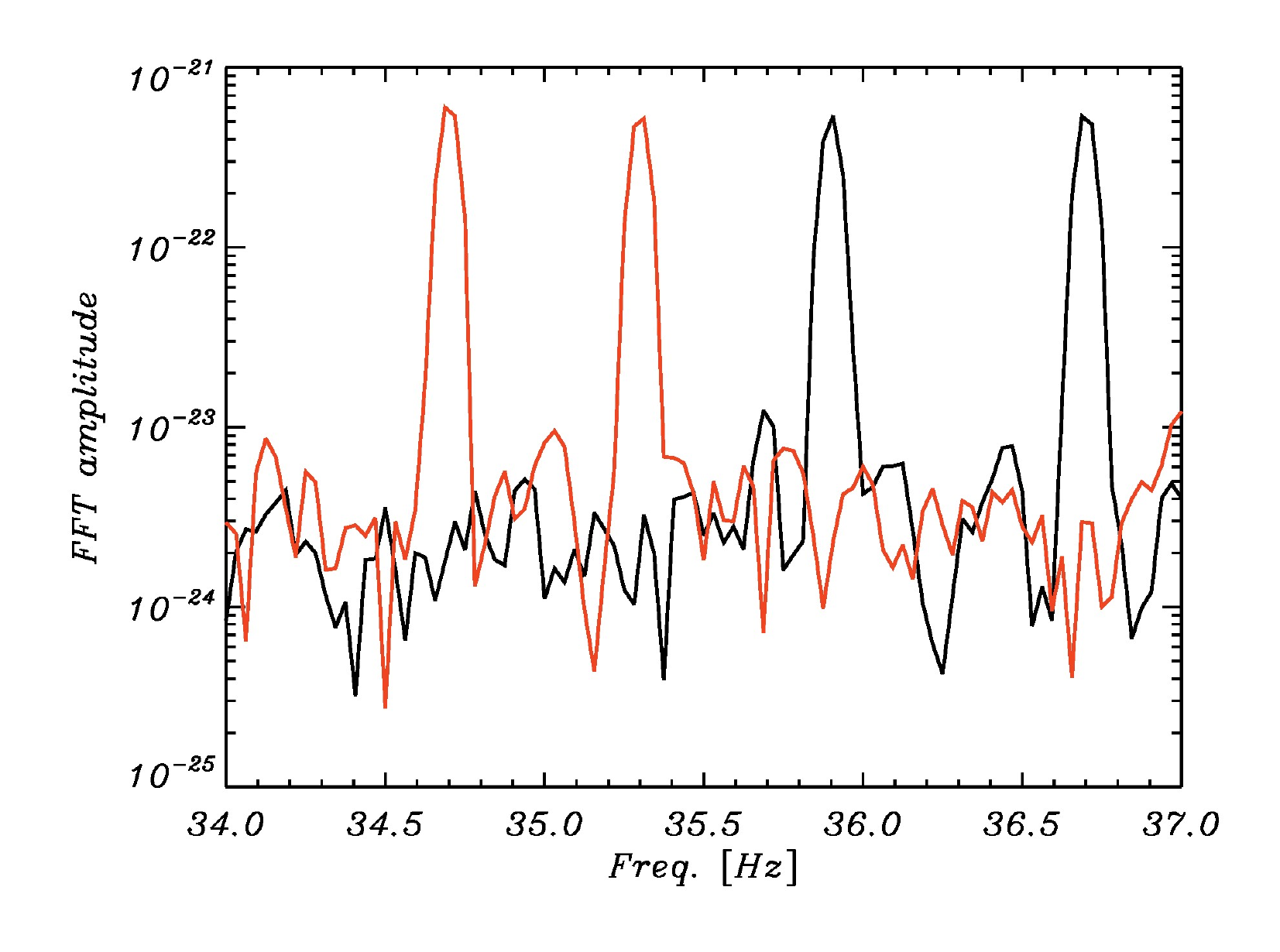}
 \caption{Fourier amplitudes for the raw 32\,s Hanford (black) and Livingston (red) data.  The dashed vertical lines indicate the frequency domain $30 < f < 350$\,Hz that will be retained after band-pass filtering.  The right panel plots a limited frequency domain $34 < f < 37$\,Hz in order to show the strong calibration lines for later reference.}
 \label{fig:fourieramplitudes}
\end{figure}

Fig.~\ref{fig:fourieramplitudes} shows the Fourier amplitudes for the Hanford and Livingston detectors as a function of frequency as obtained from the 32\,s TOD.  Narrow resonances related to AC power are clearly visible at frequencies of 60, 180, and 300\,Hz in both records.  In the right panel of this figure we focus on the strong calibration lines at 34.7 and 35.3\,Hz (Livingston) and at 35.9 and 36.7\,Hz (Hanford) for later reference.  While these amplitudes are interesting, it is important to bear in mind that the morphology of events in the time record is largely determined by the phases of the Fourier coefficients and, in particular, by the coupling between the phases of different frequencies.  It is difficult to estimate the range and nature of frequency correlations from the power spectrum alone.  It is necessary to look at the phases directly.\\

\begin{figure}[h!]
\centering
\includegraphics[width=0.32\textwidth]{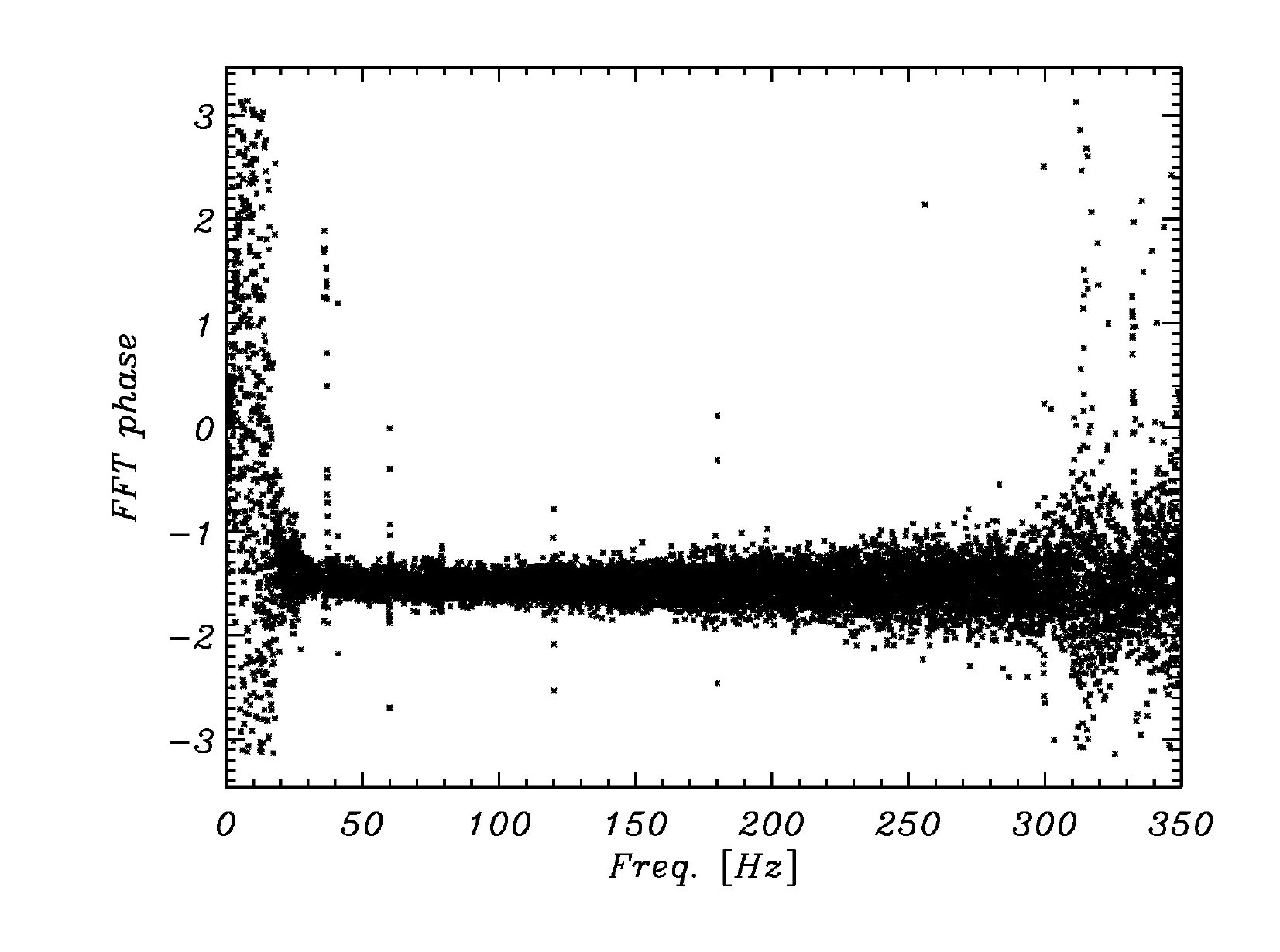}
\includegraphics[width=0.32\textwidth]{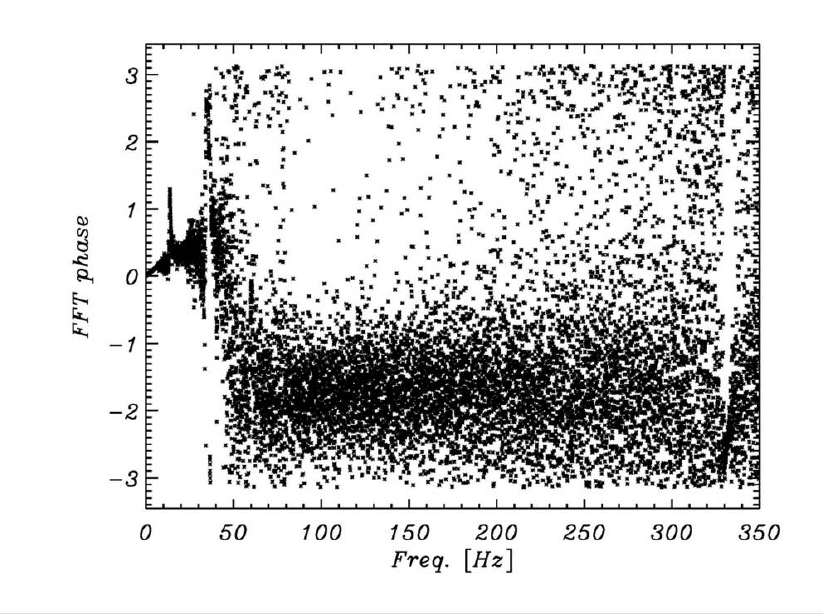}
\includegraphics[width=0.32\textwidth]{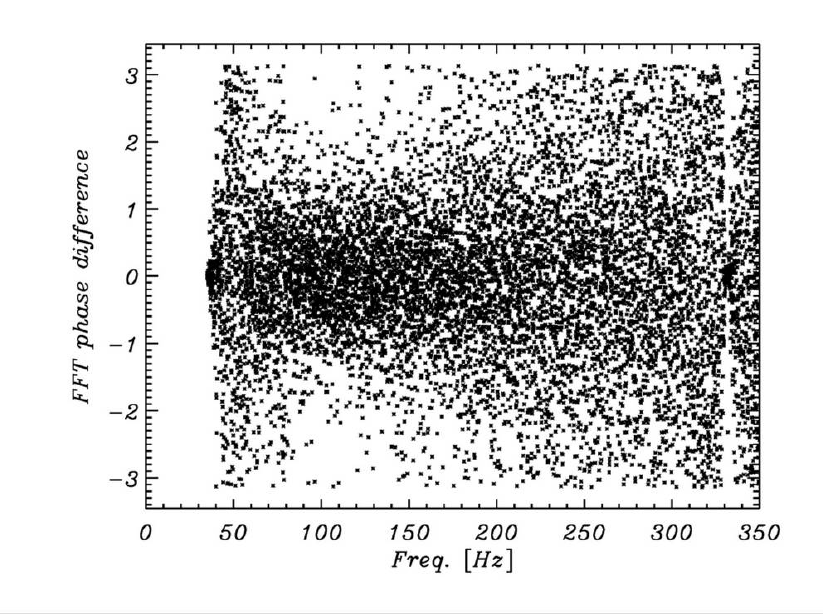}

\includegraphics[width=0.32\textwidth]{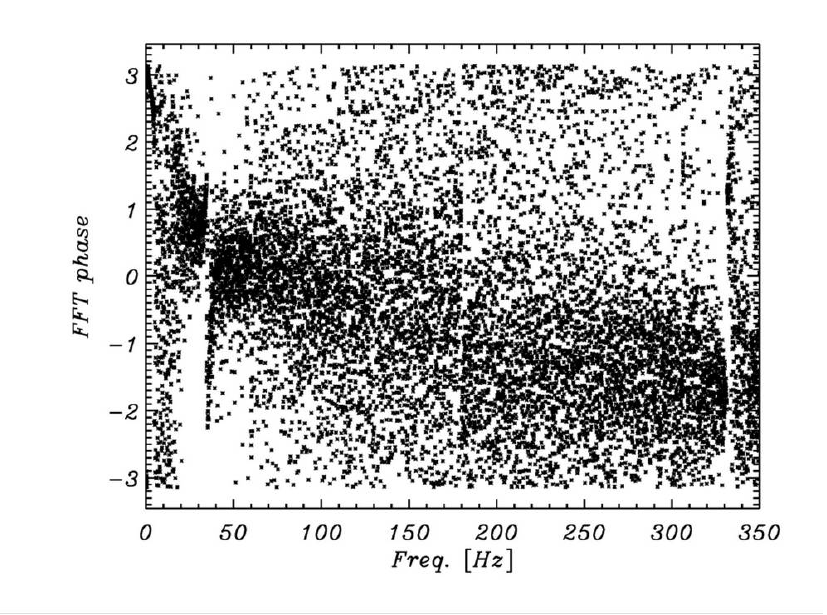}
\includegraphics[width=0.32\textwidth]{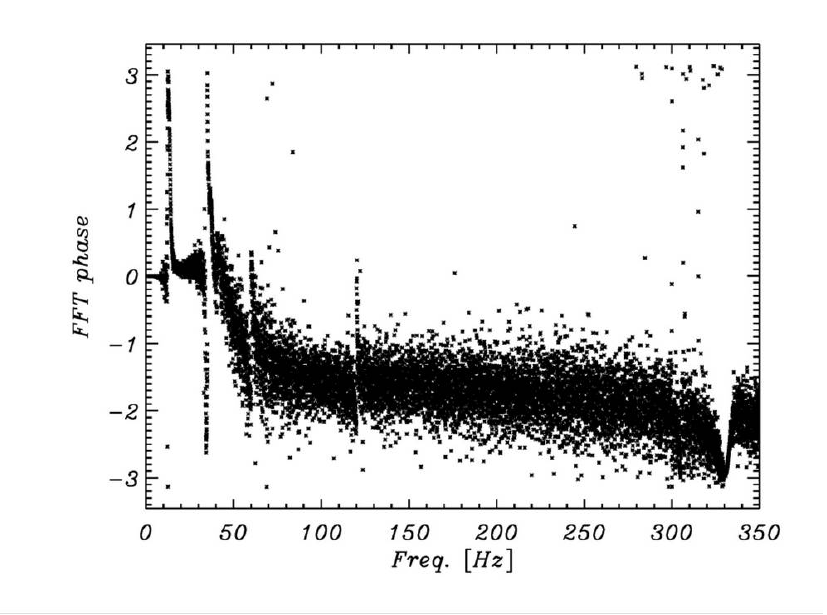}
\includegraphics[width=0.32\textwidth]{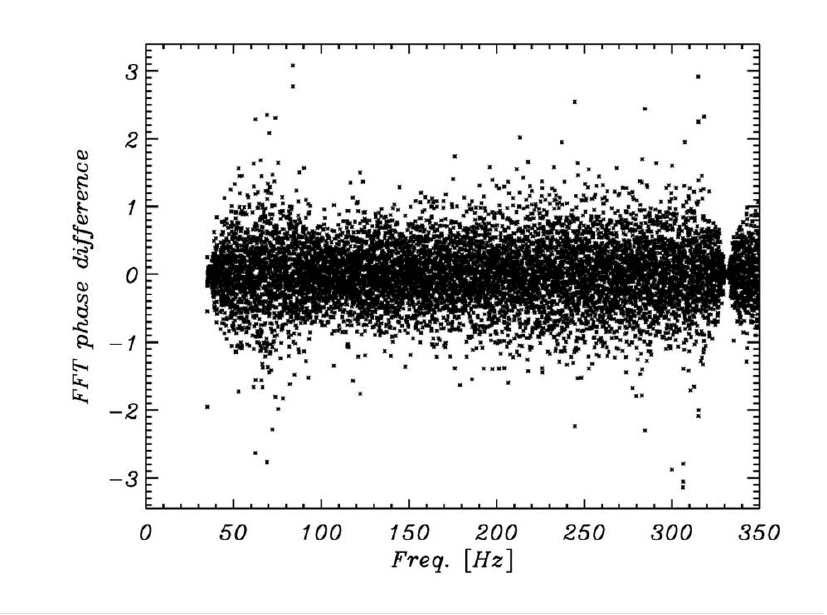}

\includegraphics[width=0.32\textwidth]{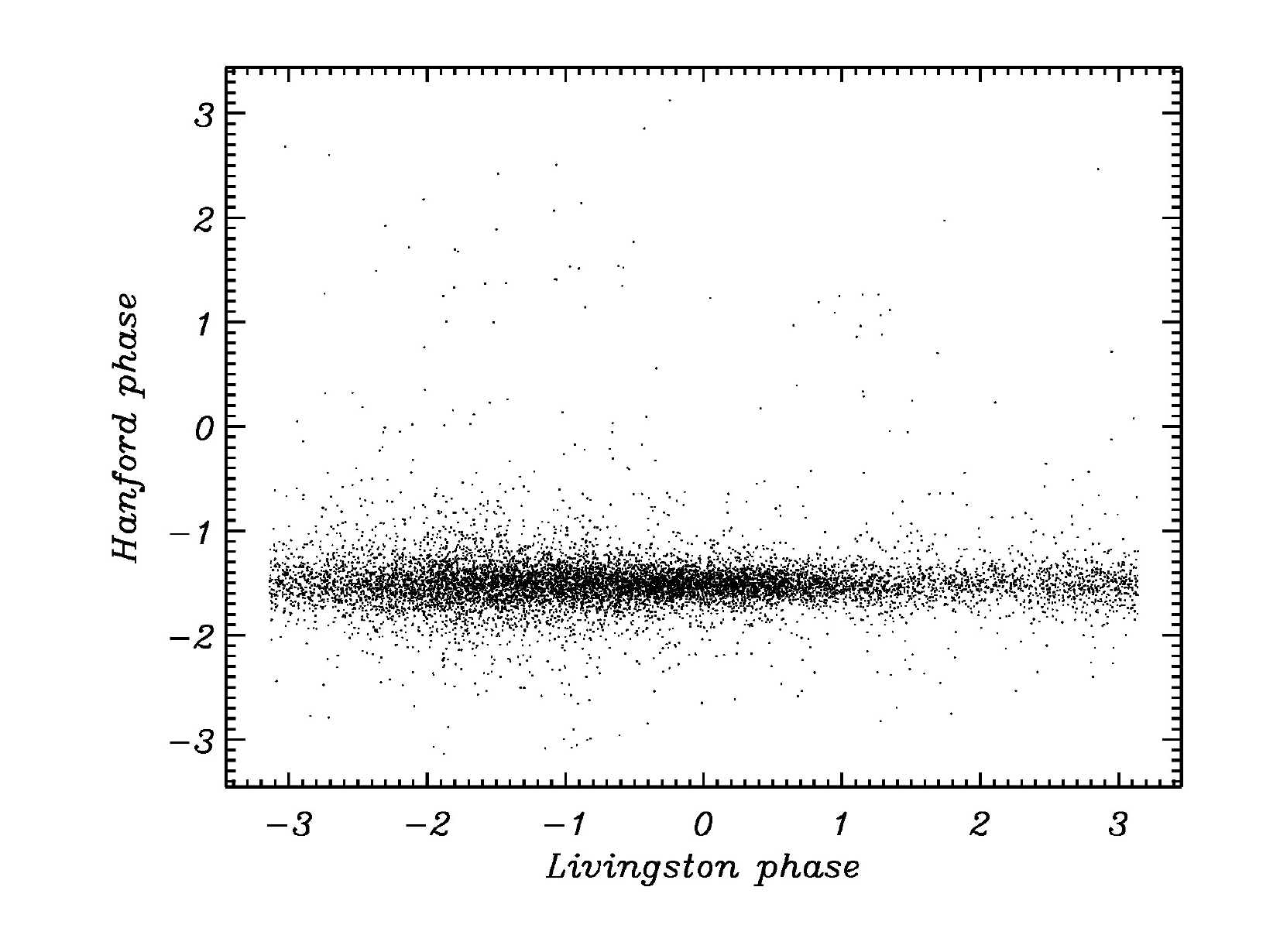}
\includegraphics[width=0.32\textwidth]{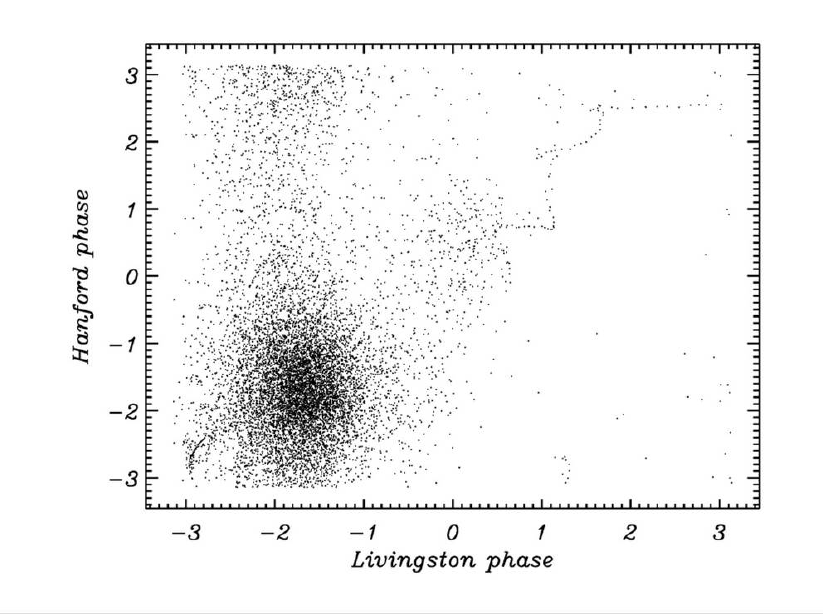}
\includegraphics[width=0.32\textwidth]{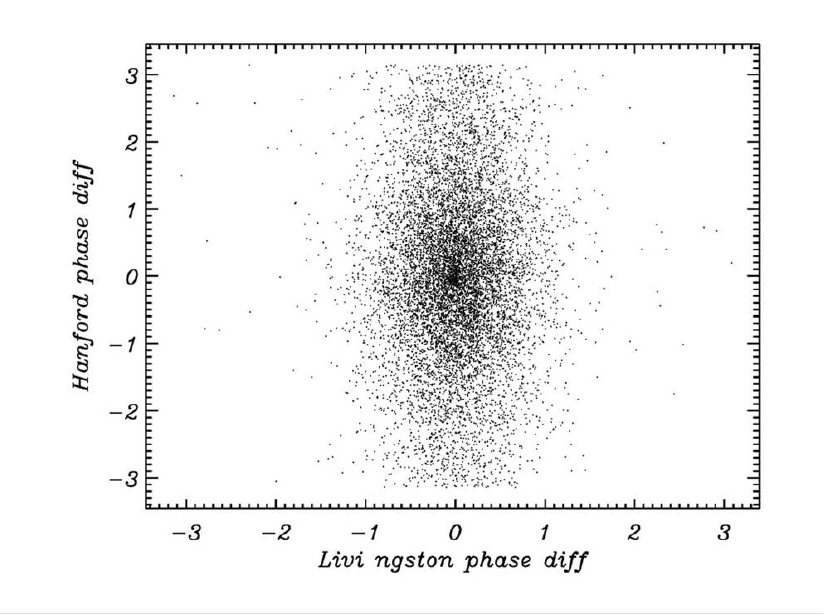}
\caption{Plots of the Fourier phases and their differences as a function of frequency from the 32\,s GW150914 data.  Top panels: The Fourier phases for the raw (left) and cleaned (center) Hanford data, and the phase differences (Eq~(\ref{equ:phase_diff})) for the cleaned Hanford data.  Middle panels: Same as top, only for Livingston.  Bottom panels: Scatter plots of the Hanford (y-axis) and Livingston (x-axis) Fourier phases for the raw data (left), the clean data (center), and the phase differences of the cleaned data.}
\label{fig:fourierphases}
\end{figure}

In order to investigate the Fourier amplitudes and phases we use the 32\,s data, cleaned by use of LIGO's approach as mentioned above.  As shown in Fig.~\ref{fig:fourierphases}, even the cleaned data reveals surprisingly strong phase correlations.\footnote{For the use of phase correlations as a measure of information in a cosmological context see Refs.~\citep{Coles2000,Chiang2000,Chiang2002}.}  Also shown are similarly significant correlations in the differences of phases of adjacent frequencies defined as

\begin{equation}\label{equ:phase_diff}
\delta \varphi_i = \arctan{ \left( \sin(\varphi_{i+1}-\varphi_{i}),\cos(\varphi_{i+1}-\varphi_{i})\right) } .
\end{equation}

The cleaned data should contain signal and residual noise.  Although the strength of the signal and noise are comparable during the 0.2\,s of the GW150914 event, noise is completely dominant for the remainder of this 32\,s record.  In other words, the Fourier amplitudes shown here are noise dominated.  Since random noise would have a uniform distribution of phases, it is evident that the noise is neither stochastic nor even roughly stochastic\footnote{LIGO states that ``the data is only roughly Gaussian''.~\citep{LIGO warning}}.  We also note that plots of the phases of the 4096\,s data show similar correlations.

Such correlations can either be numerical artifacts\footnote{For example, Fourier transforms performed over the finite interval $[0,T]$ employ a basis of states that are orthogonal and periodic on this interval.  If the function to be transformed does \emph{not\/} share this periodicity, the resulting expansion will not converge uniformly and phase correlations can be introduced artificially.} or a genuine consequence of the physically meaningful coupling of nearby frequencies.  We point out the rather surprising fact that the phase correlations in the Livingston detector (middle row in Fig.~\ref{fig:fourierphases}), already present in the raw data, are significantly increased by the cleaning procedure.  This suggests that their origin may be physical.

Whatever its origin, the non-stochastic behavior indicated by phase correlations in the supposedly clean data immediately presents an a priori challenge to the reliability of any significance estimates of possible GW events. We discuss this issue in Appendix~\ref{app:phase mixing}, including the phase mixing effects of the transition from a 32 s record to a 0.2 s record.

\section{Correlations in the noise}
\label{sec:correlations}

The event GW150914 is characterized by its shape and its almost simultaneous appearance in the Hanford and Livingston detectors with a time lag of only $6.9$\,ms.  In this section we briefly review a method, proposed in~\citep{Naselsky16,Liu16}, for confirming this time lag using correlations and apply it to noise components of the data in the immediate vicinity of GW150914.  This method can be used for all time sequences obtained and is independent of cleaning techniques.

We denote the strain data, $H(t)$ and $L(t)$, within a given time interval $t_a\leq t\leq t_b$ as $H_{t_a}^{t_b}$ and $L_{t_a}^{t_b}$, respectively, while the arrival time delay between the two sites is $\tau$.  The cross-correlation coefficient between the H and L strain data in a window of width $w$ starting at time $t$ is then given as (also shown in \citep{Liu16})
\begin{equation}
C(t,\tau,w) = {\rm Corr}(H_{t+\tau}^{t+\tau+w},L_{t}^{t+w}),
\label{eq:runningwindowcorrelation}
\end{equation}
where we allow for a relative shift of the Hanford and Livingston data in the time domain.\footnote{We note that the GW150914 signal appeared first at the Livingston site and was seen at the Hanford site approximately $6.9$\,ms later~\citep{Ligo1}.  Thus, Eq.~(\ref{eq:runningwindowcorrelation}) has been written so that $\tau$ is positive for GW150914.  We will restrict the time delay to $-10\,\text{ms}\leq\tau\leq10\,\text{ms}$ since this is the only region of interest for gravitational wave detection.}  Here, ${\rm Corr}(x,y)$ is the standard Pearson cross-correlation function between records $x$ and $y$.  In order to reveal the relationship of the correlation function in Eq.~(\ref{eq:runningwindowcorrelation}) with the Fourier phases of H and L and their time lag, we make use of the cross correlation theorem.  For simplicity, we assume that the H and L data have been normalized so that their mean value is zero and their standard deviation is unity\footnote{Note that the last term in Eq.~(\ref{equ:cc_and_phase}) illustrates the importance of the phase correlations. The full range of investigation has been done in the TOD domain and presented in Appendix~\ref{app:chaning the lower limit of bp}}:
\begin{eqnarray}\label{equ:cc_and_phase}
C(t,\tau,w) = && \int_t^{t+w} H(t'+\tau)L(t')dt' \nonumber\\
= && \int_t^{t+w} dt'\int d\omega \int d\omega' \cdot H(\omega)e^{i\omega (t'+\tau)}  L^*(\omega')e^{-i\omega' t'}\\
\approx && \sum_k |H(\omega_k)|\cdot|L(\omega_k)|\cos(\theta_k-\psi_k+\omega_k \tau),\nonumber
\end{eqnarray}
where $k$ is the index of the frequency components, and $\theta_k$ and  $\psi_k$ are the Fourier phases of $H(\omega_k)$  and $L(\omega_k)$, respectively. One can see from Eq.~(\ref{equ:cc_and_phase}) that, if the H and L phases $\theta_k$ and $\psi_k$ are uncorrelated, the terms in the sum will tend to average to zero.  A correlation in $\theta_k$ and $\psi_k$ will significantly increase the probability of a high cross-correlation coefficient.

There are two distinct questions that must be addressed in considering a gravitational wave event.  The first involves the demonstration that morphologically similar events have been seen by the two detectors and that the time delay for their observation is within the allowed $\pm 10$\,ms window, e.g., as noted by LIGO, the strongest correlation (and the largest magnitude of $C(t,\tau,w)$) for GW150914 is found for $\tau=6.9$\,ms.  Once the possible astrophysical origin of the event has been verified, comparisons with the predictions of General Relativity can provide guidance regarding the physical nature of the event.  The analysis performed by the LIGO team has tended to merge these questions by its use of templates in the identification of the event.  As we have shown, however, identical signals can be extracted without the use of templates \citep{Naselsky16}.  The LIGO collaboration has made tremendous efforts in searching for and quantifying possible terrestrial or astrophysical influences on their instruments, specifically for the time of GW150914~\citep{LigoTransientNoise}.  The effects of sources with known or unknown physical couplings to the instruments have been recorded in ``over 200,000 auxiliary channels" in order to exclude correlations with the event's characteristics.  Their analysis revealed that there ``is no evidence for instrumental transients that are temporally correlated between the two detectors"~\citep{Ligo1}.  However, an additional point must be considered.  As a consequence of this result, it is assumed that, in the absence of a gravitational wave event, the records of the Livingston and Hanford detectors will be uncorrelated.  Therefore, the possibility that there are any external or internal mechanisms, manifestly independent of gravitational waves, that can produce ``accidental" correlations must also be completely excluded as a potential source of observed correlations.

The following subsections will offer indications that such non-GW correlations are present in the GW150914 data.  While we do not claim that there is a connection between these ``accidental" correlations and the GW event itself, their mere presence challenges the integrity of the claim of gravitational wave detection.

\subsection{Calibration lines}
\label{sub:calibrationlines}

It is obvious that narrow resonances in the LIGO detectors --- due to mechanical resonances, power line harmonics, or calibration lines --- are unrelated to gravitational wave signals.  Since there is no causal connection between the properties of narrow resonances in the Hanford and Livingston detectors, it might seem reasonable to expect that correlations between them cannot be enhanced by introducing a time delay.   However, in this section we will show that correlations between calibration lines in the vicinity of 35\,Hz can be enhanced significantly by introducing a $\approx$7\,ms time delay.

\begin{figure}[h!]
\centering
\includegraphics[width=0.45\textwidth]{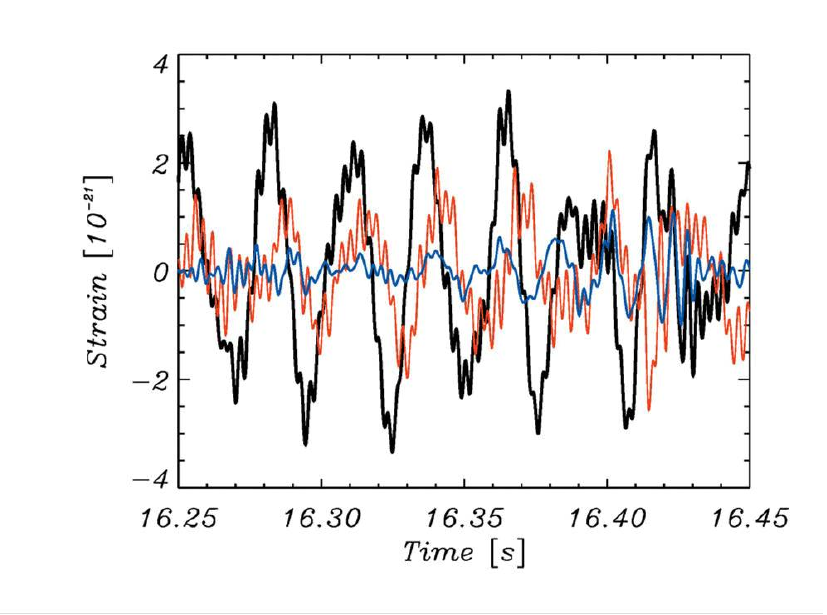}
\includegraphics[width=0.45\textwidth]{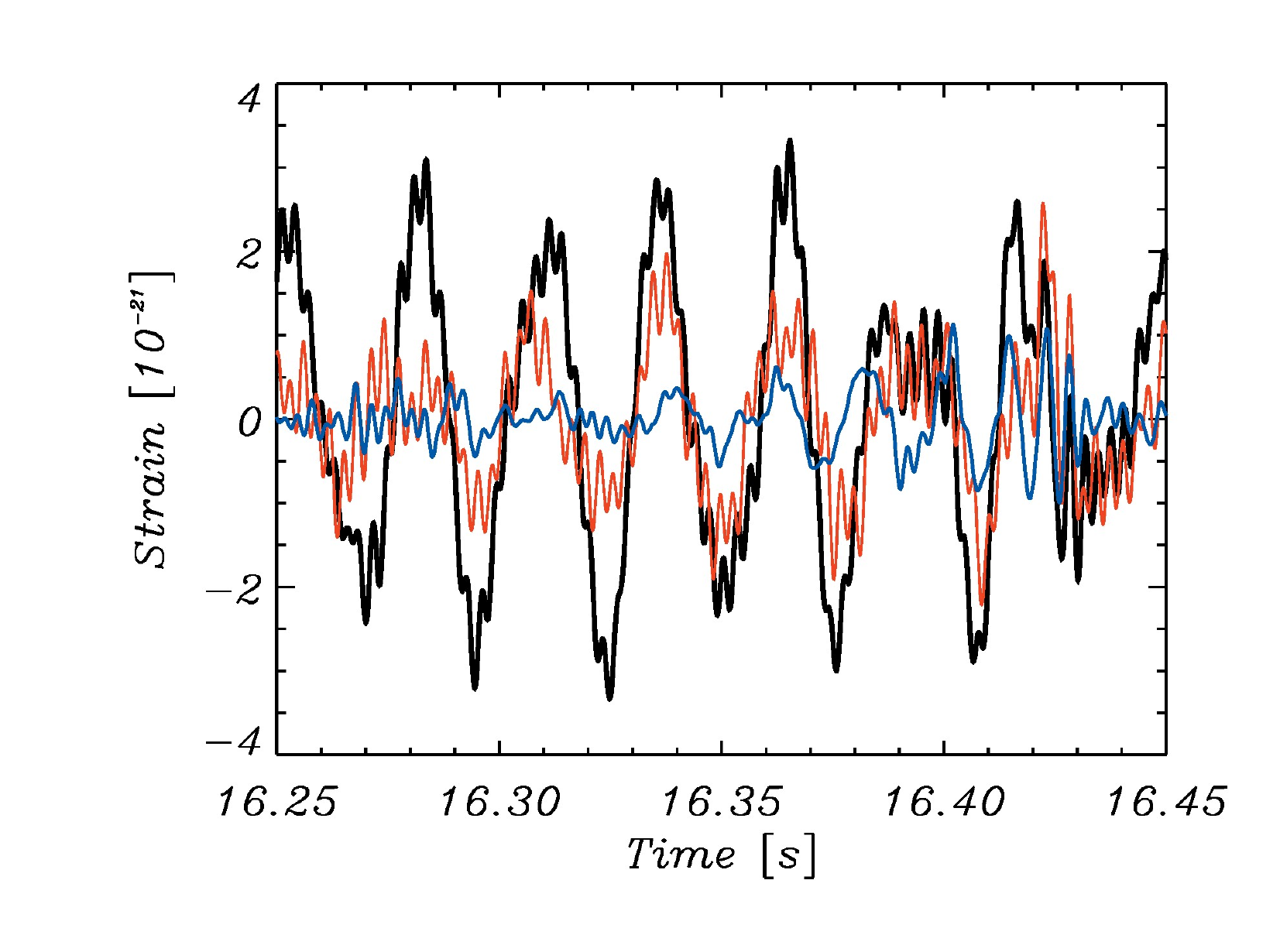}
\includegraphics[width=0.45\textwidth]{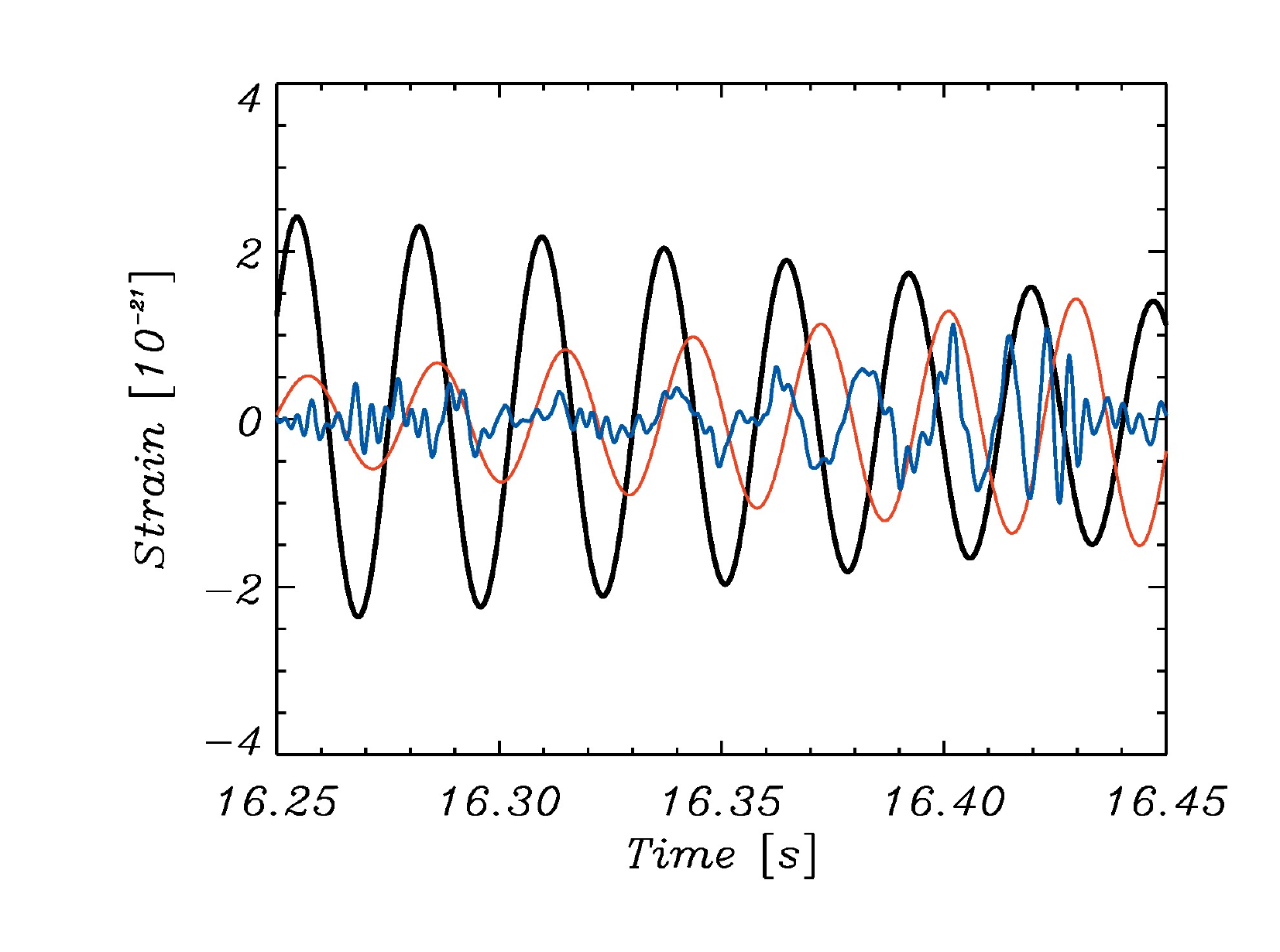}
\includegraphics[width=0.45\textwidth]{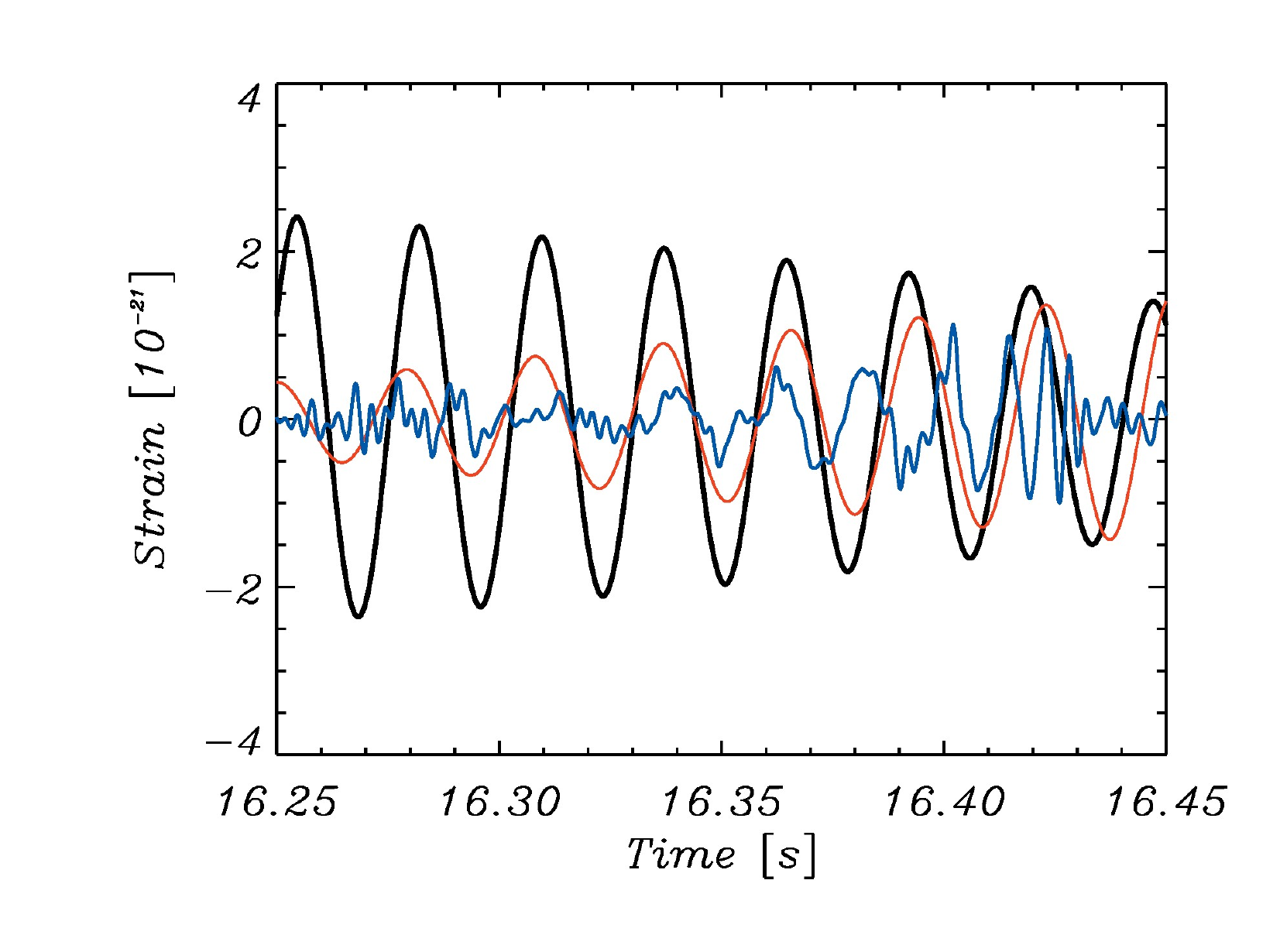}
\caption{Comparison of the calibration line signals from Hanford (black) and Livingston (red) in the vicinity of GW150914 (blue) before (left panels) and after (right panels) shifting the Livingston record by 7\,ms and inverting it in the same way as done for the identification of GW150914.  Top panels: Band-passed in the range $30 < f < 350$\,Hz.  Bottom panels: Only the four narrow resonances (calibration lines) shown in the right panel of Fig.~\ref{fig:fourieramplitudes}.}
\label{fig:calibrationslinesevent}
\end{figure}

Bearing in mind the effects of non-stationarity, we consider the band-passed TOD for the Hanford and Livingston detectors in the 0.2\,s time window of the GW150914 event.  We deviate slightly from the approach recommended by LIGO in that we decrease the lower limit of the band-pass filter from 35 to 30\,Hz in order to retain the contribution from the lowest frequency calibration line (at 34.7\,Hz) of the Livingston detector to the filtered TOD stream. (In this regard, see the right panel in Fig.~\ref{fig:fourieramplitudes}.)  No other filters have been applied to the data.  The top left panel of Fig.~\ref{fig:calibrationslinesevent} shows the reconstructed records for the Hanford and Livingston detectors following band-pass filtering.  The GW150914 event is also shown for comparison purposes.

\begin{figure}[h!]
\centering
\includegraphics[width=0.45\textwidth]{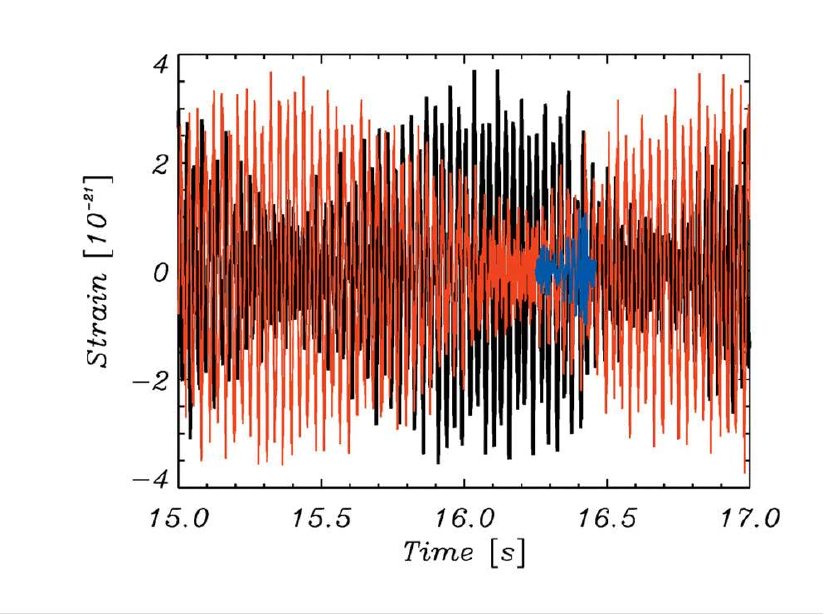}
\includegraphics[width=0.45\textwidth]{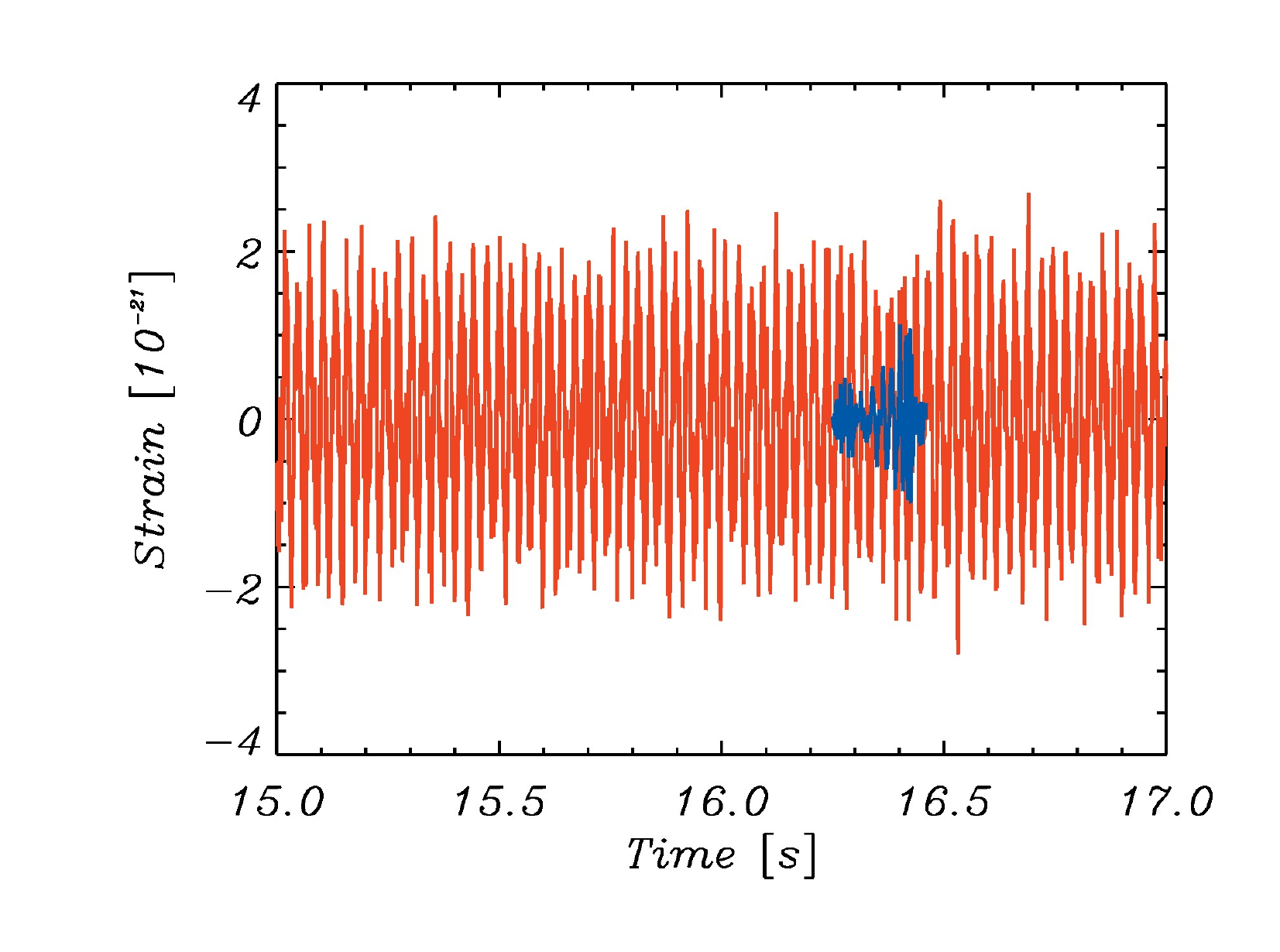}
\caption{Left: Same as top left panel of Fig.~\ref{fig:calibrationslinesevent}, here showing a broader time range for illustration of the beats formed by the calibration lines.  Right panel: Same as left, however the lower limit of the band-pass was moved from 30 to 35\,Hz.  The Hanford record was removed for better visibility of the absence of the beats in the Livingston record due to exclusion of the 34.7\,Hz line.}
\label{fig:calibrationlinesbeats}
\end{figure}

The most striking feature in Fig.~\ref{fig:calibrationslinesevent} is the dramatically improved agreement in the morphology of the Hanford and Livingston records when we perform precisely the same $\approx$7\,ms shift (and inversion) that led to the identification of the GW150914 event, see top right panel of the same figure.  As we shall indicate below, this result is remarkably robust.  Both signals are dominated by strong, low frequency oscillations, which, due to their strong prominence (see Fig.~\ref{fig:fourieramplitudes}), not surprisingly turn out to be the two calibration lines in each of the detectors.  The bottom panels of Fig.~\ref{fig:calibrationslinesevent} show only their contribution, both before and after shifting and inverting the Livingston record.  Likewise unsurprisingly, the left panel of Fig.~\ref{fig:calibrationlinesbeats} shows that the segment shown so far is actually only part of a somewhat more complex interference effect with modulation frequencies 0.8\,Hz (for Hanford) and 0.6\,Hz (for Livingston).  Setting the lower limit of the band-pass filter at 35\,Hz as suggested by LIGO would result in a reduced contribution from the 34.7\,Hz line in the Livingston detector and would thereby tend to alter the interference pattern, as shown in the right panel of Fig.~\ref{fig:calibrationlinesbeats}.  We further note that the four calibration frequencies involved are all rationally related (with an accuracy limited only by line width) and, curiously, each frequency is one-tenth of a prime number.  This means that the results shown in bottom panels of Fig.~\ref{fig:calibrationslinesevent} will be periodic with a period of 10\,s.

\begin{figure}[h!]
\centering
\includegraphics[width=0.45\textwidth]{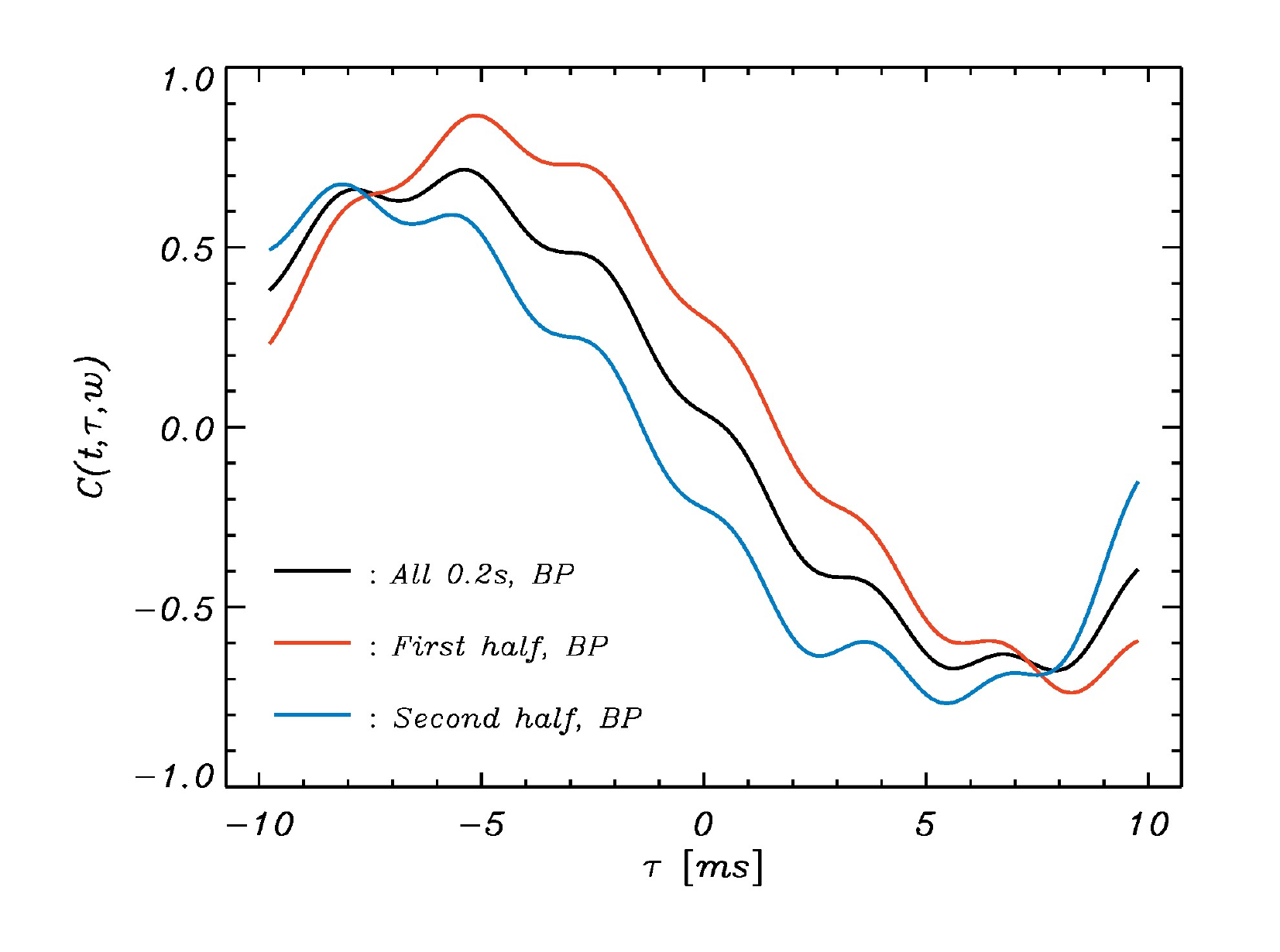}
\includegraphics[width=0.45\textwidth]{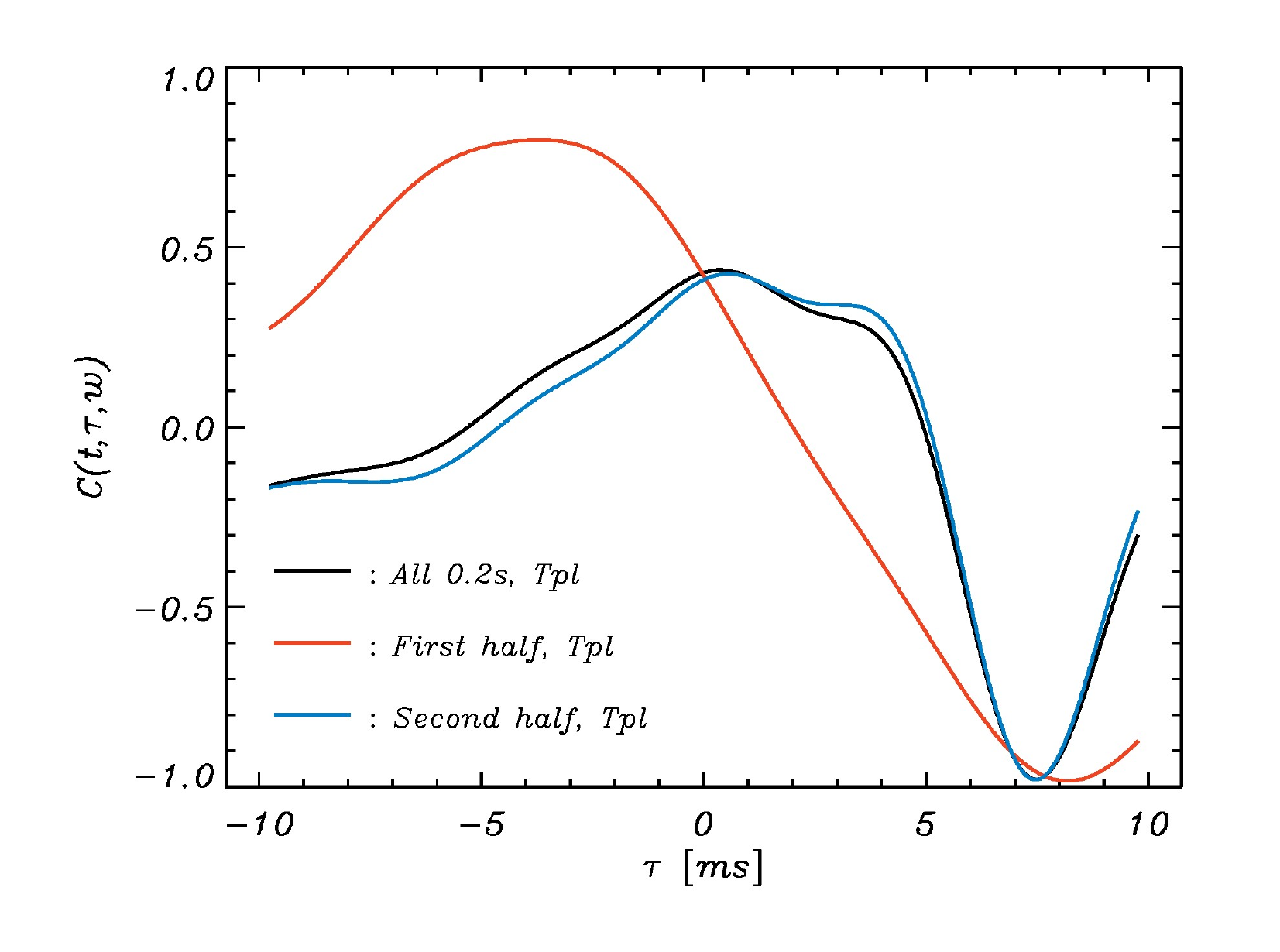}
\caption{Left: The cross-correlation $C(t,\tau,w)$ for the band-passed signal presented in the top panels of Fig.~\ref{fig:calibrationslinesevent} versus time lag $\tau$ for various windows within the 0.2\,s shown in the same figure.  Since the amplitude of GW150914 is small compared to the full signal, subtraction of GW150914 before computing the correlation function has effectively no impact.  Right: The cross-correlation of the Hanford and Livingston GW150914 templates in the same time windows for comparison.}
\label{fig:calibrationlinescorrelation}
\end{figure}

In order to investigate these interference effects more carefully, we calculate the cross-correlation $C(t,\tau,w)$, defined in Eq.~(\ref{eq:runningwindowcorrelation}), of the band-passed H and L data ($30 < f < 350$\,Hz) for various window sizes within the overall 0.2\,s window of GW150914, see Fig.~\ref{fig:calibrationlinescorrelation}.  It is important to note that the event GW150914 was not removed from the strain data.  However, due to its relatively small amplitude, its removal does not affect the result significantly.  For purposes of comparison, the right panel shows the cross-correlation between the theoretical H and L templates in the same time windows.  Due to the almost perfectly harmonic oscillations of the band-passed signal in the 0.2\,s domain, the cross-correlation is almost antisymmetric under the replacement $\tau \to -\tau$.  This effect is not seen for the GW templates.  We note that we have also performed (but do not show) a similar calculation retaining \emph{only} the narrow resonances listed in Section~\ref{sec:data}.  The resulting cross-correlations are again very similar to those shown in Fig.~\ref{fig:calibrationlinescorrelation}.\\

\subsection{Residual noise}
\label{sub:residual}

\begin{figure}[h!]
\centering
\includegraphics[width=0.42\textwidth]{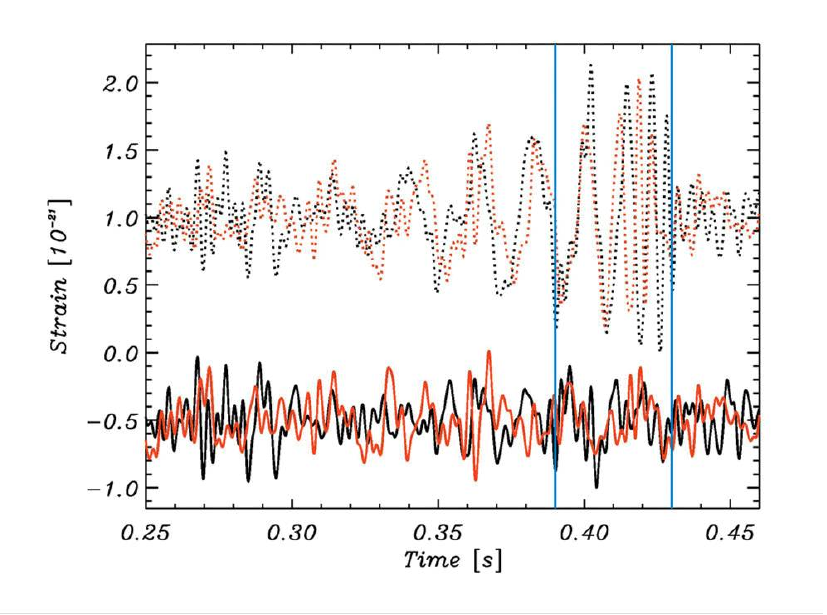}
\includegraphics[width=0.42\textwidth]{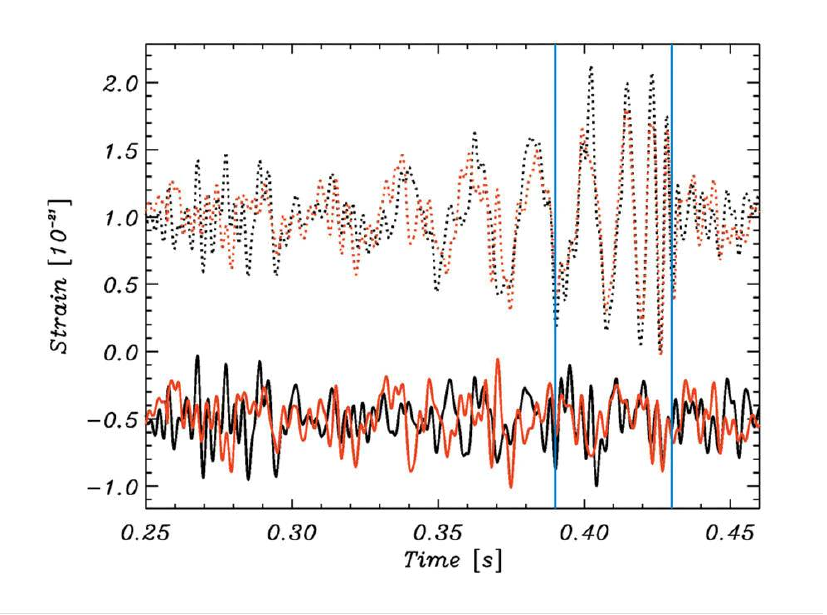}
\includegraphics[width=0.42\textwidth]{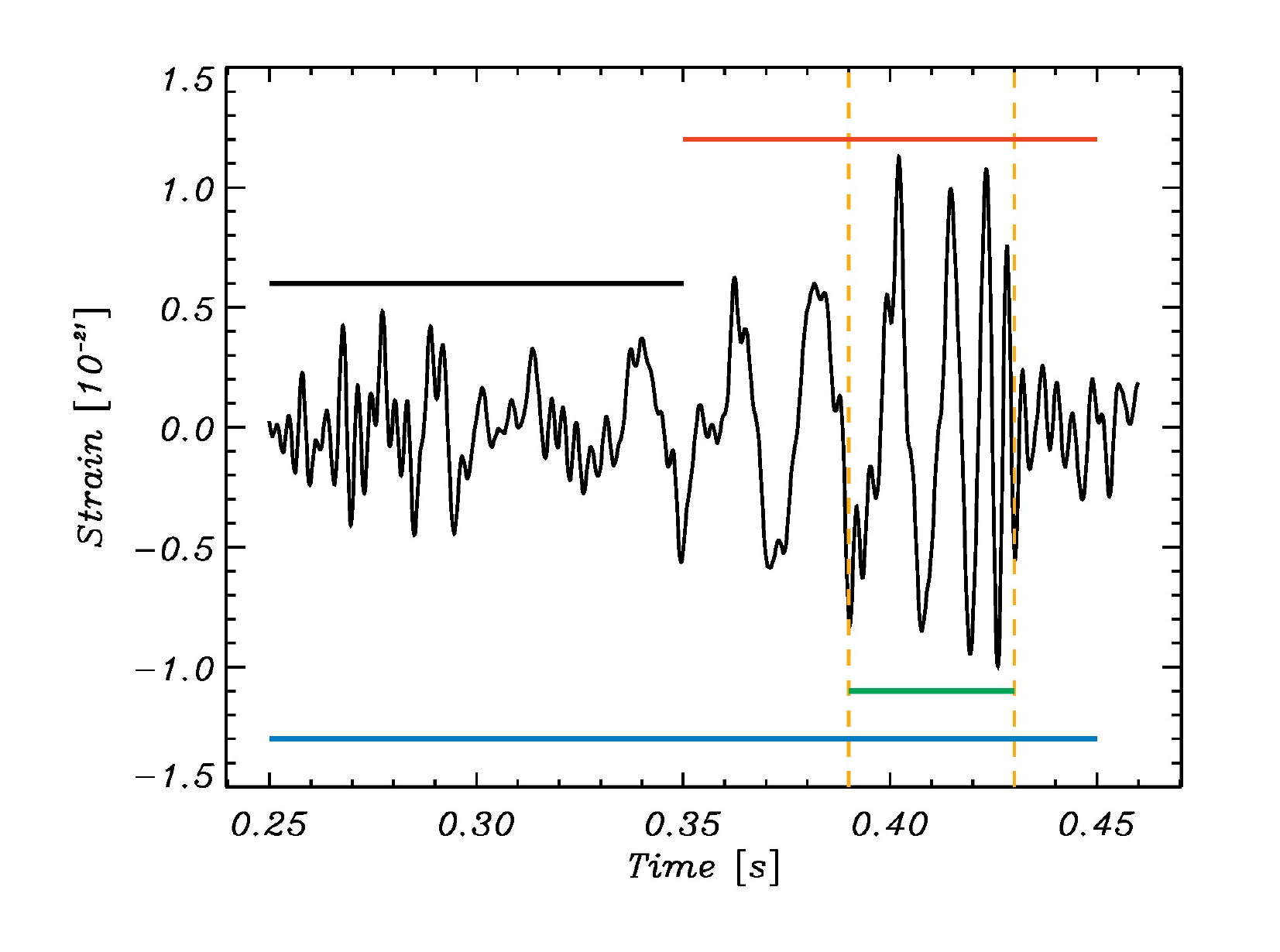}
\includegraphics[width=0.42\textwidth]{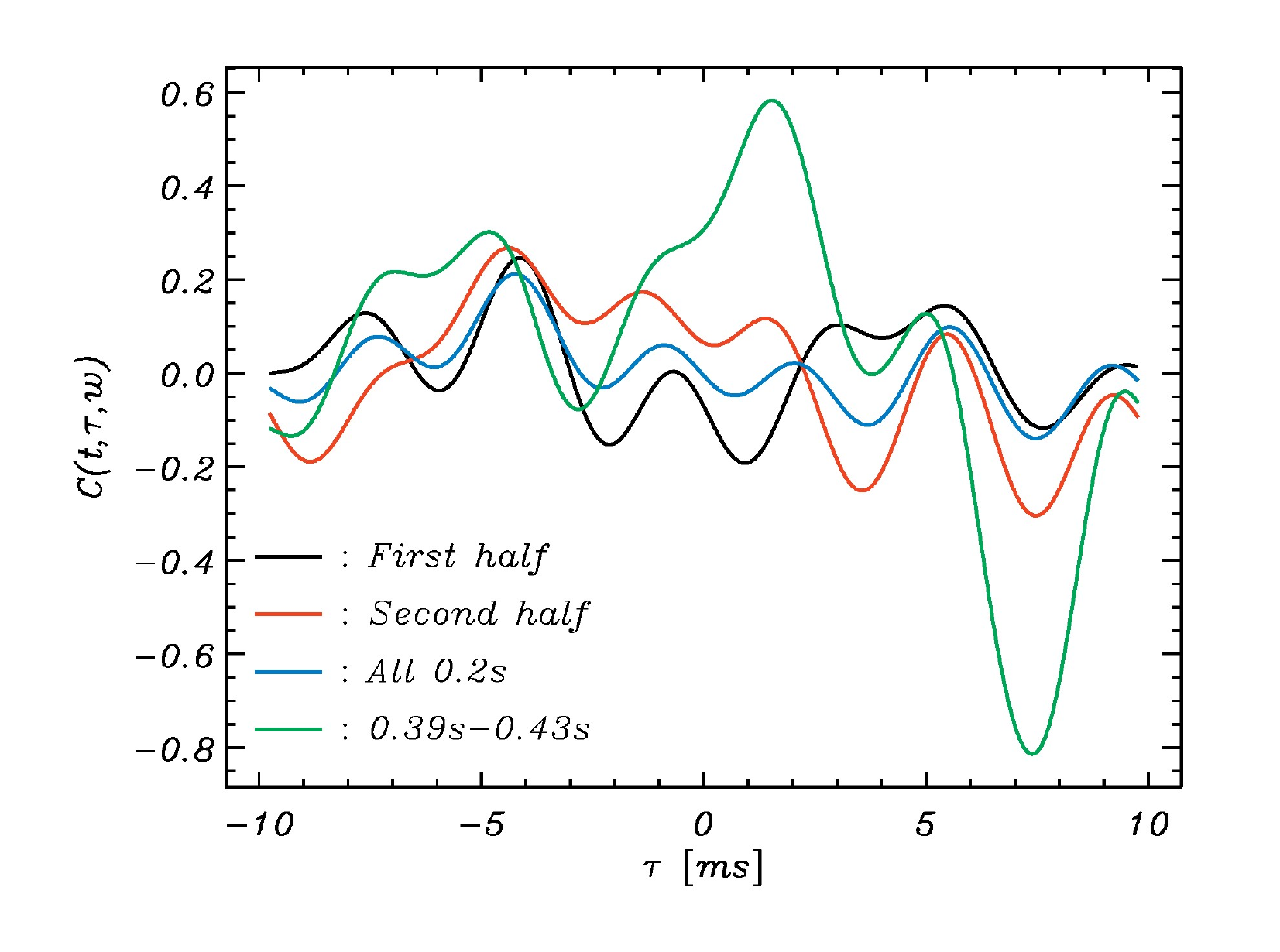}
 \caption{Top panels: The Hanford (black) and Livingston (red) records, $H$ and $L$, as well as their residuals, $H_n$ and $L_n$, after subtraction of the respective templates before (left) and after (right) shifting the Livingston record by 7~ms and inverting it.  Bottom right panel: The cross-correlation $C(t,\tau,w)$ for the noise records $H_n$ and $L_n$ as a function of the time delay, $\tau$, for various time windows as indicated in the panel to its left.}
\label{fig:residualcorrelation}
\end{figure}

We begin our search for correlations in the residual noise with a brief test using the data made publicly available provided by the LIGO collaboration (see Appendix~\ref{app:chaning the lower limit of bp} for details): In the 0.2\,s window including the GW event, the noise components are calculated as $H_n=H-H_{tpl}$ and $L_n=L-L_{tpl}$.  Here, $H_{tpl}$ and $L_{tpl}$ are the templates cleaned in the same manner as the raw data is treated to yield the clean records, $H$ and $L$.  The cross-correlation of $H_n$ and $L_n$ is calculated according to Eq.~(\ref{eq:runningwindowcorrelation}) as a function of the time delay~$\tau$.  The results of this calculation are shown in Fig.~\ref{fig:residualcorrelation} for a variety of time intervals, $w$ (as marked in the figure).  While there was no reason to expect that there would be any significant correlation between these noise signals, they reveal a significant cross correlation for a time delay\footnote{The negative cross-correlation accounts for the inversion to be performed in the Livingston record.} of approximately 7\,ms for all time windows considered.  For purposes of comparison, we note that the cross-correlation between the cleaned Hanford and Livingston data lies between $-0.12$ and $-0.81$ for the four ranges shown in Fig.~\ref{fig:residualcorrelation}.  It must be noted, that the template used here is the maximum-likelihood waveform.  However, a family of such waveforms can be found to fit the data sufficiently well (see e.g. panels in the second row of Fig.\,1 in Ref.~\citep{Ligo1}).  To provide a rough estimate of this uncertainty, we have also considered the possibility of a free $\pm10\%$ scaling of the templates, so that  $H_n=H-(1 \pm 0.1)H_{tpl}$ and $L_n=L-(1 \pm 0.1)L_{tpl}$.  The results are nearly identical to those of Fig.~\ref{fig:residualcorrelation}.  Considering that the residual noise is significantly greater than the uncertainly introduced by the family of templates, this is not surprising.  It would appear that 7\,ms time delay associated with the GW150914 signal is also an intrinsic property of the noise.\newline

\subsection{A null output test}
\label{sub:nulloutput}

We now consider a similar but somewhat more elaborate approach in the form of a so-called \textit{null output test}.  We define this null output test to be an analysis of a data set that does not contain the input signal, with the aim of demonstrating that the behavior of the output has nothing in common with the desired signal.

\begin{figure}[h!]
\centering
\includegraphics[width=0.42\textwidth]{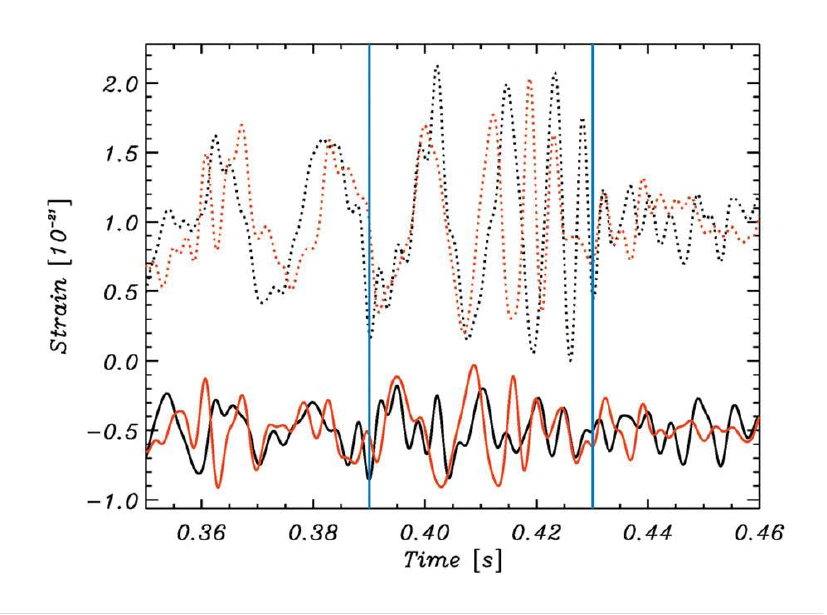}
\includegraphics[width=0.42\textwidth]{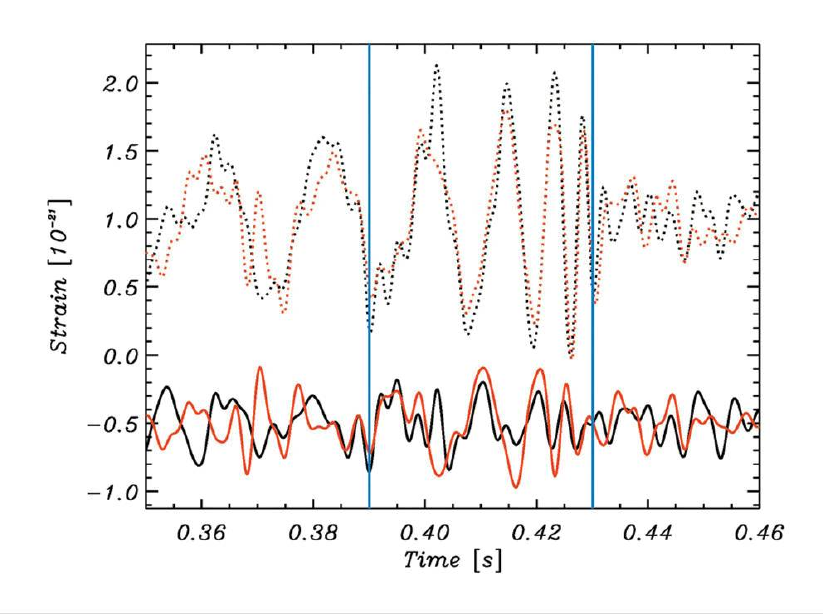}
\includegraphics[width=0.42\textwidth]{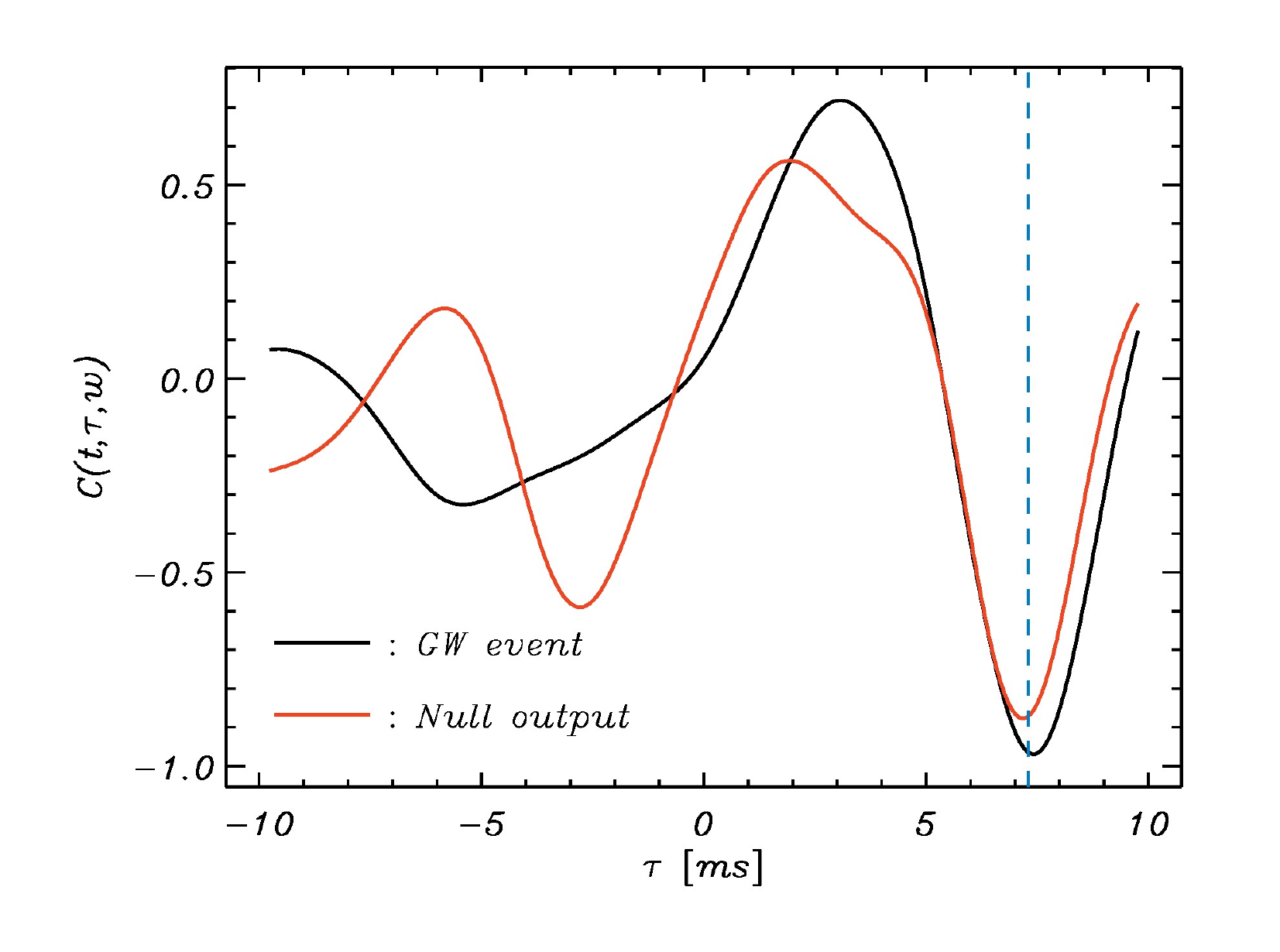}
\includegraphics[width=0.42\textwidth]{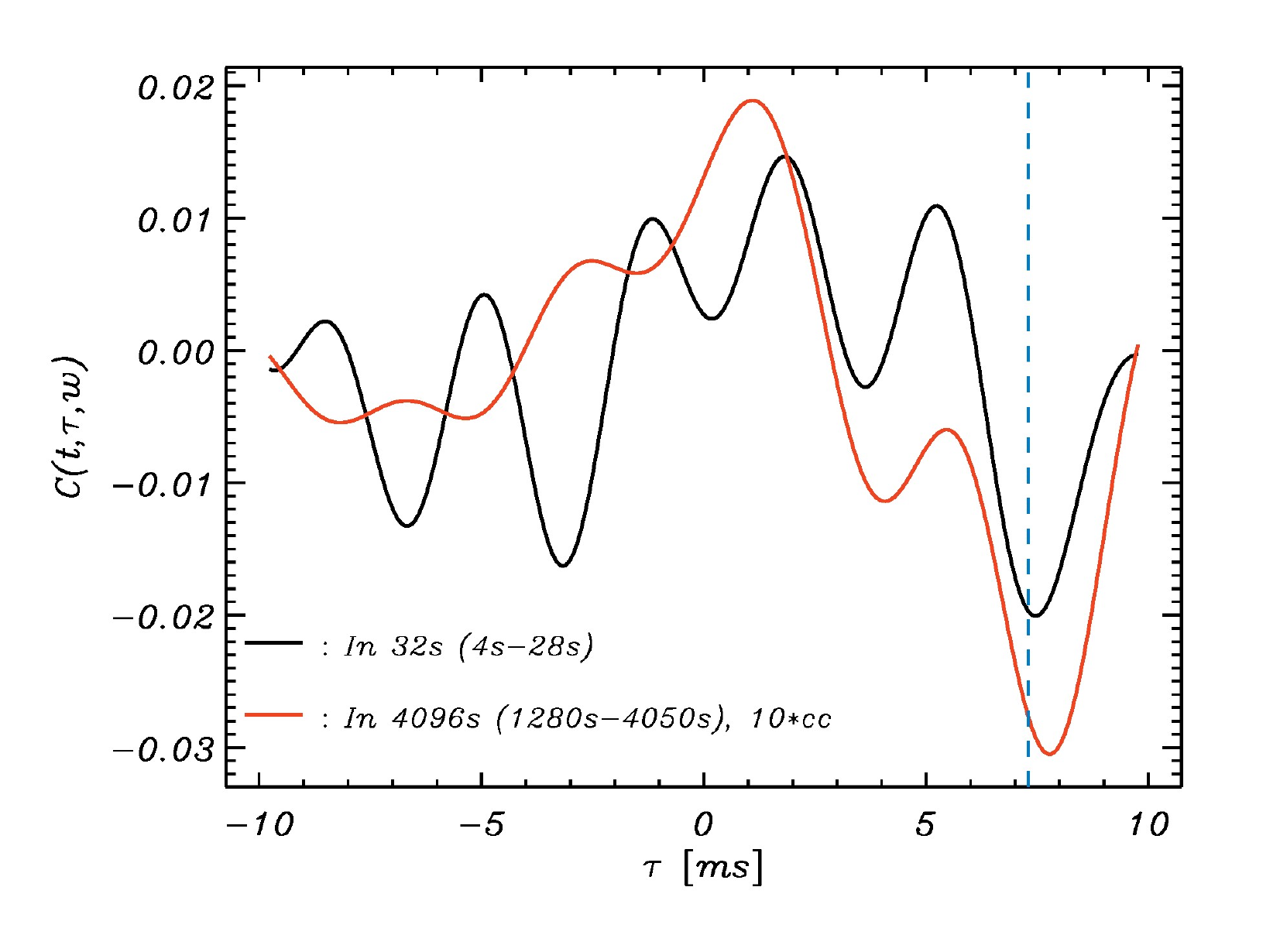}
\caption{Top panels: As top panels in Fig.~\ref{fig:residualcorrelation}, however here the residuals are the null outputs $H_{no}$ (black) and $L_{no}$ (red).  Bottom left panel: The cross-correlation $C(t,\tau,w)$ for the null outputs (red) compared with that of the GW events in the same time range (black).  The time window is marked by the blue vertical lines.  Bottom right panel: The cross-correlation $C(t,\tau,w)$ for the 32\,s null output (see footnote~\ref{edge}) compared with that of a nearly 45 minutes record of cleaned data (at 3.2$\sigma$, see~\citep{Liu16}).  The latter was amplified by a factor of 10 for better comparison.}
\label{fig:nulloutputcorrelation}
\end{figure}

We begin by subtracting the unfiltered theoretical GW150914 template from the 32\,s unfiltered Hanford and Livingston data in order to construct suitable null input data sets, $H_{ni}$ and $L_{ni}$, respectively.  We then clean $H_{ni}$ and $L_{ni}$ and thereby produce the null output data sets, $H_{no}$ and $L_{no}$, which are shown and compared with the GW signal (desired output) in Fig.~\ref{fig:nulloutputcorrelation}.  Ideally, the GW signal should have nothing in common with $H_{no}$ and $L_{no}$.  We then select the central portion of the GW event from 0.39 to 0.43\,s (marked by the blue vertical lines in Fig.~\ref{fig:nulloutputcorrelation}) for which the GW signal is strongest and calculate $C(t,\tau,w)$ for $H_{no}$ and $L_{no}$ by use of Eq.~(\ref{eq:runningwindowcorrelation}) in this time interval, see lower left panel of Fig.~\ref{fig:nulloutputcorrelation}.  It is evident that there is a strong cross correlation between the null output data sets and that this correlation is extremized for the by now familiar inversion and 7\,ms shift of the Livingston data, with the CC-value being $-0.88$.  For comparison, the correlation function of the cleaned data (including the event) of the Hanford and Livingston detectors in the same time range is -0.96 for the same time lag.

For further comparison we also show the correlation function computed for the full 32\,s $H_{no}$ and $L_{no}$ data sets.\footnote{\label{edge}4\,s were removed from both the beginning and end of the records in order to avoid the influence of edge effects.}  These results, shown in the lower right panel of Fig.~\ref{fig:nulloutputcorrelation}, resemble those of the lower left panel obtained with a far smaller time window and provide further support for the findings of our previous work~\citep{Liu16}. We have also investigated the dependence of the cross-correlation function on the low frequency boundary of the band pass filter, which is summarized in Appendix~\ref{app:chaning the lower limit of bp}

We have seen in several different ways that the 7\,ms time lag is not a characteristic unique to the GW150914 event.  Since the time lag analysis plays a central role in confirming that the Hanford and Livingston detectors have seen the same genuine GW event, the results of this section raise the possibility that this confirmation may not be completely reliable.

\section{GW151226 and GW170104}
\label{sec:otherevents}

About two weeks before the end of its first observing run with the Advanced LIGO detectors, the LIGO collaboration discovered GW151226~\citep{Ligo2}, their second gravitational wave detection.  This event had a lower signal to noise ratio but a significance similar to that of the first event.  The time lag at which it was found in the two detectors was $\tau_{GW2}\approx1$\,ms.  Recently, after improving the instruments for the second observing run, they were able to claim a third event~\citep{Ligo3} with a time delay of $\tau_{GW3}\approx-3$\,ms.\footnote{Signs of $\tau_{GW2}$ and $\tau_{GW3}$ following the convention of Eq.~(\ref{eq:runningwindowcorrelation}).}  The question immediately arises whether features similar to those described above can also be found for these events.  We shall focus on the residual noise test described in Sec.~\ref{sub:residual}.

\begin{figure}[h!]
\centering
\includegraphics[width=0.42\textwidth]{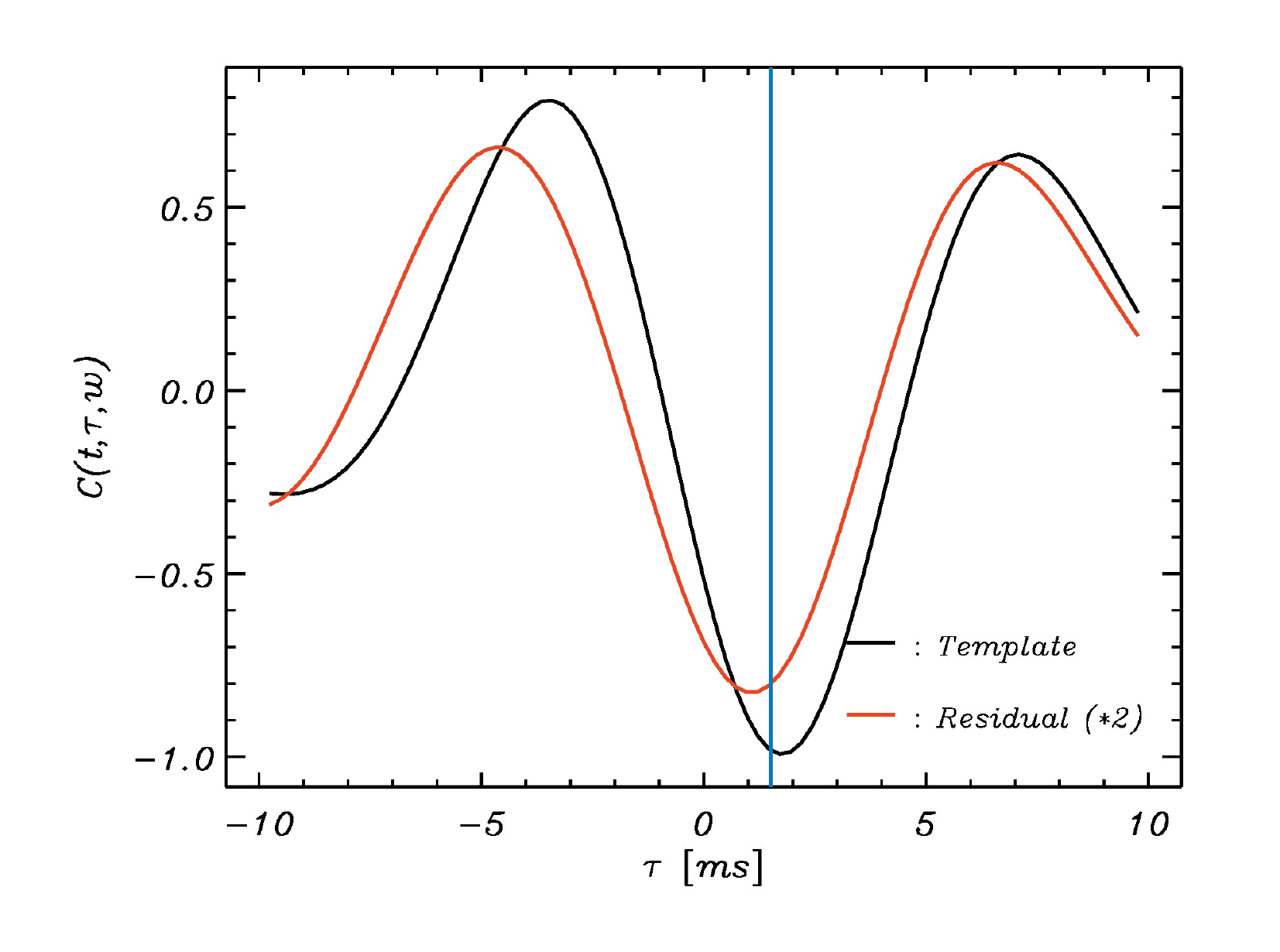}
\includegraphics[width=0.42\textwidth]{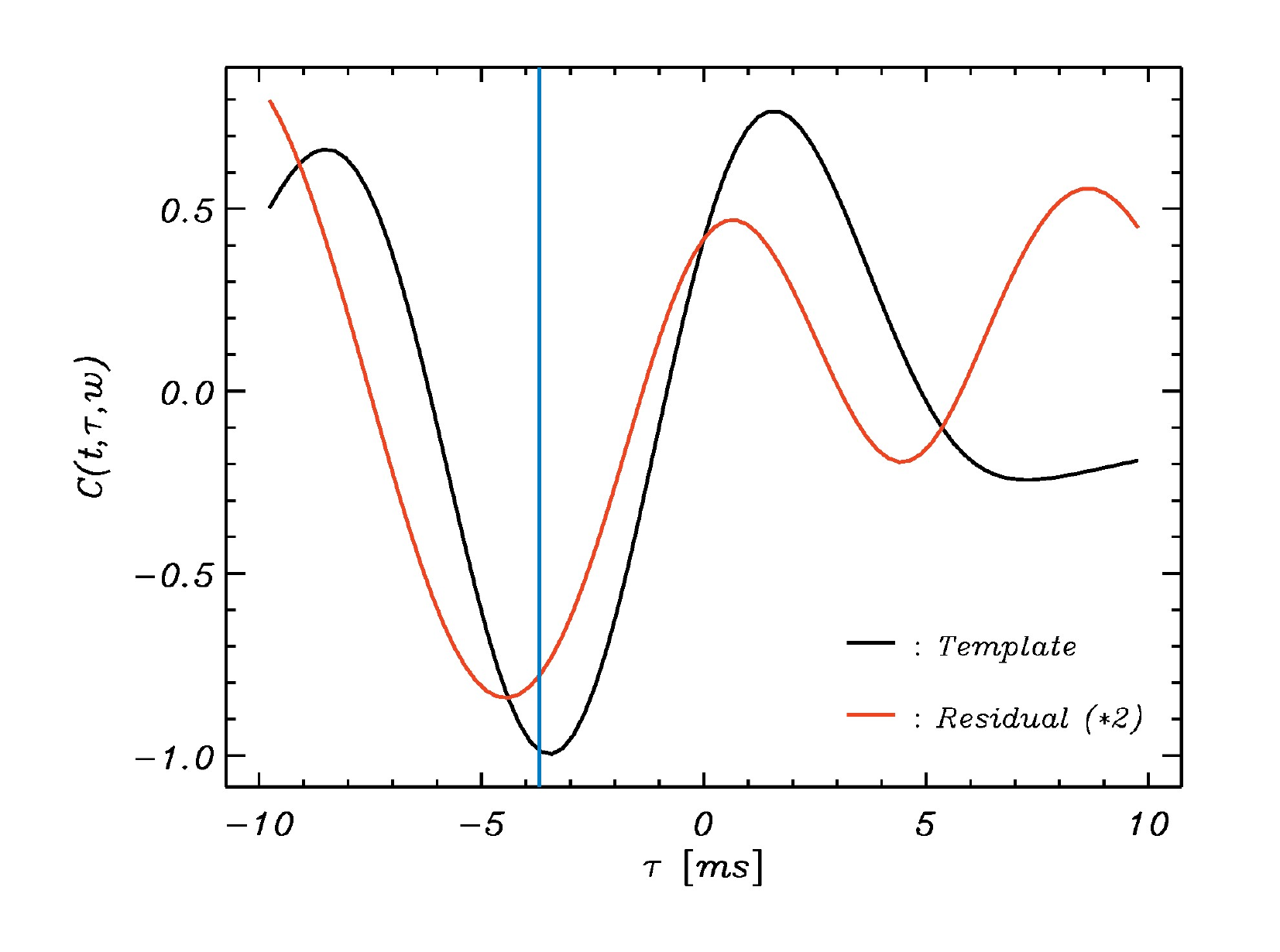}
\includegraphics[width=0.42\textwidth]{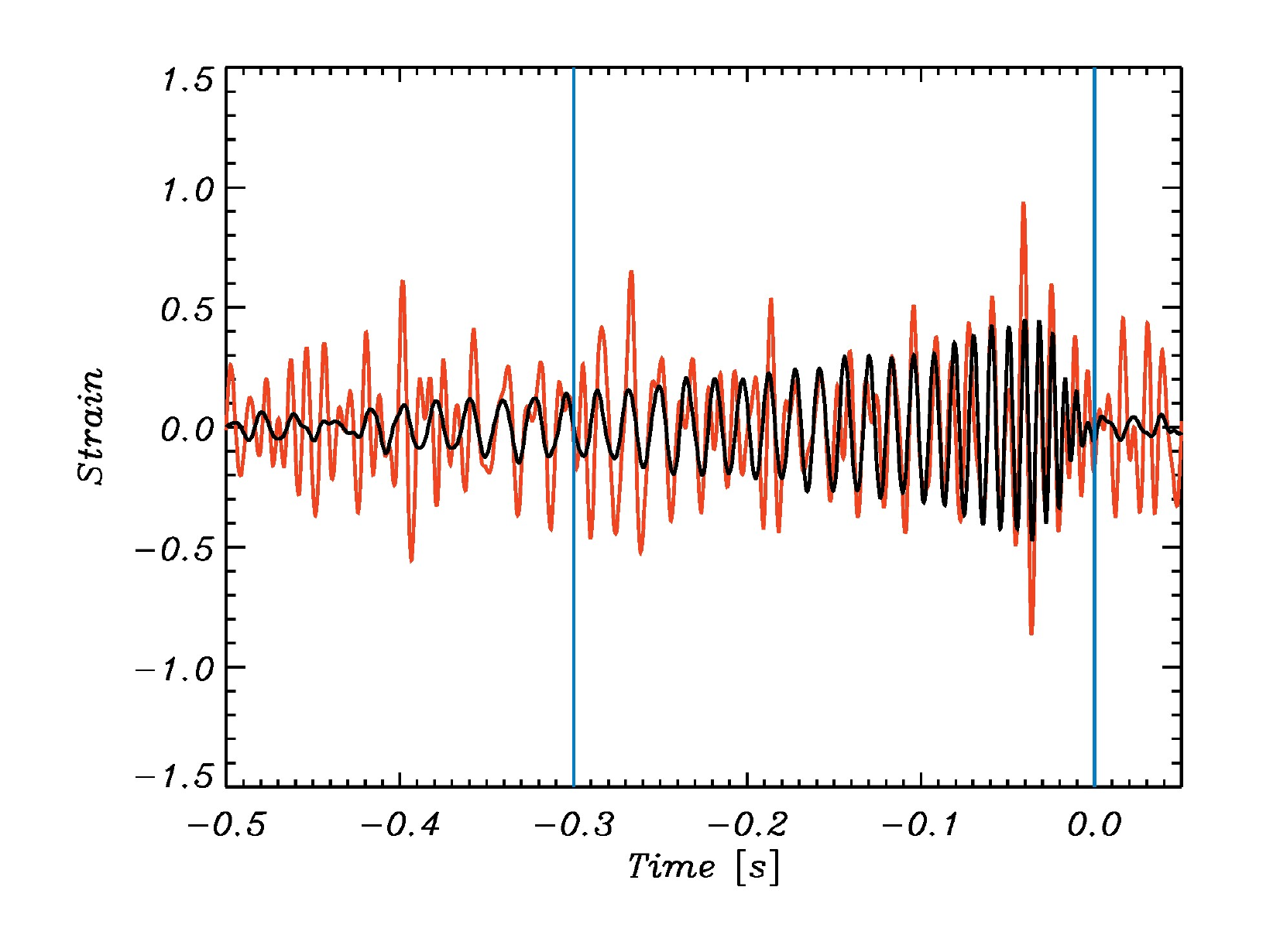}
\includegraphics[width=0.42\textwidth]{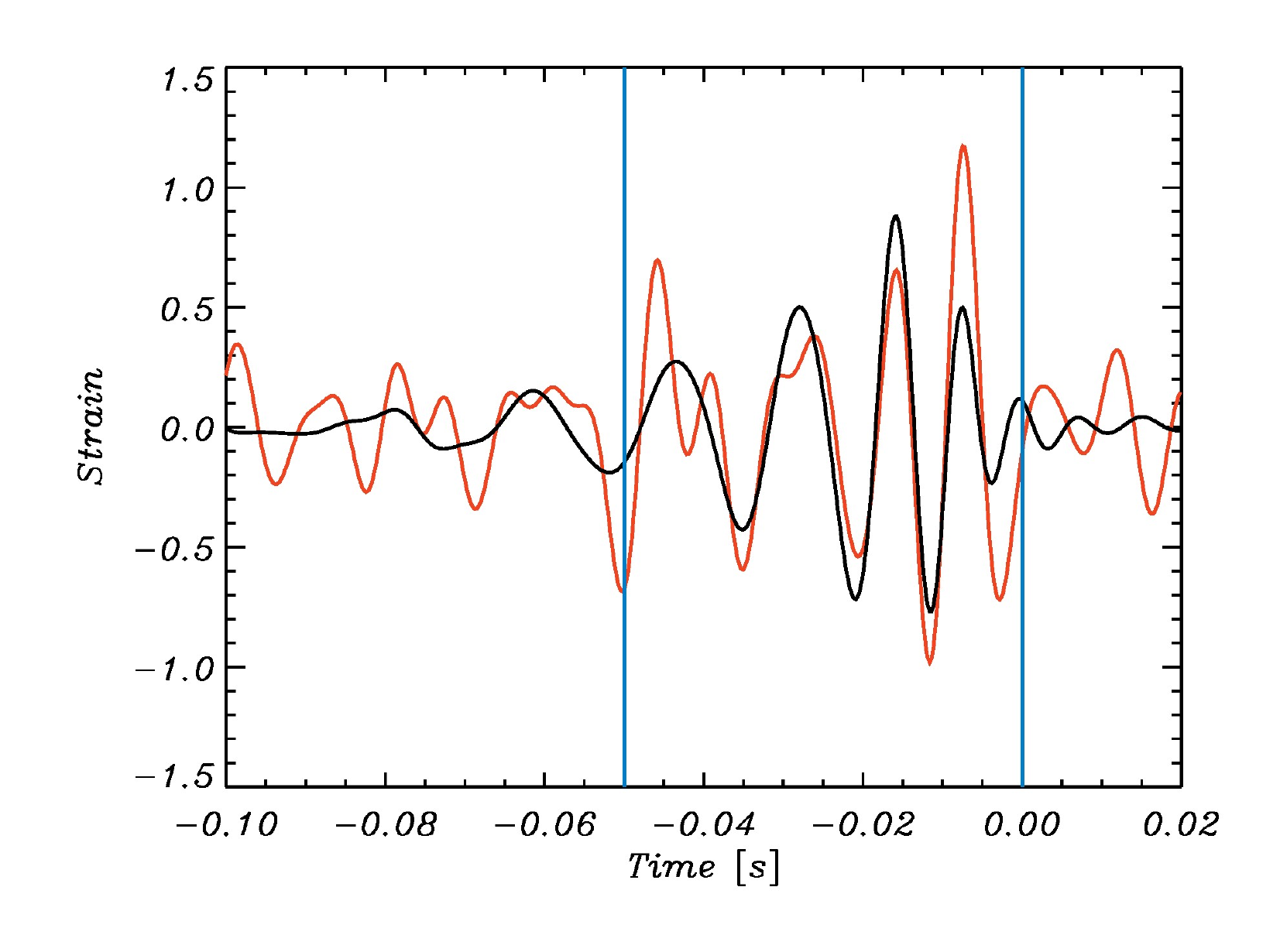}
\caption{Top panels: The cross-correlation function $C(t,\tau,w)$ for the residuals (red) and the theoretical templates (black) calculated for GW151226 (left) and GW170104 (right) in the ranges indicated by the vertical lines in the panels below.  Please note that the correlation functions for the residuals have been multiplied by a factor of two for better comparison.  Bottom panels: Cleaned data (red) compared to the cleaned best-fit templates (black).}
\label{fig:residualotherevents}
\end{figure}

As a first step, we must clean the data surrounding the two events.  We again choose to follow the suggestion by LIGO~\citep{LIGO soft1}.  However, it should be remarked that here both GW151226 and GW170104 have been cleaned by a combination of whitening and band-passing instead of the time-domain filter used above for GW150914.\footnote{Whitening does not generally remove narrow resonances, and thus potentially contaminates the signal.  It is remarked by the LIGO team, however, that various subtleties in their summary were abolished for simplicity.  We expect that the results in this section will remain valid when the LIGO procedure is followed more faithfully.}  As before, we subtract the filtered templates from the cleaned data to obtain estimates of the residual noise in both datasets.  In order to remove the influence of high frequencies that do not play a significant role in the GW templates, we also band pass the residual strains by $50-150$\,Hz.  These strains are then used to compute the correlation function $C(t,\tau,w)$, Eq.~(\ref{eq:runningwindowcorrelation}), in a range indicated in Fig.~\ref{fig:residualotherevents} along with the results.  The correlation functions for the respective events are compared to those obtained from the filtered theoretical templates.\footnote{Note that we previously compared the correlation function of the residual noise with that of the cleaned data.  Here, due to lower signal-to-noise ratios even after cleaning, we choose to use the theoretical templates instead.}

The results of Sec.~\ref{sub:residual} concerning the GW150914 event are now seen to hold for all three events.

\section{Concluding remarks}\label{endgame}

Ideally, the search for gravitational waves should be separated into two independent phases.  An initial template-free step should identify candidate events and demonstrate that they are of astrophysical  origin.  The second step, which will inevitably involve comparison with general relativistic calculations, should attempt to determine the physical nature of the event.  Fortunately, the LIGO GW150914 event is sufficiently strong that it can be seen in both the Hanford and Livingston detectors without templates and that the cross correlation is high.  The evidence that this event is astrophysical lies primarily in the fact that this cross correlation is maximized by an inversion of the Livingston data and a 7\,ms shift of the record that is within the allowed $\pm 10$\,ms window.  The results of Section~\ref{sec:correlations} suggest, however, that similarly strong agreement between the Hanford and Livingston detectors can be obtained from time records constructed exclusively from narrow resonances in the Fourier transform of the data.  In spite of efforts to ``cleanse'' the data of the effects of these resonances, their strength renders it difficult to be certain that there has not been significant ``leakage'' of these effects to neighboring frequencies.  The strong and unexpected correlations in the phases of the Fourier coefficients noted in Section~\ref{sec:data} may be indicative of such leakage.

It has been reported for both GW150914 and for GW170104 that the residual noise following from the subtraction of the template from the cleaned data is consistent with Gaussian noise and does not contain features characteristic of gravitational wave signals.  (See~\citep{Ligo1} and~\citep{Ligo3}, respectively)  This is taken to imply that there are no biases in the modeling of the waveforms.  While our findings do not contradict the previous statement about near Gaussianity during the time of the events, this is to be contrasted with
the present demonstration that the residuals show apparent correlations between the detectors.  It is striking that these correlations are maximized by applying nearly the same time shifts as found for the GW events themselves  --- for all three GW events reported to date.

The purpose in having two independent detectors is precisely to ensure that, after sufficient cleaning, the only genuine correlations between them will be due to gravitational wave effects.  The results presented here suggest this level of cleaning has not yet been obtained and that the detection of the GW events needs to be re-evaluated with more careful consideration of noise properties.

\acknowledgments
%The authors thank Ole Ulfbeck and Viatcheslav Mukhanov for numerous valuable discussions.  This research has made use of the LIGO software package and data.  This work was funded in part by the Danish National Research Foundation (DNRF) and by Villum Fonden through the Deep Space project.  Hao Liu is supported by the National Natural Science Foundation for Young Scientists of China (Grant No. 11203024) and the Youth Innovation Promotion Association, CAS.

The authors thank Ole Ulfbeck and Viatcheslav Mukhanov for numerous valuable discussions as well Peter Coles, whose critical reading of the manuscript was greatly appreciated. We are particularly indebted to the anonymous JCAP referees whose insights, questions and critical comments led us to add the Appendix to the present version of this article. This work has made use of the LIGO software package and data. Our research was funded in part by the Danish National Research Foundation (DNRF) and by Villum Fonden through the Deep Space project. Hao Liu is supported by the National Natural Science Foundation for Young Scientists of China (Grant No. 11203024) and the Youth Innovation Promotion Association, CAS. We are grateful to A.Waananen for help in design of our Web site: \url{http://www.nbi.ku.dk/gravitational-waves/}

\newpage

\appendix

\begin{flushleft}
\textbf{\Large{Appendix}}
\end{flushleft}

\section{General remarks}
We are grateful to the referees for very important remarks that allow a better understanding of the properties of the LIGO data analysis as well as the cross-correlations of the residuals discovered in our paper. Actually, many of the points raised by the referees should be addressed to the LIGO team, and the corresponding details of the data analysis need to be made publicly available. Unfortunately, this has not yet been done. This is why we present our analysis of the problem of the band pass filter and the calibration lines, the phase correlations, the issue of apodization etc. An important note is that we have used the cleaning technique proposed by LIGO. We use this approach for the 32 s record and tested it for the 4096 s record. In the GW150914 domain our reconstructed signals in Hanford and Livingston are characterized by a relative error $3\times10^{-4}$ in respect to the publicly available cleaned data for the same domain. We would like to point out that our detection of the Hanford and Livingston cross-correlations is based on the official LIGO data, and the usage of the LIGO templates has to be validated by the LIGO team. We also note that, for GW150914, the signal is sufficiently strong to be detected by various cleaning techniques. However, even for this event the theoretical template was used for matched filtering and detection of the GW-event from the time ordered  4096 s data (see Ref.~[1]).

The use of theoretical templates is not the best way to understand the properties of detections and residuals. However, for the remaining 3 events detected by LIGO, these theoretical templates are "vital" not only for investigation of the residuals, but for the detection of the events.
The present discussion of Hanford-Livingston noise cross-correlations is based on the LIGO data presented in:

\textbf{\url{https://losc.ligo.org/s/events/GW150914/P150914/fig1-residual-H.txt}}

\textbf{\url{https://losc.ligo.org/s/events/GW150914/P150914/fig1-residual-L.txt}}

\begin{figure}[tbh!]
  \centering
  \includegraphics[width=0.48\textwidth]{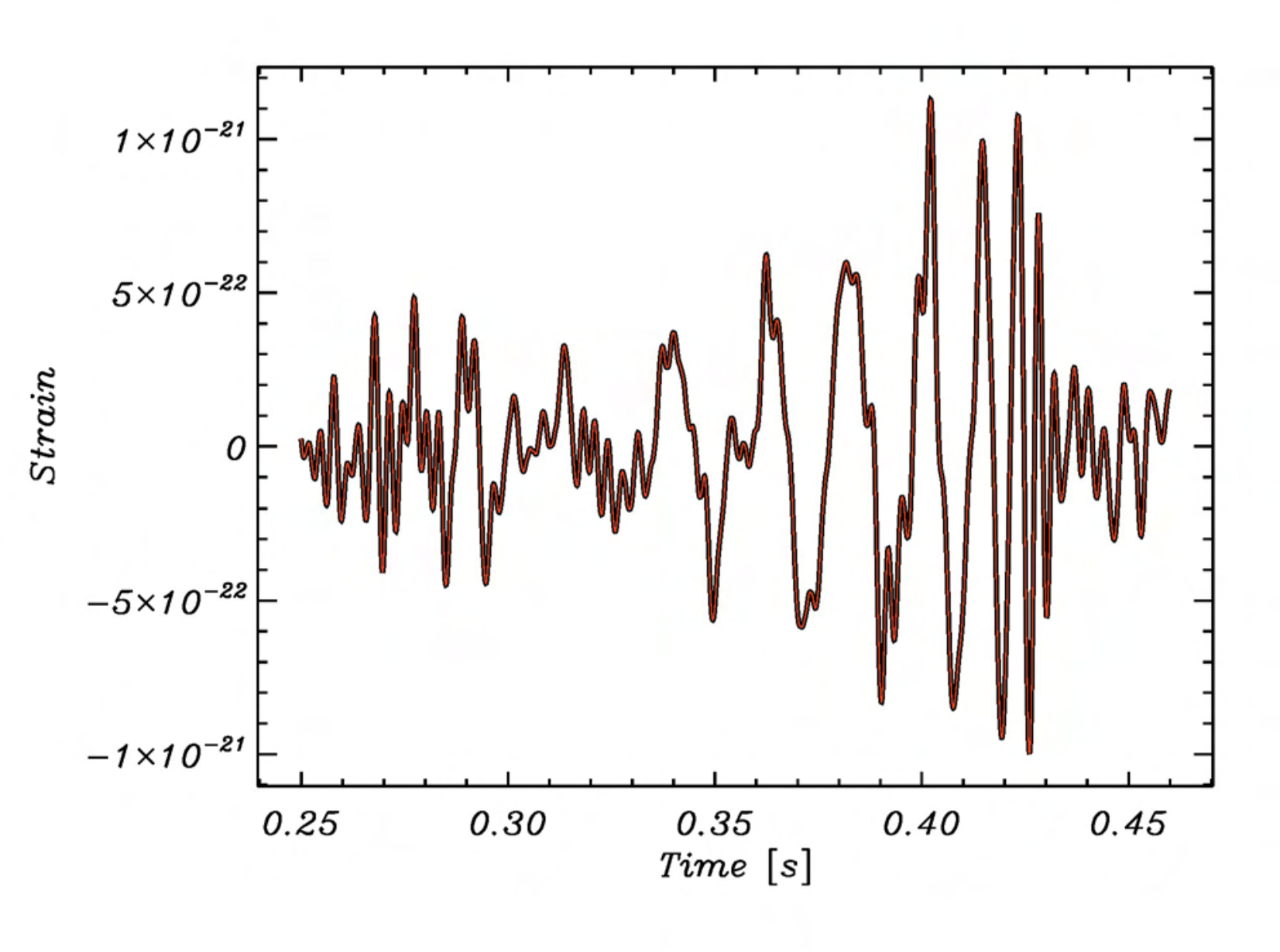}
  \includegraphics[width=0.48\textwidth]{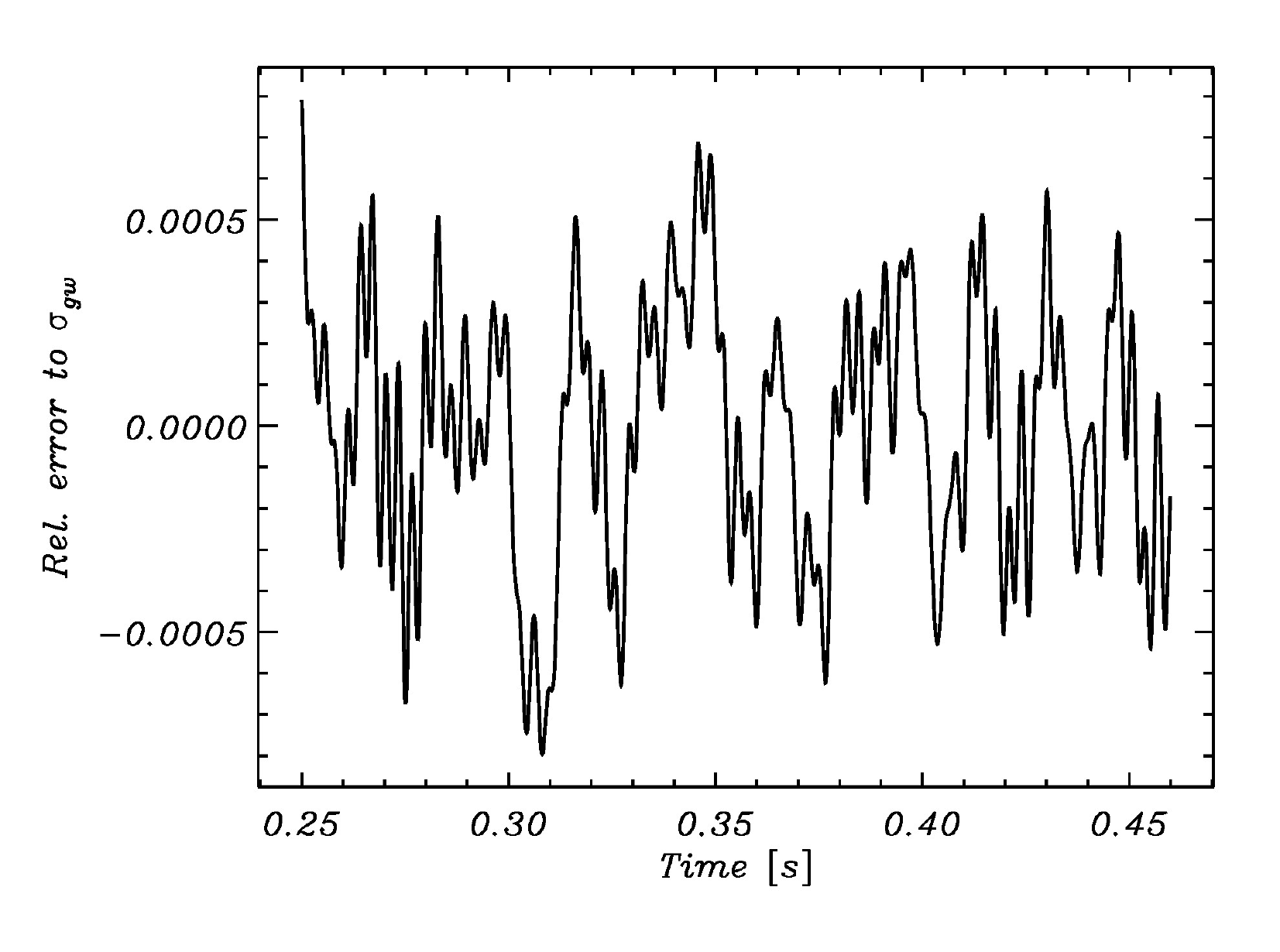}
  \caption{Left: The comparison of the 0.2 s GW signal for the Hanford detector reconstructed by us from the cleaned 32 s record (red) and by LIGO (black). Right: the error between the two curves relative to $\sigma_{gw150914}$. We have $\sigma_{error}/\sigma_{gw150914}=3.3\times10^{-4}$.}
\label{fig0}
\end{figure}

Some of the topics to be considered require cleaned 32 s data, for which LIGO's results are not publicly available. This is why, in Fig.~\ref{fig0}, we show the 0.2 s cleaned GW150914 signal constructed from our cleaned 32 s record and compare it with the publicly available 0.2 s LIGO data. The band pass filter is applied for 35-350 Hz.

\section{Some issues in cleaning the 32 s record}\label{app:phase mixing}

\begin{figure}[tbh!]
  \centering
  \includegraphics[width=0.42\textwidth]{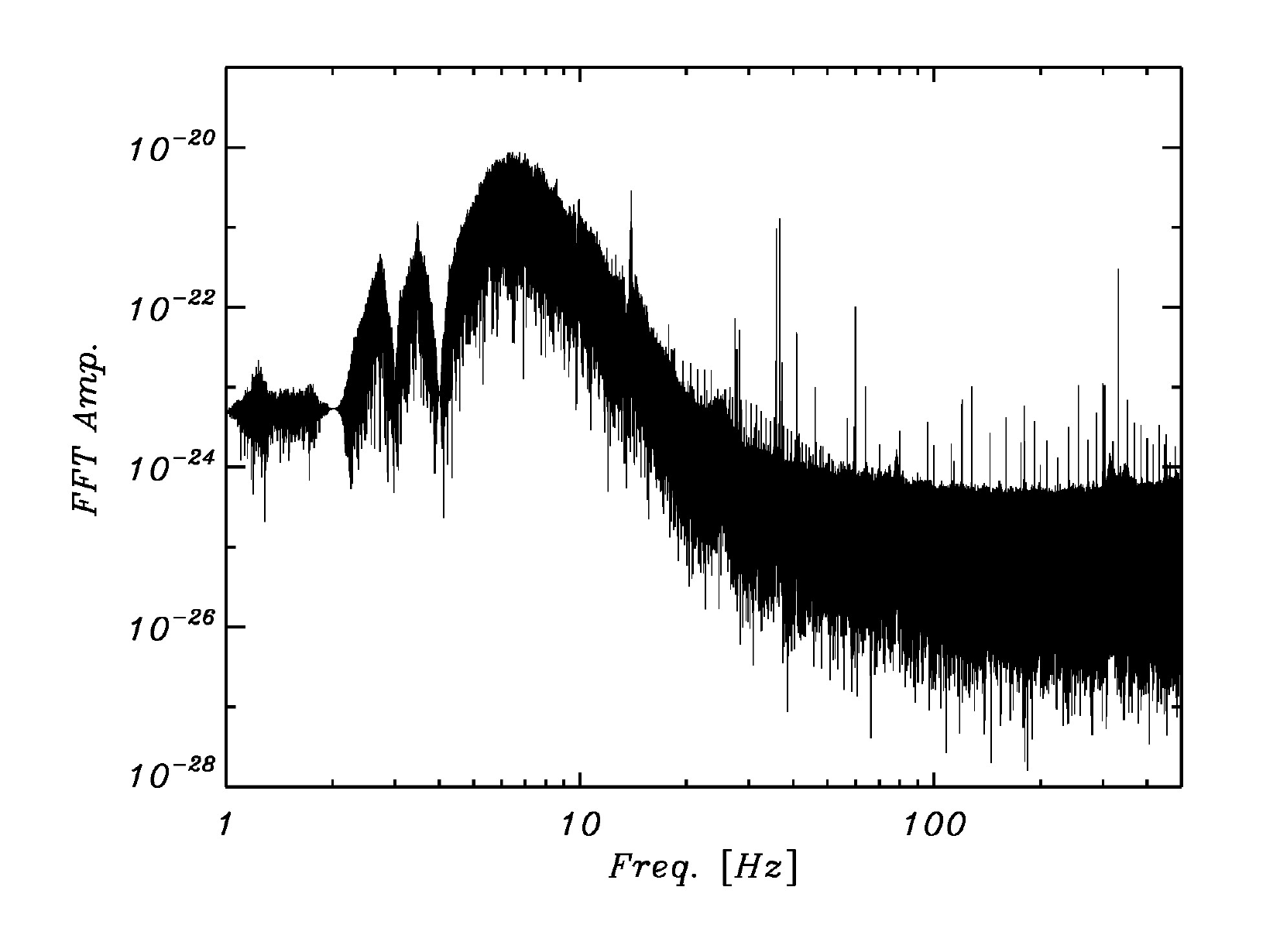}
  \includegraphics[width=0.42\textwidth]{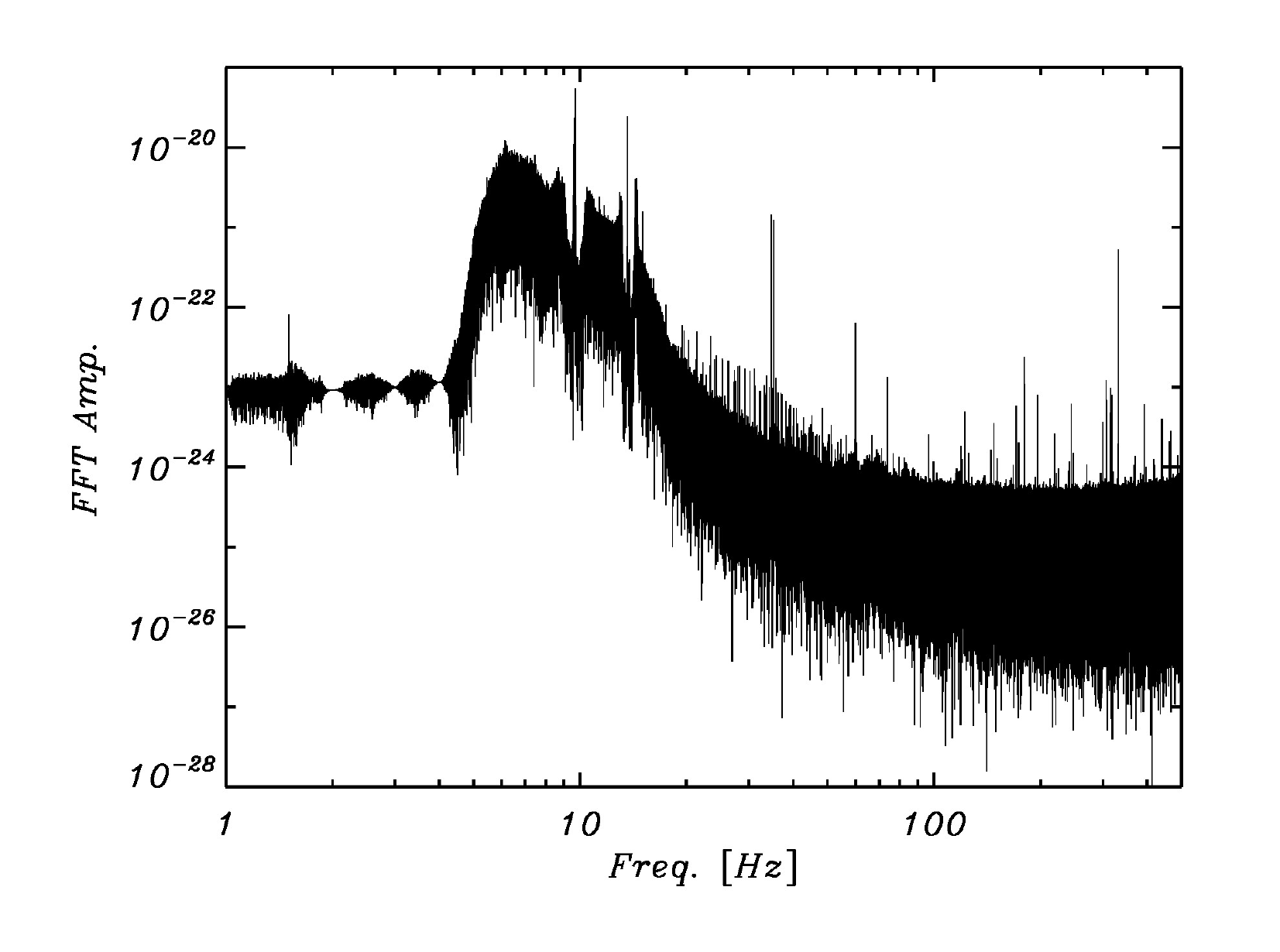}
  \caption{The FFT amplitude for the Hanford 4096 s raw record (left) and Livingston (right)}
  \label{fig2}
\end{figure}

In Fig.~\ref{fig:straindata} of the main text, we showed the 32 s raw data taken from the LIGO archive (\textbf{\url{https://losc.ligo.org/events/GW150914/}}). The LIGO team used a 4096 s record for cleaning the data sets, which reveals the same peak-like structure in the power spectra for Hanford and Livingston detectors. In Fig.~\ref{fig2} we show the power spectra for the 4096 s raw data from \textbf{\url{https://losc.ligo.org/events/GW150914/}}, for comparison with the 32 s power spectra in Fig.~\ref{fig:fourieramplitudes} of the main text.

Note that all spectra were obtained without any apodization or whitening, and they directly represent the structure of the raw data. In the following, we discuss all these factors one by one.

\subsection{Phase lock, edge effects and apodization}
The properties of the signals presented in Fig.~\ref{fig:straindata} of the main text allow us to propose the following model of the raw data:
\begin{eqnarray}
S_{H,L}=\sum_jA_j\delta(t-t_j)+n_{H,L}(t)
\label{eq1}
\end{eqnarray}
where the first term is the sum over index $j$ for the peak-like signals with amplitudes $A_j$ (positive and negative), the $n_{H,L}(t)$ term represents the residual noise as well as a possible GW signal.
Note that at the points $t=t_j$ the amplitude of the resonant signal is much greater than $n(t_j)$. The direct convolution of Eq.~(\ref{eq1}) to the Fourier domain gives:
\begin{eqnarray}
S(\omega)=\sum_jA_je^{-i\omega t_j} +n(\omega)
\label{eq2}
\end{eqnarray}
So, according to Eq.~(\ref{eq2}) the power spectrum $|S(\omega)|^2$ and the phases $\Psi(\omega)$ of the signal are given by:
\begin{eqnarray}
&&|S(\omega)|^2= |n(\omega)|^2 + \sum_{j,k}A_jA_ke^{-i\omega (t_j-t_k)} + n^*(\omega)\sum_jA_j e^{-i\omega t_j}+n(\omega)\sum_jA_j e^{i\omega t_j}=\nonumber\\
&&=|n(\omega)|^2 + \sum_{j}A^2_j +\sum_{j\neq k}A_jA_k\cos[\omega(t_j-t_k)]+ 2|n(\omega)|\sum_jA_j\cos[\omega t_j-\phi(\omega)]
\label{eq3}
\end{eqnarray}
where $n(\omega)=|n(\omega)|e^{-i\phi(\omega)}$ and the $\phi(\omega)$ are the phases of the residuals\footnote{Here we use $-\phi(\omega)$ for the definition of the phases of $n(\omega)$}.
\begin{eqnarray}
\tan\Psi(\omega)=-\frac{\sum_jA_j\sin(\omega t_j)+|n(\omega)|\sin[\phi(\omega)]}{\sum_jA_j\cos(\omega t_j)+|n(\omega)|\cos[\phi(\omega)]}
\label{eq4}
\end{eqnarray}

We would like to emphasise  the existence of the terms $\propto A_j|n(\omega)|$  in Eq.~(\ref{eq3}), which will propagate from the low frequency domain to higher frequencies. This interference between the high amplitude signal and the low amplitude "noise" is familiar in CMB science when the kinematic dipole in the time ordered data (TOD) is present along with $f^{-1}$ noise, foregrounds and the CMB signal. It is known that the extraction of  the kinematic dipole from the TOD is one of the key problems of CMB data analysis. The structure of the LIGO signal is much more complicated due to the non-stationarity of the noise and non-Gaussianity of the GW-signal. The LIGO team does not address this problem.

To illustrate the phase lock effect we define
\begin{eqnarray}
\sum_j A_j\sin(\omega t_j)=&&A\sin[\phi_{lock}(\omega)] \\ \nonumber
\sum_j A_j\cos(\omega t_j)=&&A\cos[\phi_{lock}(\omega)],
\end{eqnarray}
where $A$ is the amplitude. If $A\sin[\phi_{lock}(\omega)]\gg |n(\omega)|\sin[\phi(\omega)]$ and  $A\cos[\phi_{lock}(\omega)]\gg |n(\omega)|\cos[\phi(\omega)]$, then the phase lock is given by:
\begin{eqnarray}
\tan\Psi(\omega)\simeq-\tan[\phi_{lock}(\omega)]\left[1+2\frac{|n(\omega)|}{A}\frac{\sin[\phi(\omega)-
\phi_{lock}(\omega)]}{\sin[2\phi_{lock}(\omega)]}\right]
\label{eq5}
\end{eqnarray}
where the second term in Eq.~(\ref{eq5}) is significantly smaller than the first. Finally, from Eq.~(\ref{eq5}) we get:
\begin{eqnarray}
\Psi(\omega)\simeq -\phi_{lock}(\omega) +\pi m  +(-1)^{m+1}\frac{|n(\omega)|}{A}\sin[\phi(\omega)-\phi_{lock}(\omega)]\hspace{0.5cm} m=0,1,...
\label{eq6}
\end{eqnarray}
where $|n|/A\ll 1$.
Obviously, the phase lock depends on the particular realisation of the peaks and, especially, the average of them, and will change, if some of the present assumptions regarding the balance between A and
$|n|$ are violated. For instance, if $\sum_jA_j\cos(\omega t_j)\ll |n(\omega)|\cos(\phi(\omega))$, but $A\sin(\phi_{lock}(\omega))\gg |n(\omega)|\cos(\phi(\omega))$,
we  get:
\begin{eqnarray}
\Psi(\omega)\simeq -\frac{\pi}{2}+\frac{|n(\omega)|}{A}\cdot\frac{\cos[\phi(\omega)]}{\sin[\phi_{lock}(\omega)]} +\pi m, \hspace{0.5cm} m=0,1,...
\label{eq7}
\end{eqnarray}
From Eq.~(\ref{eq6})~-~(\ref{eq7}) we can see that the phase lock effect depends on the properties of the major component of the contaminants, ($A\sin(\phi_{lock}(\omega))\gg |n(\omega)|\sin(\phi(\omega))$ or $A\cos(\phi_{lock}(\omega))\gg |n(\omega)|\cos(\phi(\omega))$, as seen from Fig.~\ref{fig:fourierphases} above.
In this figure we show the phases for 32 s records  for Hanford and Livingston detectors before and after cleaning. %removal of the calibration lines, power lines etc., but without implementation of the band pass filter (BPF) 35-350 Hz.
This allows us to investigate the dependence of the cleaned record on the frequencies of the band pass filter (BPF) even outside the domain of 35-350 Hz recommended by LIGO. Thus, the phase lock effect seen in Fig.~\ref{fig:fourierphases} of our paper illustrates the phase correlations for the entire frequency domain, including components with
$f\le 35$ Hz. From Fig.~\ref{fig:fourieramplitudes} of the main text and Fig.~\ref{fig2} of this appendix it is clearly seen that the phase lock effect is related to the low frequency signals, which exceed the GW150914 event by some 3 orders of magnitude.

From Fig.~\ref{fig:straindata} one can see straightforwardly that edge effect and apodization are insufficient, since the main structure of the signal is determined by the large amplitude peaks almost uniformly distributed along the 32 s record. These peaks are at least  of 3 orders of magnitudes greater
then GW signal and the corresponding noise.

\subsection{Phase mixing due to the transition from 32 s record to the 0.2 s record}
A very important step of the LIGO data analysis is the transition from the 4096 s record to the \mbox{0.2 s} record after BPF and cleaning. Below we will show that this transition acts as a strong and non-linear phase filter that significantly distorts the phases of the 32 s record to those of the 0.2 s record. We denote $S_{32}(t)$ as the data sets for the cleaned %but not band passed
32 s records for Hanford and Livingston with calibration lines and corresponding peaks removed, as presented in the LIGO tutorial and applied in our paper. The signal $S_{32}(t)$  is characterized by the Fourier amplitudes and phases, given in Fig.~\ref{fig2} and Fig.~\ref{fig:fourierphases} above. We consider a small interval of length T within the $S_{32}(t)$-record
and get the Fourier amplitudes $|S_{T}(\nu)|$ and phases $\Phi_T(\nu)$:
\begin{eqnarray}
S_T(\nu)=\frac{1}{T}\int_{-T/2}^{T/2}dt S_{32}(t)e^{-i\nu t}=|S_{T}(\nu)|e^{i\Phi_T(\nu)},
\label{eq8}
\end{eqnarray}
keeping in mind that
\begin{eqnarray}
S_{32}(t)=\int_{-\infty}^{\infty}d\omega S_{32}(\omega)e^{i\omega t}.
\label{eq9}
\end{eqnarray}
Substitution of Eq.~(\ref{eq9}) into Eq.~(\ref{eq8}) gives us the following relationship between the phases $\Phi_T(\nu)$ and   the phases $\Phi(\omega)$ for the 32 s record:
\begin{eqnarray}
\tan\Phi_T(\nu)=\frac{\int d\omega |S_{32}(\omega)|W(\omega,\nu)\sin(\Phi(\omega))}{\int d\omega |S_{32}(\omega)|W(\omega,\nu)\cos(\Phi(\omega))},
\label{eq10}
\end{eqnarray}
where:
\begin{eqnarray}
W(\omega,\nu)=\frac{\sin Z}{Z}, \hspace{0.5cm} Z= \frac{T}{2}(\omega-\nu)
\label{eq11}
\end{eqnarray}

\begin{figure}[tbh!]
  \centering
  \includegraphics[width=0.48\textwidth]{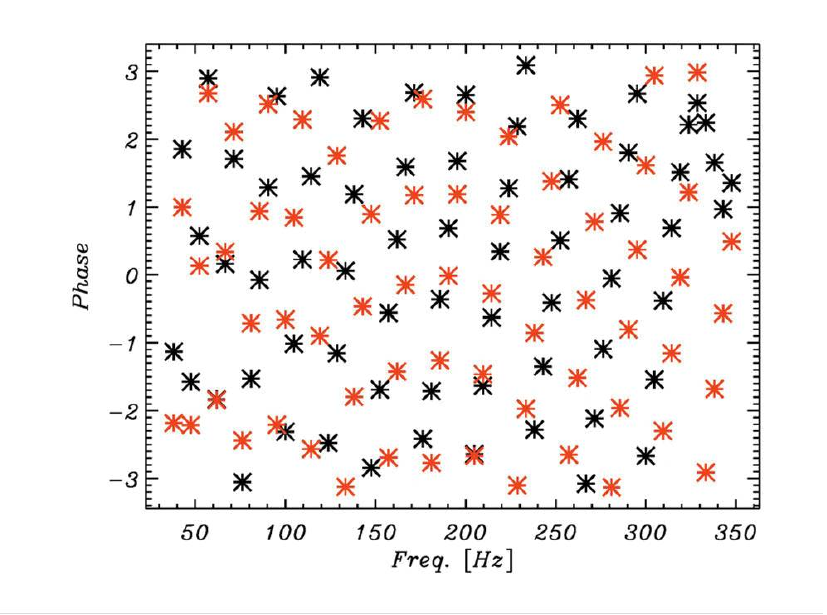}
  \includegraphics[width=0.48\textwidth]{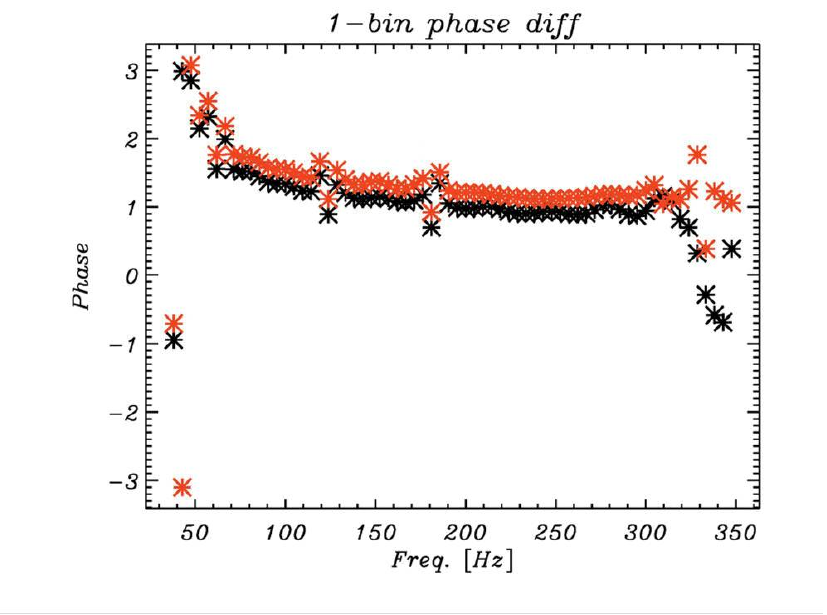}
  \caption{The phases of the 0.2s GW150914 template (left) and the difference between adjacent phases for the same template (right). The black stars correspond to the Hanford template, the red stars are for Livingston. The BPF domain is 35-350 Hz.}
  \label{fig3}
\end{figure}

From Eq.~(\ref{eq11}) one draws the following conclusion. If the phases of the 32 s record are randomly and uniformly distributed, mixing them will result in a random and uniform distribution. However, for the 0.2 s record containing the GW150914 signal, the phases are not at all random, as seen from Fig.~\ref{fig3}. Thus, to generate these correlations, the phases of the 32 s record
must have a phase lock, which would be decreased partially by the smoothing of $W(\omega,\nu)$ in Eq.~(\ref{eq10}).
%\begin{eqnarray}
%\tan\Phi_T(\nu)=\frac{\int d\omega |S_{32}(\omega)|W(\omega,\nu)\sin(\Phi(\omega))}{\int d\omega |S_{32}(\omega)|W(\omega,\nu)\cos(\Phi(\omega))}
%\label{eq10}
%\end{eqnarray}

\subsection{Whitening and phase mixing}
In addition, we note that the whitening filter, which acts in the Fourier domain, cannot change the phase correlations if the length of the record is fixed. To see this we return to Eq.~(\ref{eq2}) and rewrite this equation using a whitening operator based on the model of the noise and other signals: $F_w(\omega)=|F_w(\omega)|e^{i\xi(\omega)}$ (see the LIGO tutorial):
\begin{eqnarray}
S_w(\omega)=D(\omega)S(\omega)=\frac{S(\omega)}{|F_w(\omega)|}=\frac{|S(\omega)|}{|F_w(\omega)|}e^{i\Psi(\omega)}
\label{eq100}
\end{eqnarray}
Here $S_w$ is the signal after whitening, $|F_w(\omega)|$ and $\xi(\omega)$ correspond to the amplitudes and phases of the noise or any others signals that are present. Thus, if the whitening filter is ``phase neutral'' (e.g. if only the amplitudes are affected)  and there is no transition from the long records to the short records, it cannot change the phases of the signal.

However, an essential part of the LIGO cleaning technique is the transition from the cleaned long record (4096 s in the LIGO tutorial, and 32 s and 4096 s in our analysis) down to the short
0.2 s record. In this case the re-organisation of phases strongly depends on the implementation of the whitening filter. It follows from Eq.~(\ref{eq10}) that, after whitening of the long record, the phases for the short record are given as:
\begin{eqnarray}
\tan\Phi_T(\nu)=\frac{\int d\omega |S_{32}(\omega)||D(\omega)|W(\omega,\nu)\sin(\Phi(\omega))}{\int d\omega |S_{32}(\omega)||D(\omega)|W(\omega,\nu)\cos(\Phi(\omega))}
\label{eq101}
\end{eqnarray}
As an example, one can choose the whitening filter to be $|S_{32}(\omega)||D(\omega)|=1$, which makes the effective power spectrum constant independent of $\omega$, and the re-organisation of phases is now given only by the ``propagator'' $W(\nu,\omega)$. Obviously, the newly derived phases and their cross-correlations are different from those in Eq.~(\ref{eq10}). Thus, the whitening filter must be applied with considerable care when making the transition from long to short records.

\section{Band Pass Filtering of the data}
The band pass filter is the critical element of the data cleaning provided by the LIGO team. The LIGO team have used a fourth-order Butterworth filter for low- and high-passing of the 4096 s time ordered data with characteristic frequencies 35-350 Hz. This filter is characterised by the absence of gain ripple in the pass band and stop band and by a slow cutoff. We use the same filter for the \mbox{32 s} record in order to obtain the 0.2 s GW signal. However, unlike LIGO, we will fix the frequency of 350 Hz for the low-pass domain and vary the other frequency, setting its equal to 20, 25, 30, 35, 40, 50 and 60 Hz. In the following,  we have used the program for BPF provided by the LIGO team with these frequencies.

\begin{figure}[tbh!]
  \centering
  \includegraphics[width=0.4\textwidth]{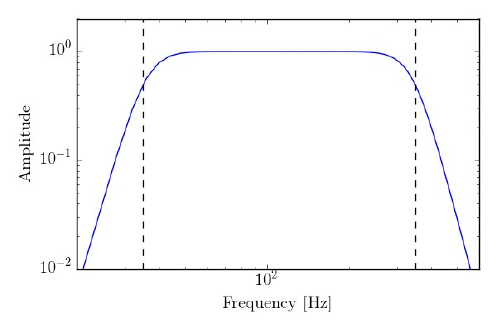}
  \includegraphics[width=0.4\textwidth]{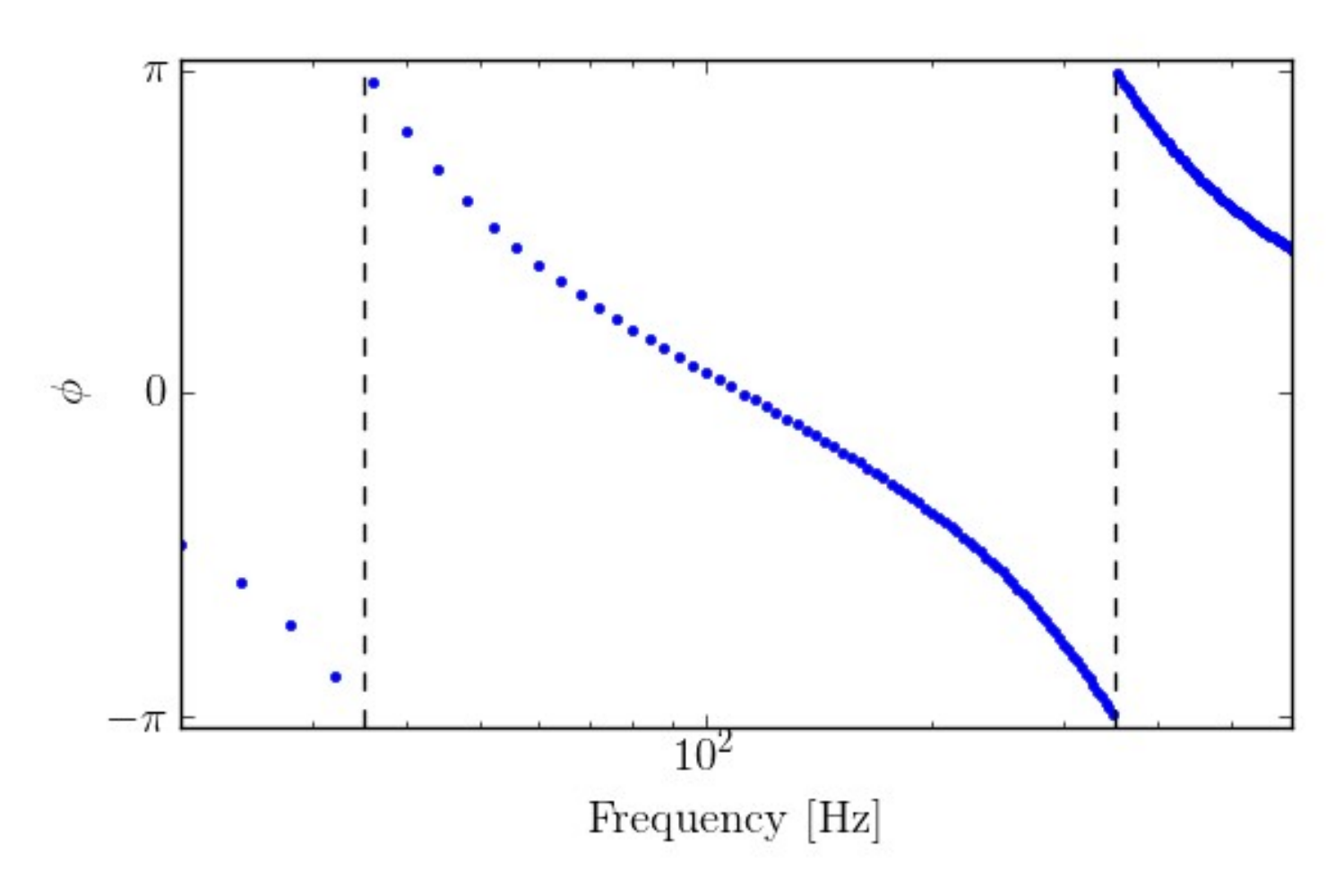}
  \caption{The gain (left) and the phase lag (right) for the Butterworth filter at   35-350 Hz domain.}
  \label{fig4}
\end{figure}

The transfer function of the  BPF filter is defined in the domain of complex frequencies $s=\gamma+j\omega$, where $\gamma$ is the real part and $\omega$ the imaginary part of the frequency. The back and forth\footnote{Note that here ``back and forth'' means to and from the Laplace space, which is different to the ``forward-backward'' filtering that is designed for minimization of the phase shift} passage to/from this domain is performed by the Laplace transform and its inverse. This is why  the term "input signal" denotes the Laplace transform of the time representation of the input signal.
The transfer function $H(s)$ of a filter is the ratio of the output signal $Y(s)$  to that of the input signal $X(s)$ as a function of the complex frequency $s$: $H(s)=Y(s)/X(s)$.
In discrete-time systems, the relation between an input signal and output  is dealt with using the z-transform. For the fourth-order Butterworth filter, we will use $G(\omega)=|H(s)|$ as the gain and $\phi_{BW}(\omega)=\arctan(H(s))$ as the phase lag, presented in Fig.~\ref{fig4}.

\section{Two tests for the Butterworth filter}
As we will show in the next section, the Butterworth filter is a critical part of cleaning the 32 s record. This is why we present two tests that illustrate the uncertainties resulting from the implementation of this filter.

\subsection{Leakage of narrow resonances}
In this section we use a toy model to show that the Butterworth filter does not completely eliminate narrow resonances lying outside of the frequency window selected for the band pass. Even though the amplitude of such a signal is reduced, its influence can be seen clearly in the residuals. Furthermore, continuous leakage into higher frequencies is also observed.

We begin by constructing an artificial 32 second signal made up of a 20 Hz sine wave and white noise with a standard deviation three orders of magnitude smaller than that of the sine wave using a sample rate of 4096 Hz. Each of the following signals:
\begin{enumerate}
  \item Sine wave only, which mimics the excess power at low frequencies (see Fig.~\ref{fig2})
  \item Noise only, which is assumed to be Gaussian white noise, and
  \item Sum of both
\end{enumerate}
is filtered  by application of a fourth-order Butterworth band pass filter characterized by a lower bound of 50 Hz and an upper bound of 350 Hz. These two bounds, together with the order of the Butterworth filter, characterize the decay of the filter's response, i.e. there may be power from frequencies outside this range which still makes a non-zero contribution to the signal. The Butterworth filter is applied via a forward-backward infinite response filter.

\begin{figure}[tbh!]
  \centering
  \includegraphics[width=0.48\textwidth]{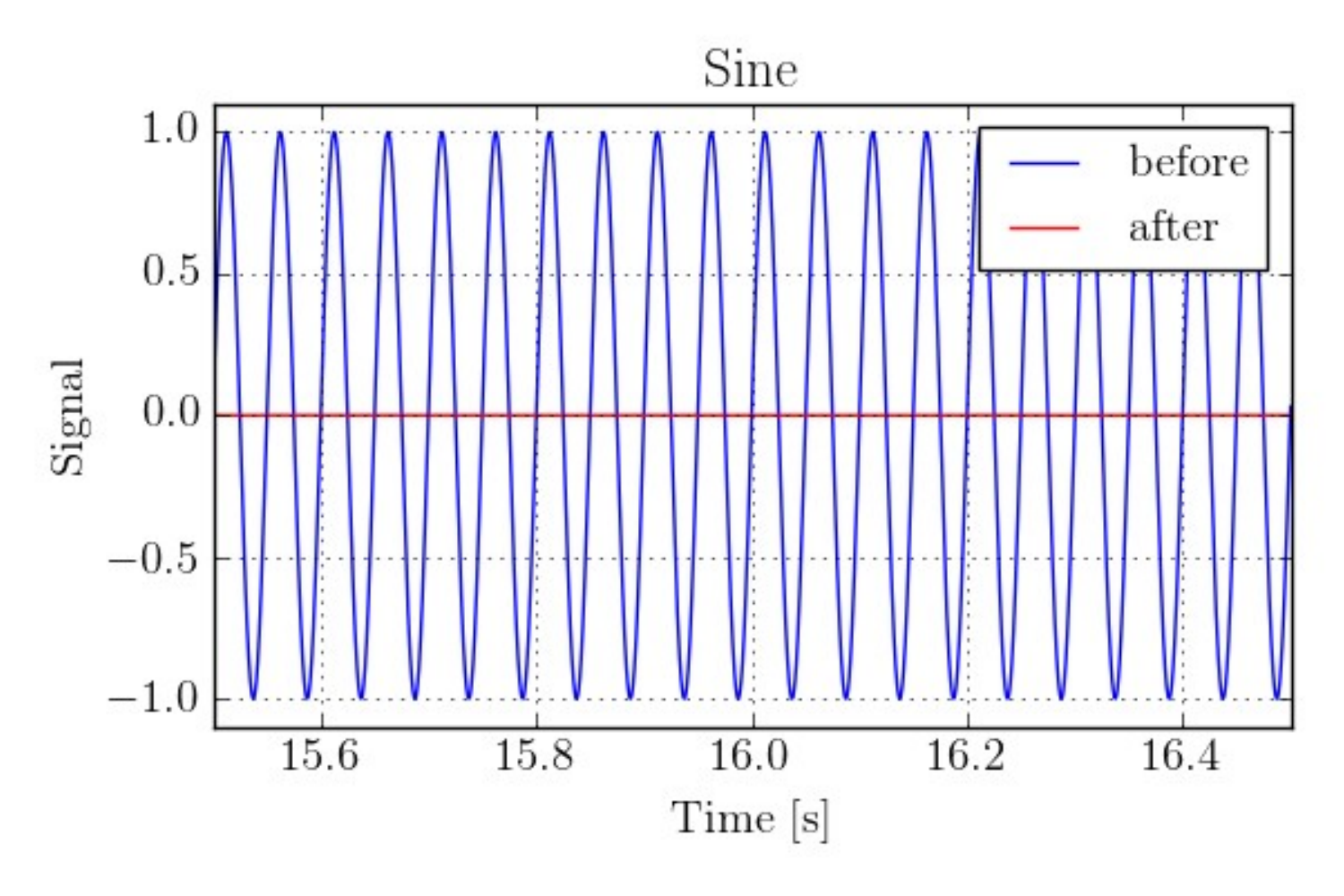}
  \includegraphics[width=0.48\textwidth]{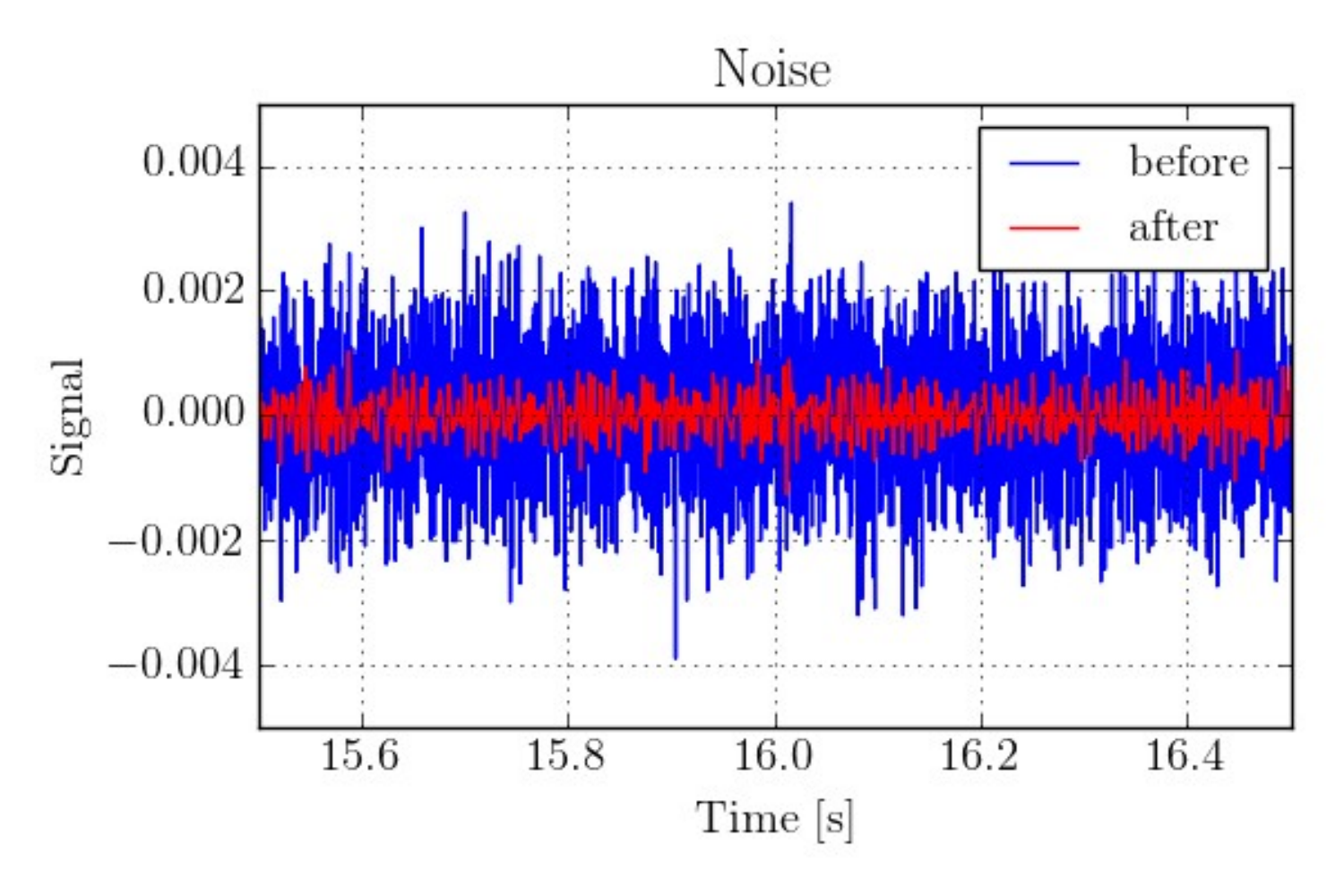}
  \caption{The left panel corresponds to the case A before and after filtering, the right panel shows the same for case B. Note that the red line on the left has a small amplitude but is not zero. }
  \label{fig3.1}
\end{figure}

Fig.~\ref{fig3.1}  shows the signals before and after cleaning. A range of 1 second in the middle of the record has been selected for better visibility of the signals.

\begin{figure}[tbh!]
  \centering
  \includegraphics[width=0.32\textwidth]{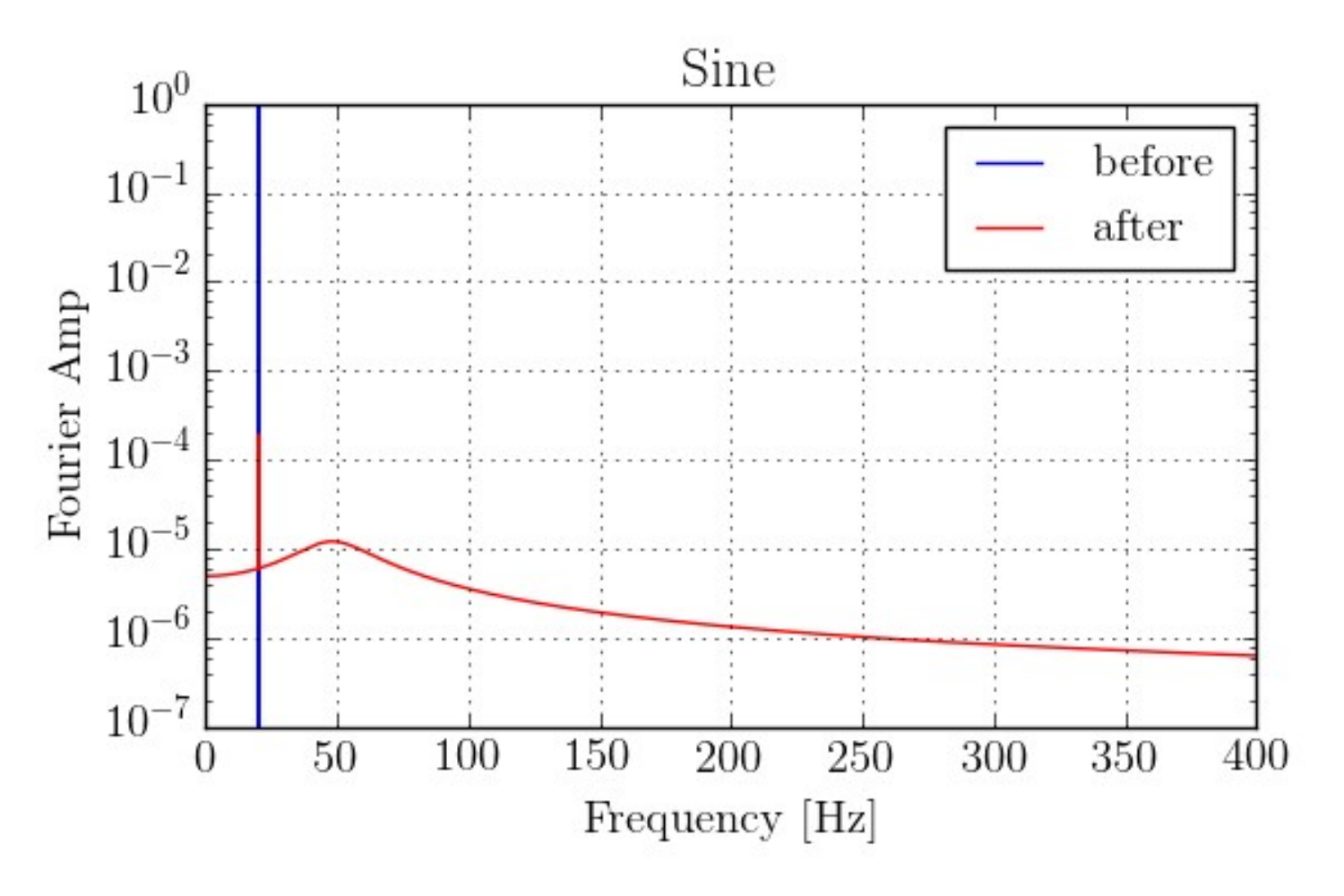}
  \includegraphics[width=0.32\textwidth]{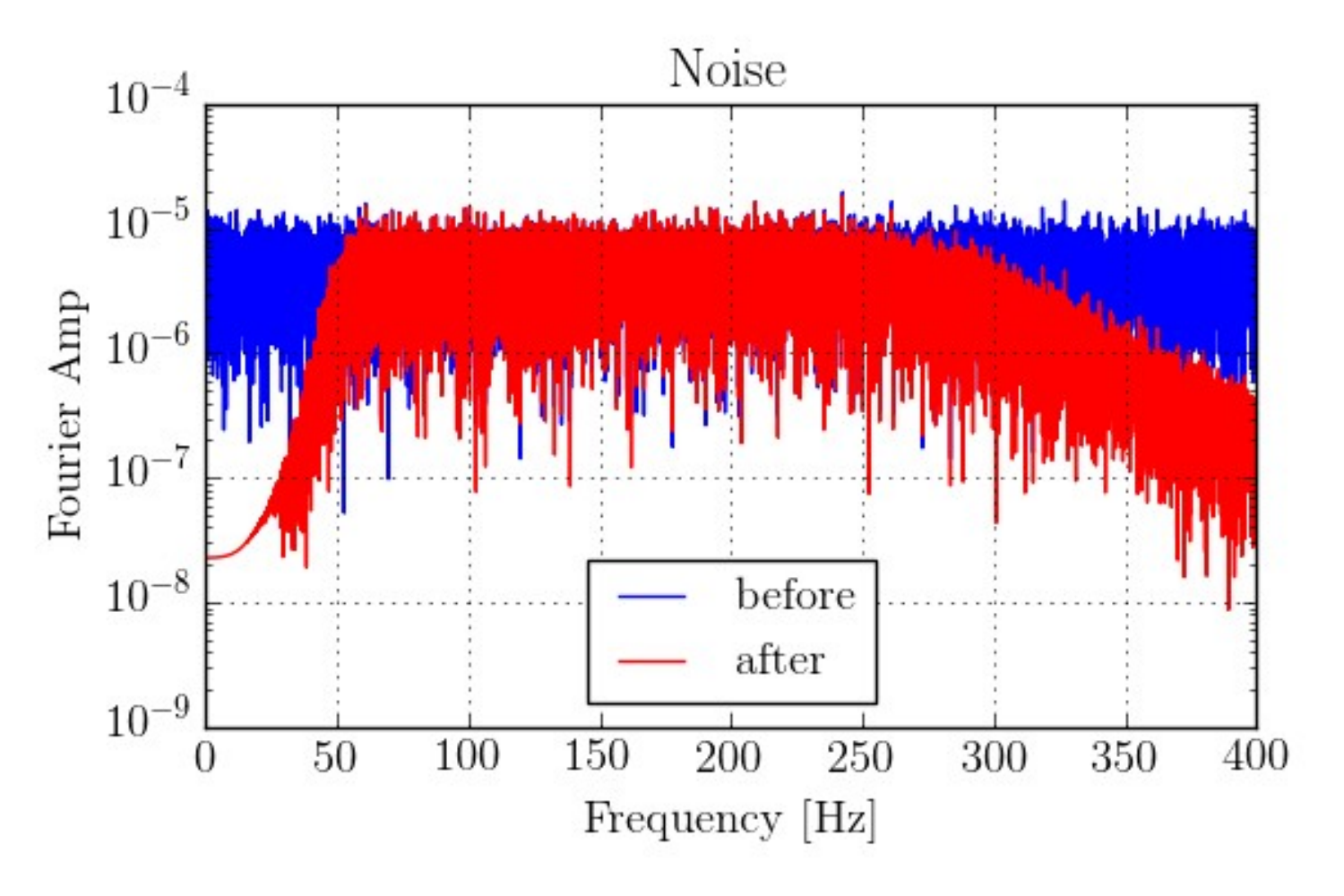}
  \includegraphics[width=0.32\textwidth]{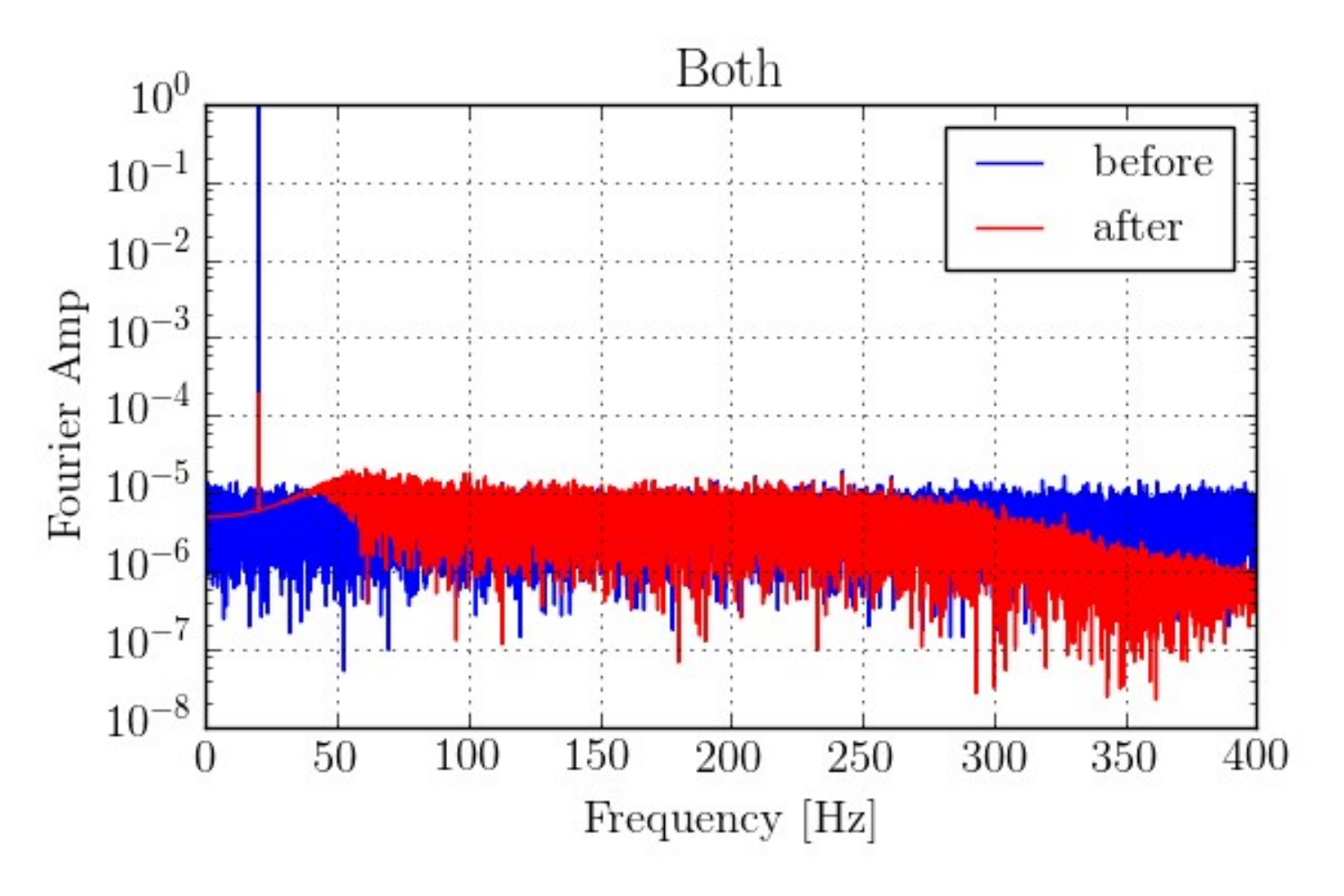}
  \caption{ Fourier amplitudes for the signals in Fig.~\ref{fig3.1}. All figures show the amplitudes of the respective signals (cases A-C) before and after linear filtering with the Butterworth filter.}
\label{fig3.2}
\end{figure}

In Fig.~\ref{fig3.2} the corresponding Fourier amplitudes are shown, clearly indicating the influence of the band pass filter. It can also be seen that the band pass filter does not remove the entire 20Hz line.

\begin{figure}[tbh!]
  \centering
  \includegraphics[width=0.48\textwidth]{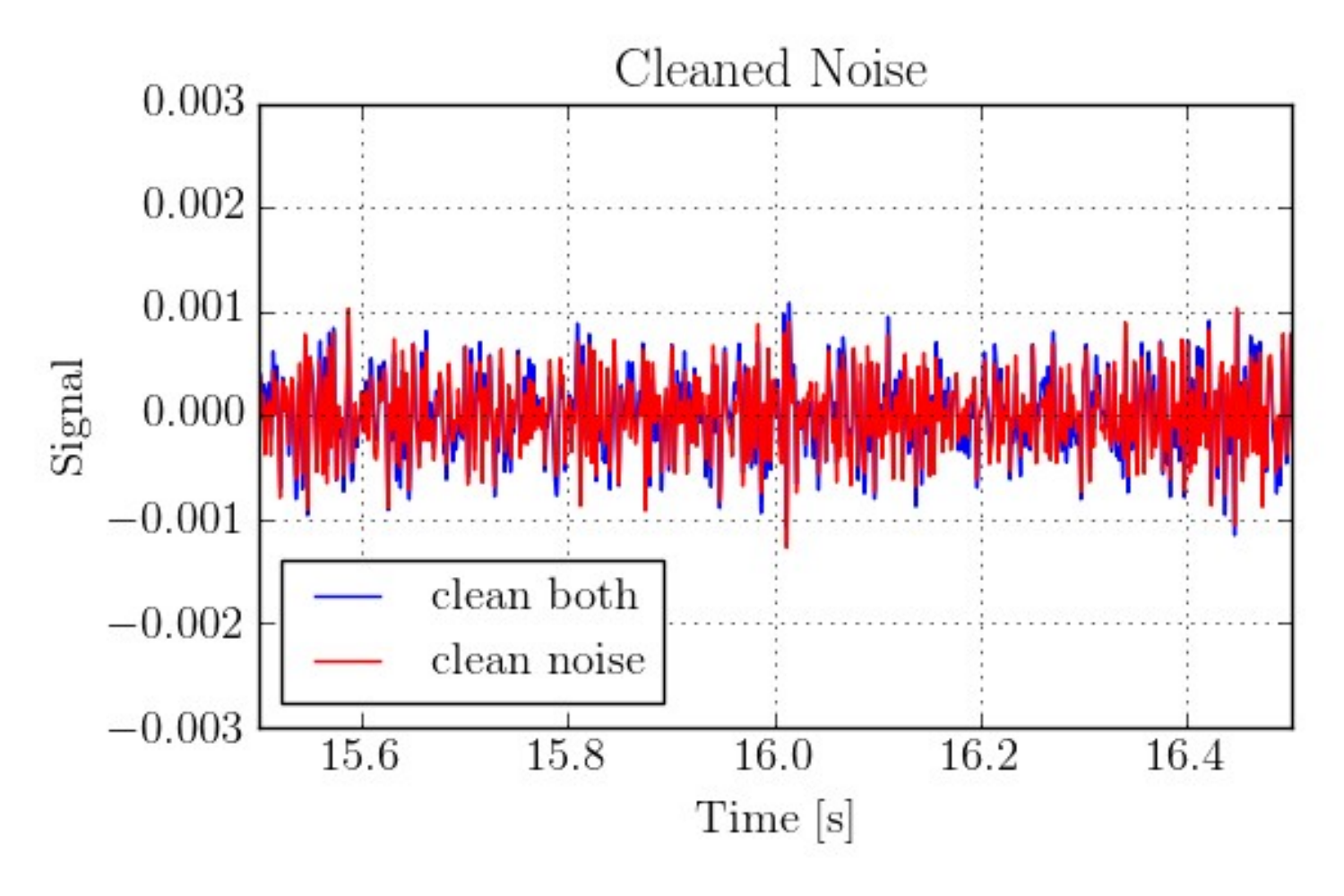}
  \includegraphics[width=0.48\textwidth]{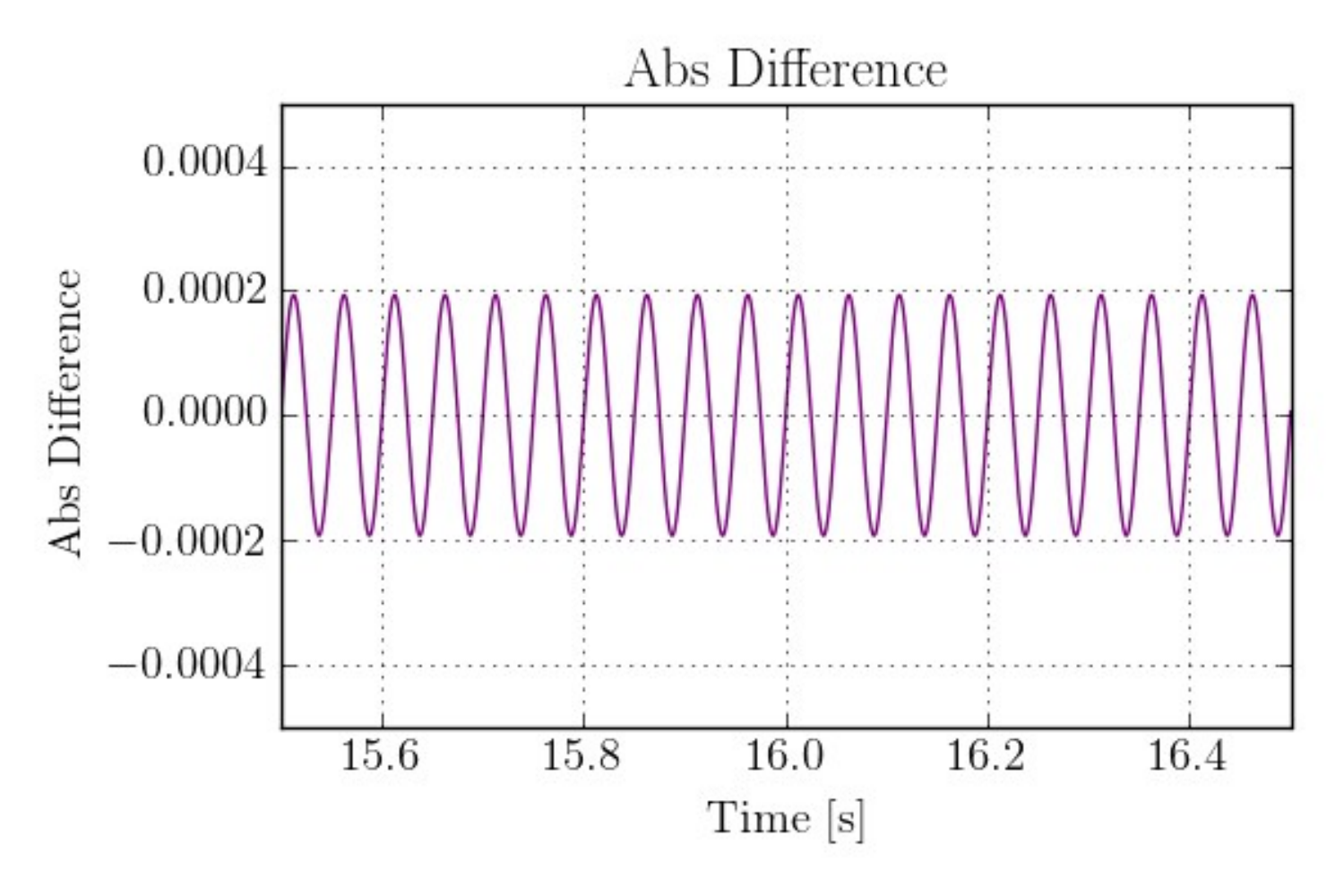}
  \caption{Left panel: The cleaned records of noise only (case B) and the sum of noise and sine wave (case C). Right panel: Their  difference (case C-case B) after cleaning. }
\label{fig3.3}
\end{figure}

Further, the filtering process introduces power in frequencies higher than the original line, visible also in the sum of sine wave and noise. It is therefore not surprising to find that the difference between the cleaned noise and the cleaned sum of noise plus sine wave is dominated by a 20 Hz modulation, see Fig.~\ref{fig3.3}. The amplitude of these residuals is lower than that of the input sine wave.
% and it also is lower than the amplitude of the cleaned records
However, this simple model demonstrates the difficulty of removing the influence of signals whose amplitudes are orders of magnitude higher than those of the desired signal.
This figure allows us to estimate the leakage from the low frequency spike of the power spectrum at 20 Hz to the domain of the Butterworth filtration 50-350 Hz. The propagated signal
in the noise domain is clearly a sin-wave with amplitude $2 \times 10^{-4}$ related to the amplitude of the spike. Obviously, decreasing the BPF frequency from 50 Hz down to 35 Hz or lower will increase the leakage dramatically.

\subsection{Errors in the phase reconstruction}
\begin{figure}[tbh!]
  \centering
  \includegraphics[width=0.48\textwidth]{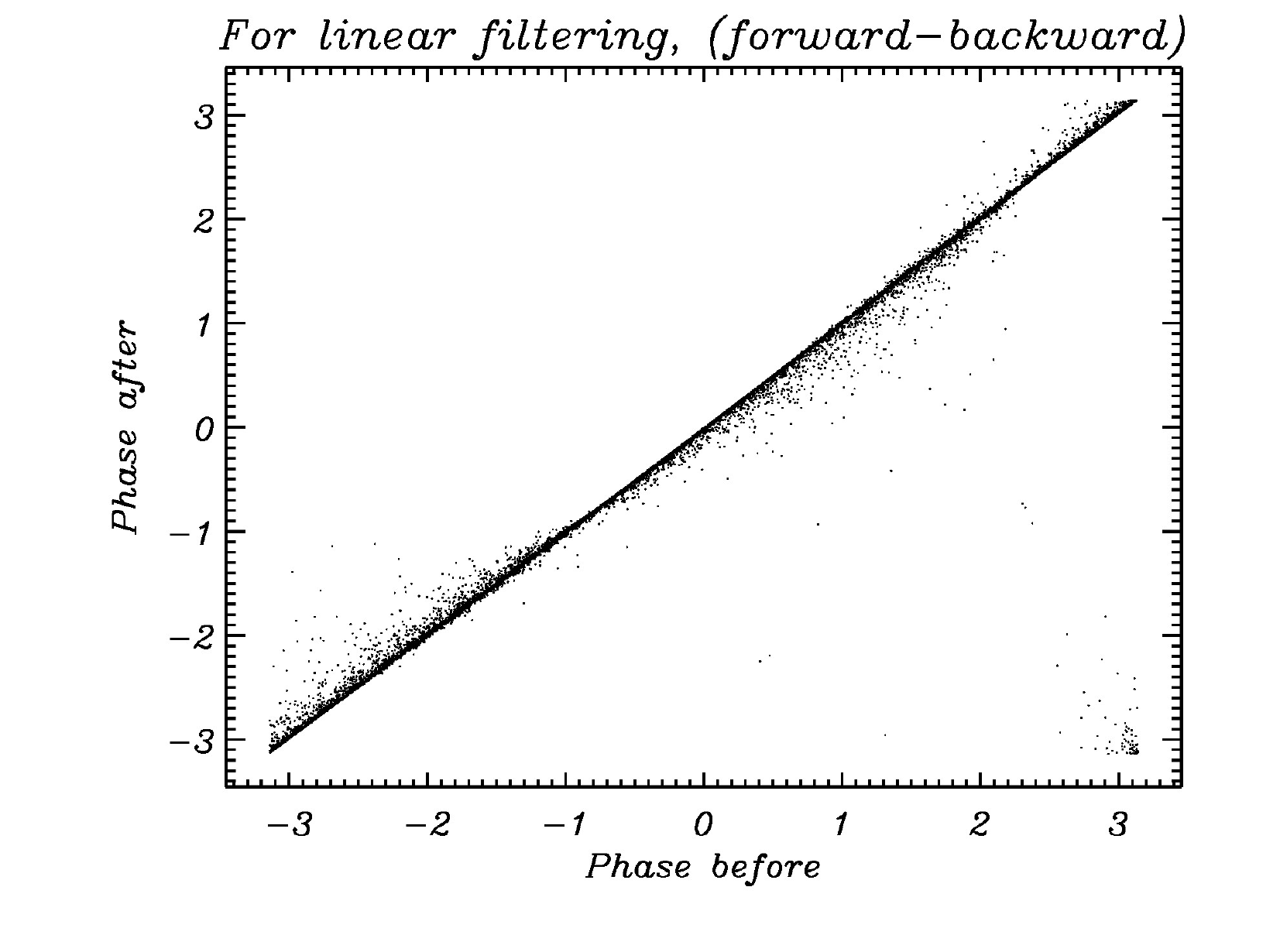}
  \includegraphics[width=0.48\textwidth]{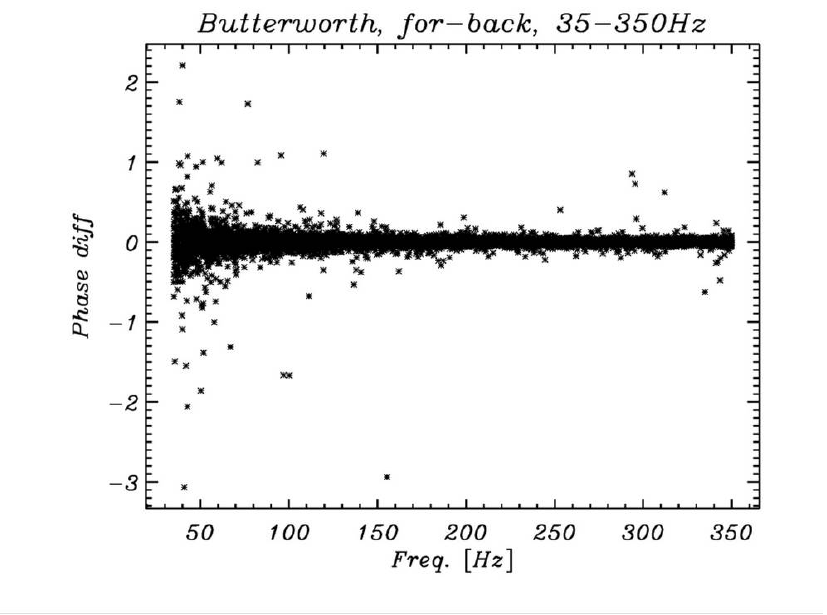}
  \caption{Left: The comparison of the input and output phases for the random white noise after the Butterworth filter with band pass 35-350 Hz. Right: The input-output phase difference for 35-350 Hz. }
\label{fig3 .4}
\end{figure}

The Butterworth filter is a \emph{phase-active} filter that affects both amplitude and phase as can be seen from Fig.~\ref{fig4}. The reconstruction of the phases is achieved by complex conjugation of the transfer function and can potentially have some residuals. In order to estimate these residuals, we use a \mbox{32 s} realization (with the same sampling as for the LIGO record) of the white noise and
convert it forward and backwards with this filter, focusing on the phase reconstruction. In Fig.~\ref{fig3 .4} we show the phase diagram for input and reconstructed phases for the Butterworth filter with 35-350 Hz.

We see clearly that even for the 32 s uniform random Gaussian noise the reconstruction of the phases is characterized by phase differences of about 0.5 radians on average. In
the low frequency domain (35 Hz) some of the modes can depart by as much as 1 to 2 radians..

\section{Extension of the BPF pass band}
Here, we would like to present the reconstruction of the GW150914 event by BPF with different low frequency cutoffs.
The same filtering will  be applied for the theoretical template for GW150914 event provided by the LIGO team. As before, the cleaning of the raw data has been done in the 32 s domain, using various windows of the Butterworth filter. Subsequently, the signal in the GW domain is extracted from the data.

As a brief summary of steps needed to reproduce the results in this section is as follows:

\begin{itemize}
\item Select raw NR template.
\item Select low frequency cutoff $f_{Low}$.
\item Construct a fourth-order Butterworth filter with low frequency cutoff $f_{Low}$ and high frequency cutoff $f_{High}=350\,$Hz.
\item Bandpass the raw NR template with a linear (back-and-forth) filter using the Butterworth filter and remove narrow resonances by linear (back-and-forth) band stop.
\item Repeat all above steps for the raw data.
\end{itemize}

\begin{figure}[tbh!]
  \centering
  \includegraphics[width=0.48\textwidth]{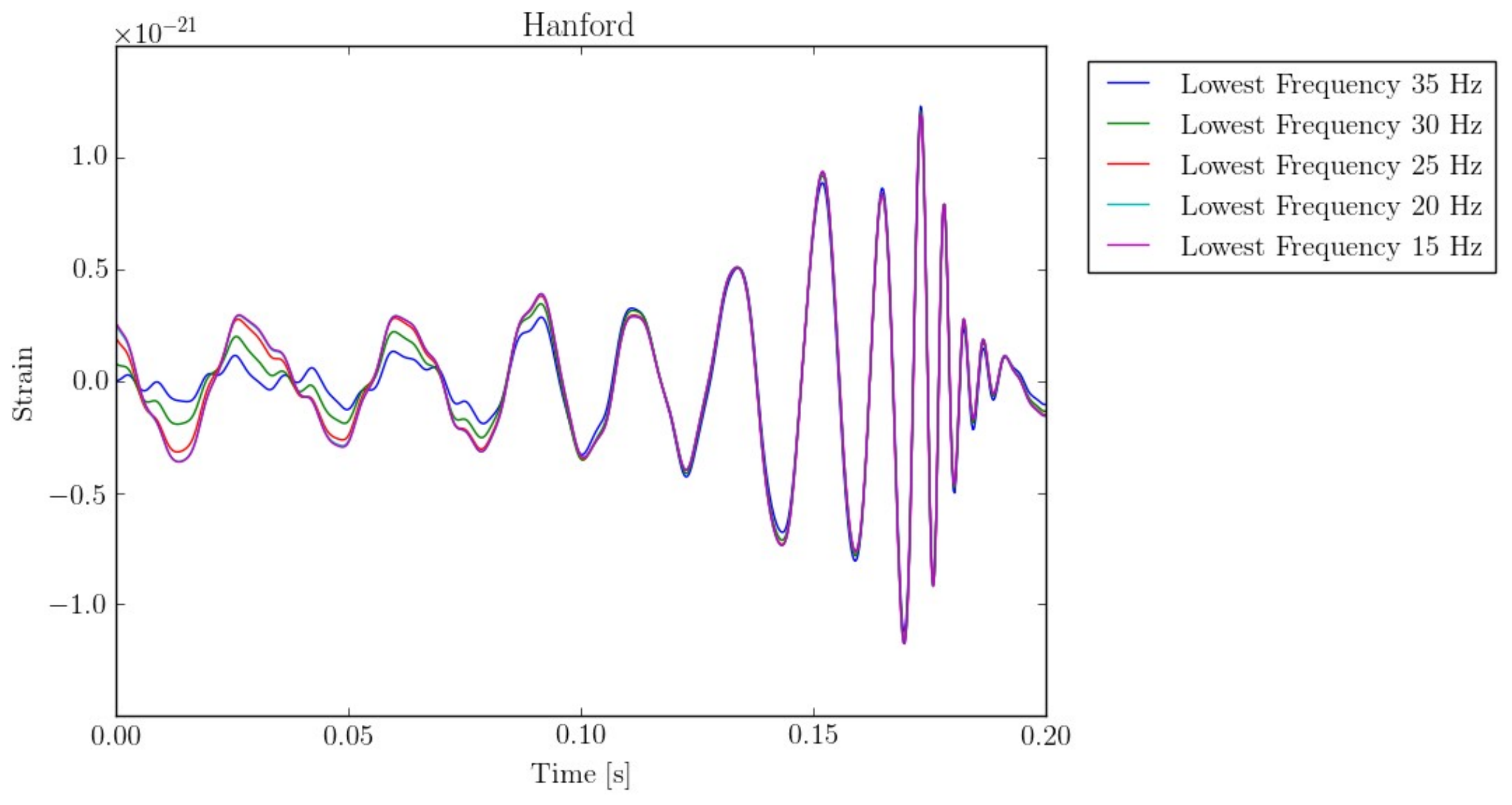}
  \includegraphics[width=0.48\textwidth]{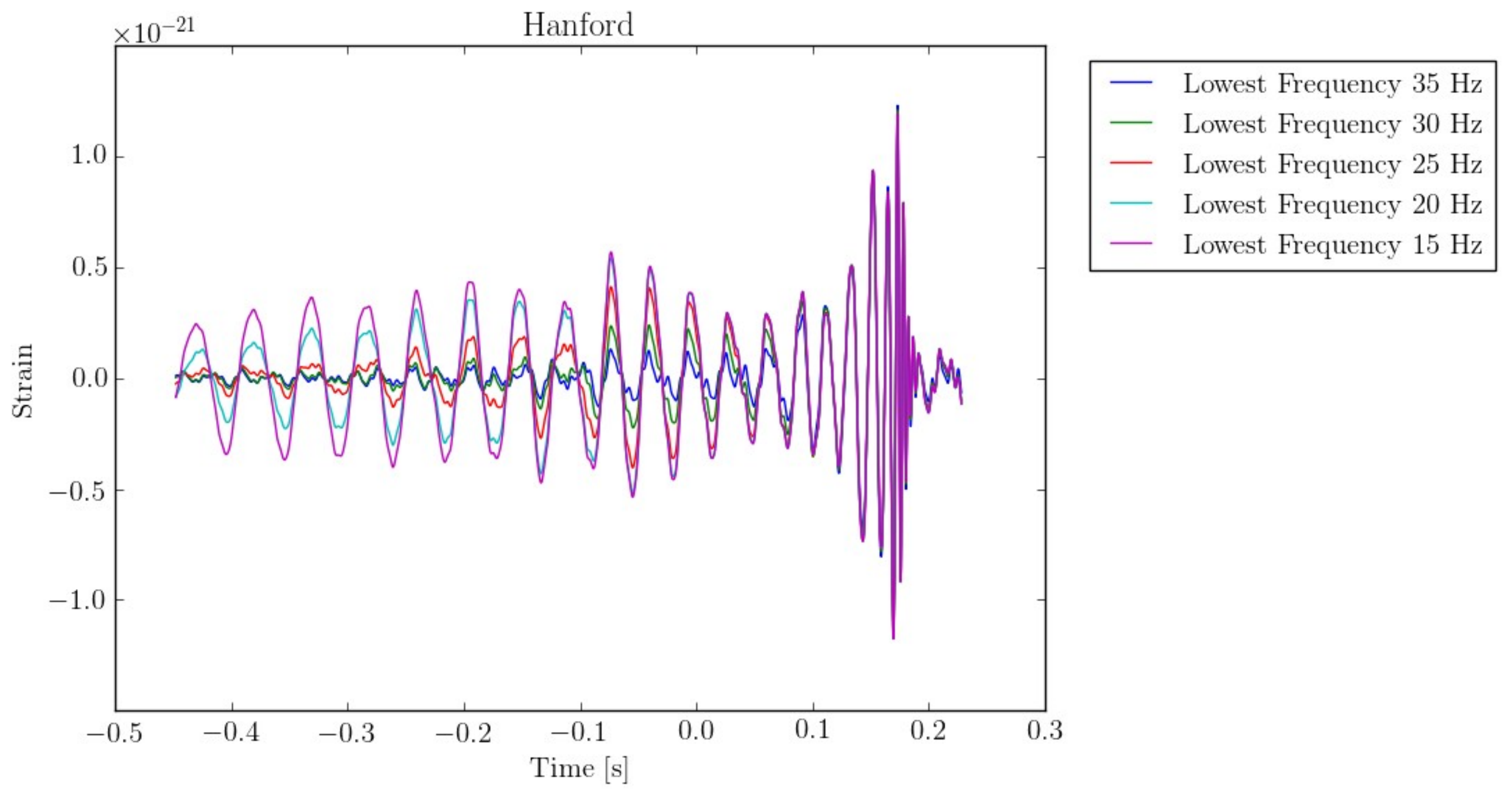}
  \caption{The numerical relativity template (NR) after cleaning with different choices of the low frequency cutoff, as noted in the legends.  Left and right panels show the same data but for different time windows.}
\label{fig4.1}
\end{figure}

\begin{figure}[tbh!]
  \centering
  \includegraphics[width=0.48\textwidth]{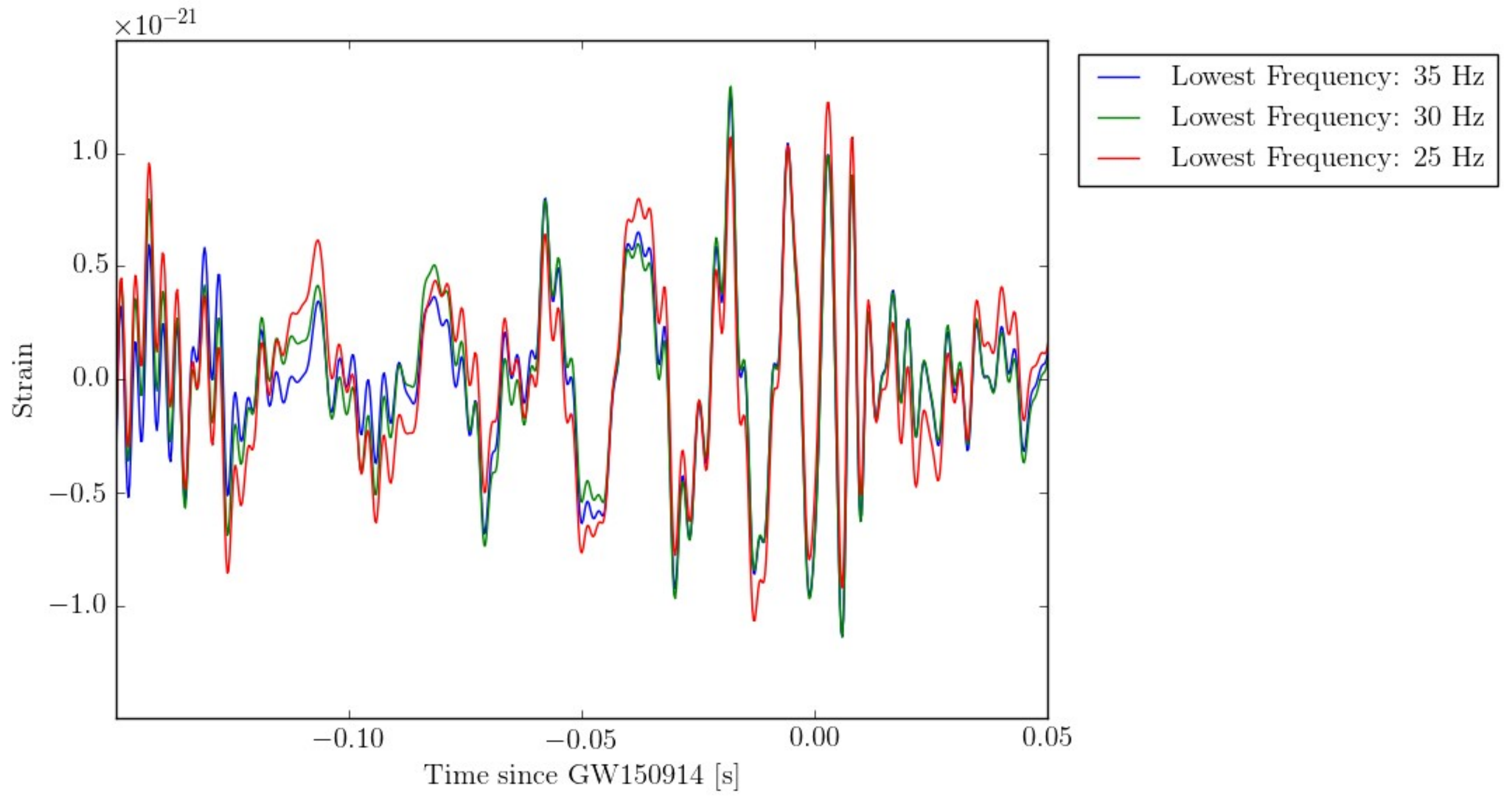}
  \includegraphics[width=0.48\textwidth]{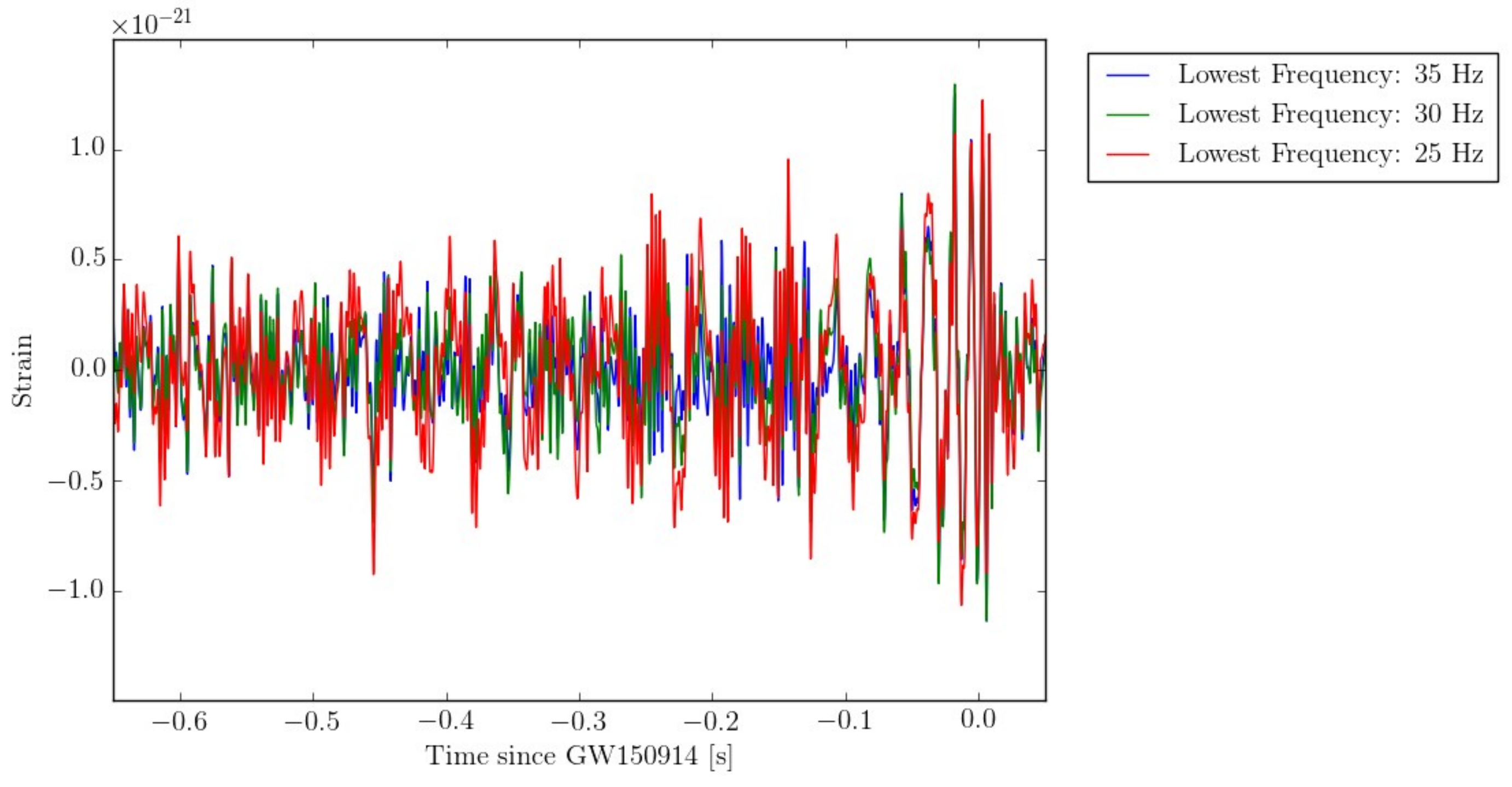}
  \caption{The Hanford data after cleaning with different choices of the low frequency cutoff, as noted in the legends.  For better visual inspection here only three choices of the low frequency cutoff were made.  Left and right panels show the same data but for different time windows.}
\label{fig4.2}
\end{figure}

In Fig.~\ref{fig4.1} we show the theoretical  numerical relativity template with the Butterworth BPF for lower filter frequencies of 15, 20, 30 and 35 Hz. The high frequency of the BPF is the same (350 Hz) for all the cases. As can be seen from this figure, as the lower frequency of the BPF decreases, the signal amplifies for the first 0.1 s is only slightly changed in the region of the chirp effect (the second half of the record). For comparison, we extend the length of the record to -0.5 s before the chirp to illustrate the properties of the Hanford template long before the time domain of interest. Fig.~\ref{fig4.2} illustrates the same behaviour of the Hanford strain.
Note the significant amplification of the amplitude of the signal before the GW150914 time domain as the BPF is lowered. This effect, does not exist for the template at the same level. For $-0.5\le t\le -0.4$ s and for $-0.3 \le t\le -0.2$ s and a 25 Hz lower frequency of the BPF, one can find peaks with amplitudes comparable to the GW-signal in the chirp domain.

\begin{figure}[tbh!]
  \centering
  \includegraphics[width=0.32\textwidth]{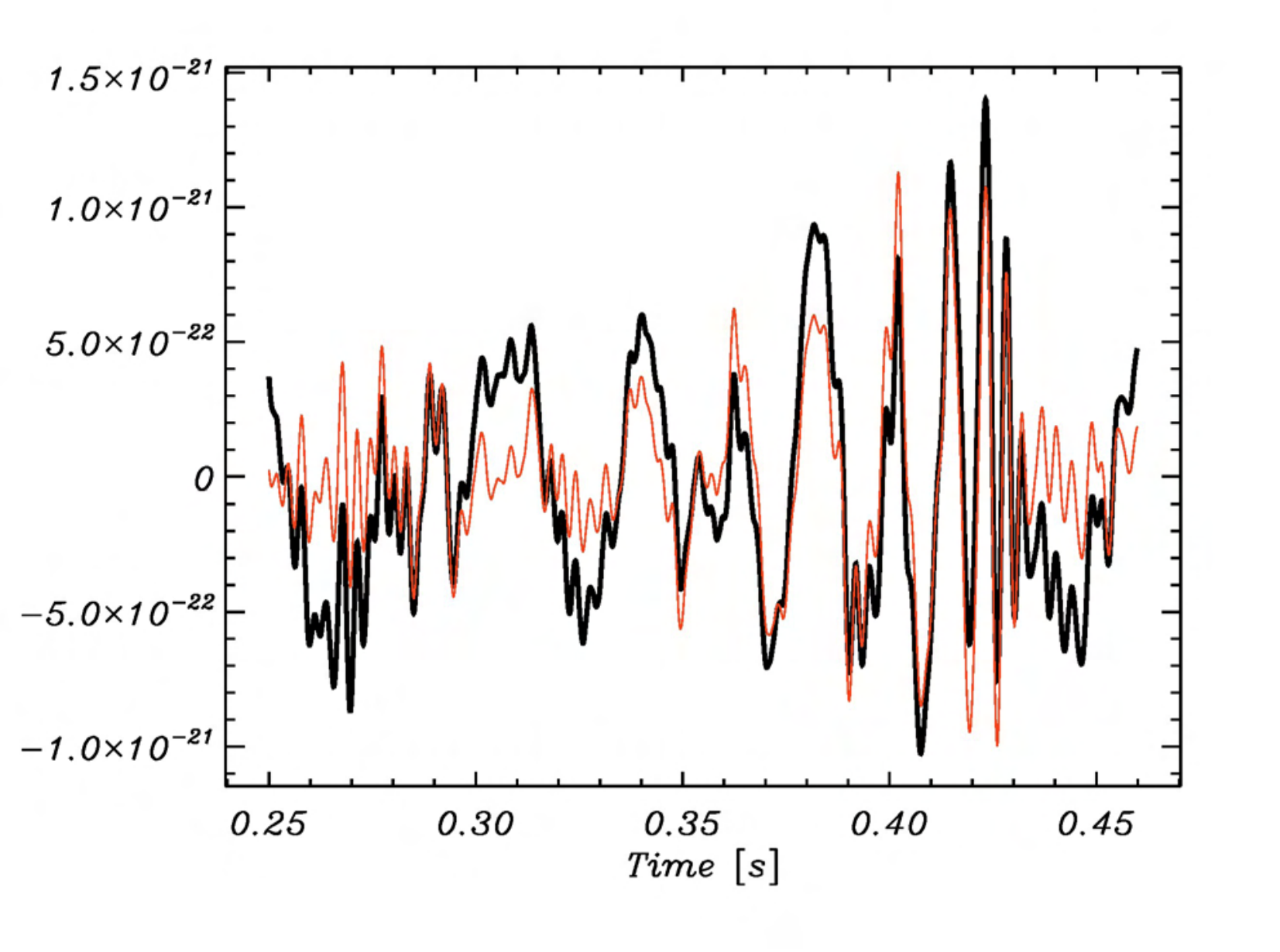}
  \includegraphics[width=0.32\textwidth]{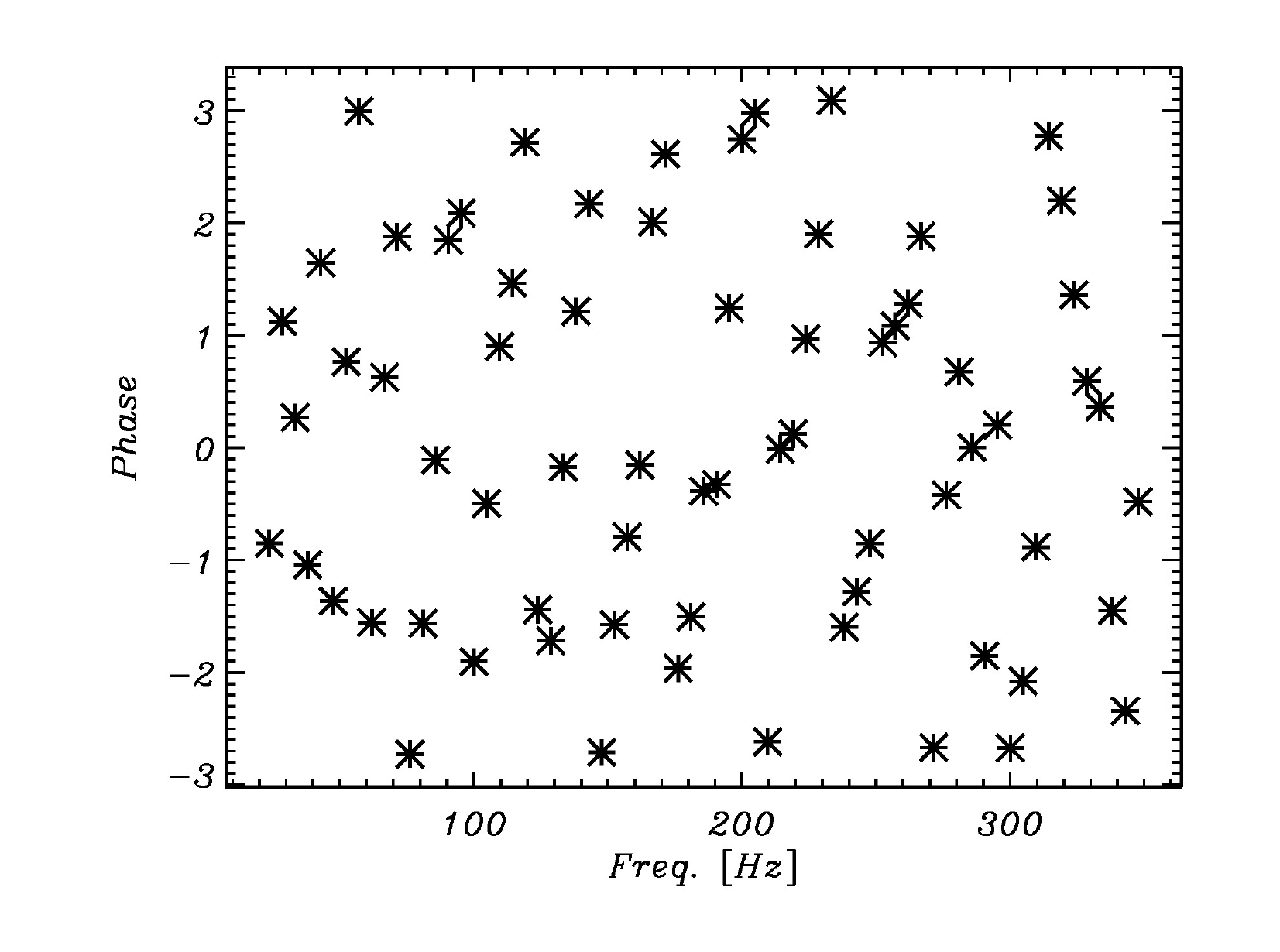}
  \includegraphics[width=0.32\textwidth]{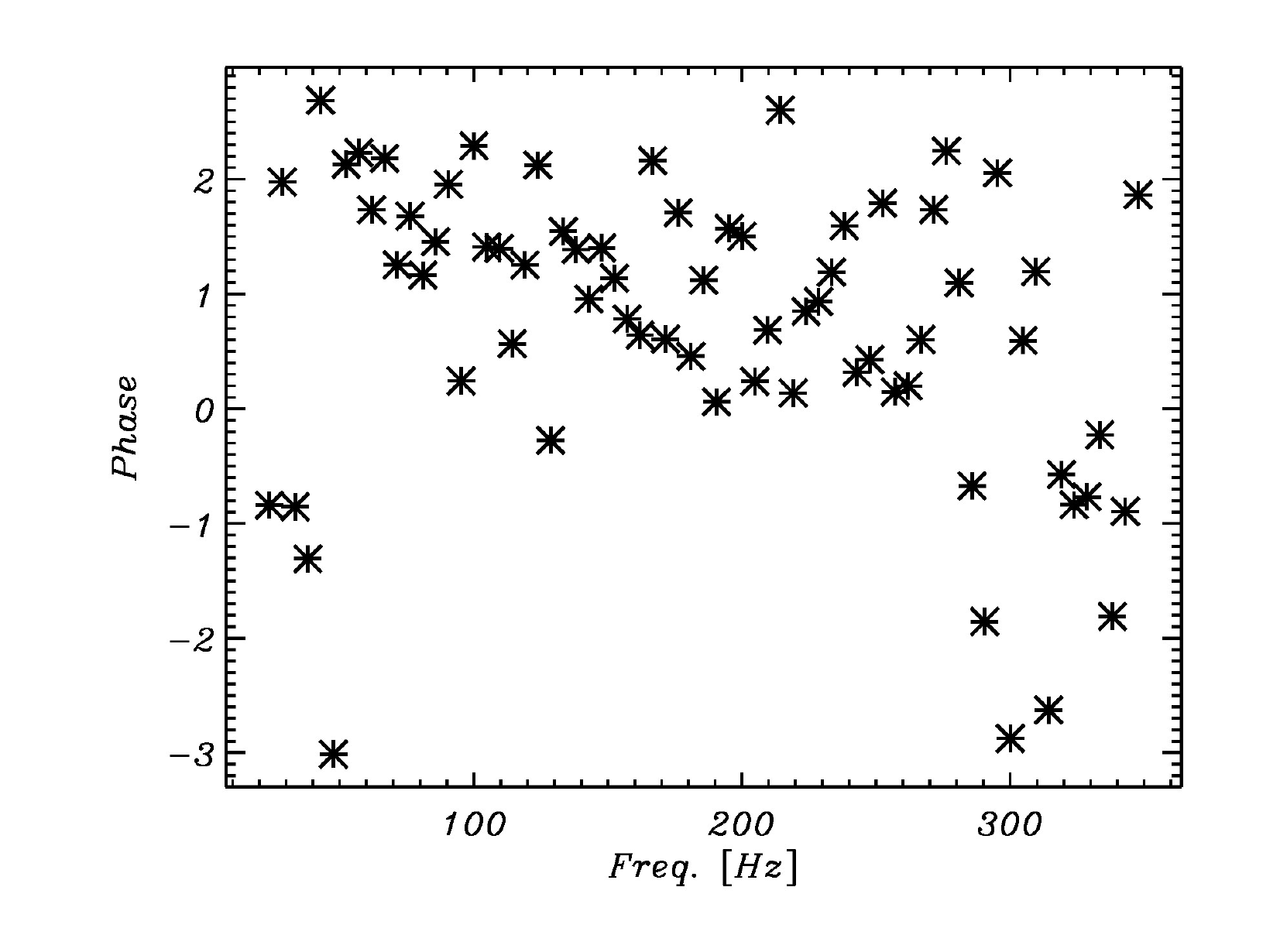}

  \includegraphics[width=0.32\textwidth]{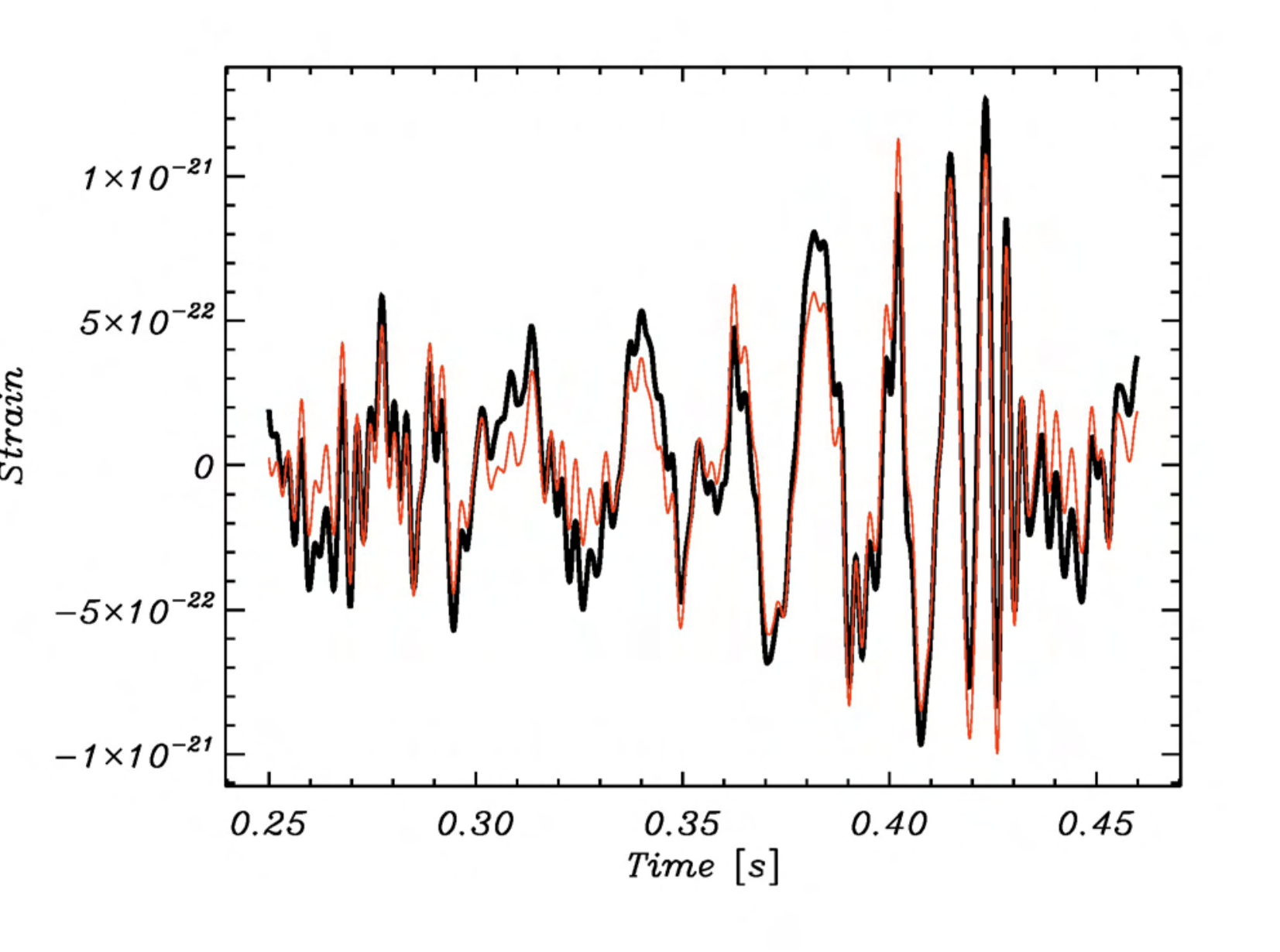}
  \includegraphics[width=0.32\textwidth]{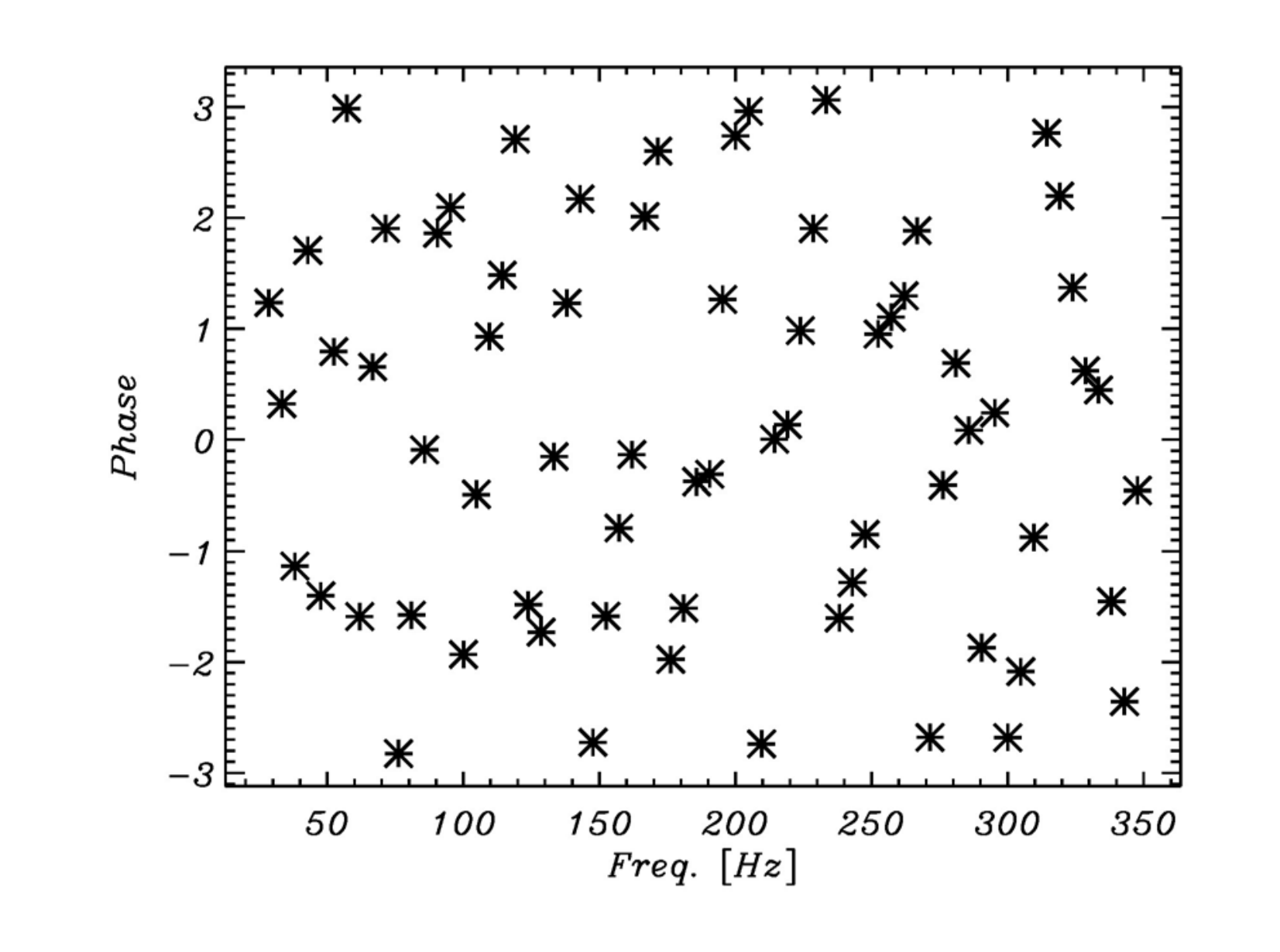}
  \includegraphics[width=0.32\textwidth]{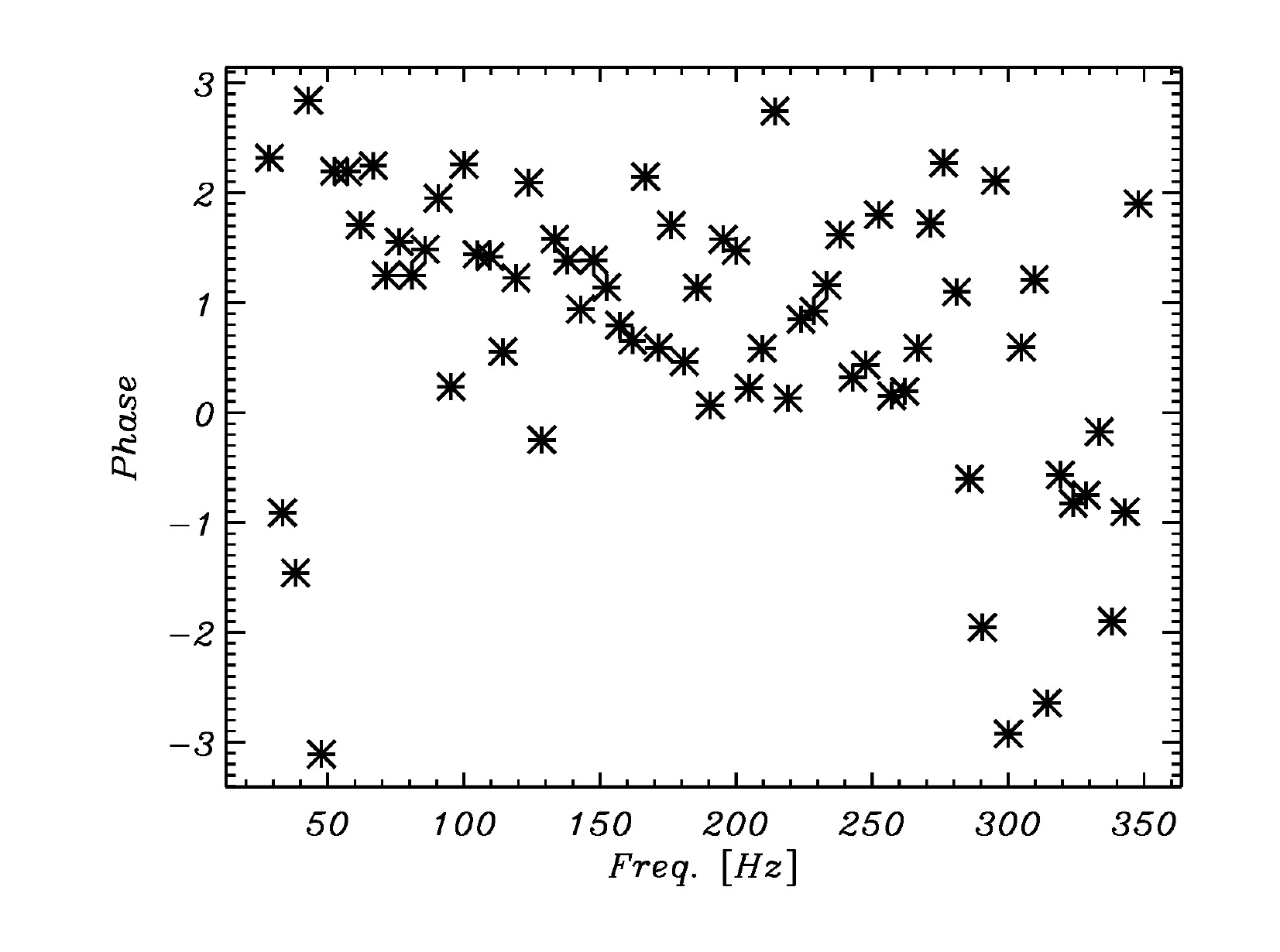}

  \includegraphics[width=0.32\textwidth]{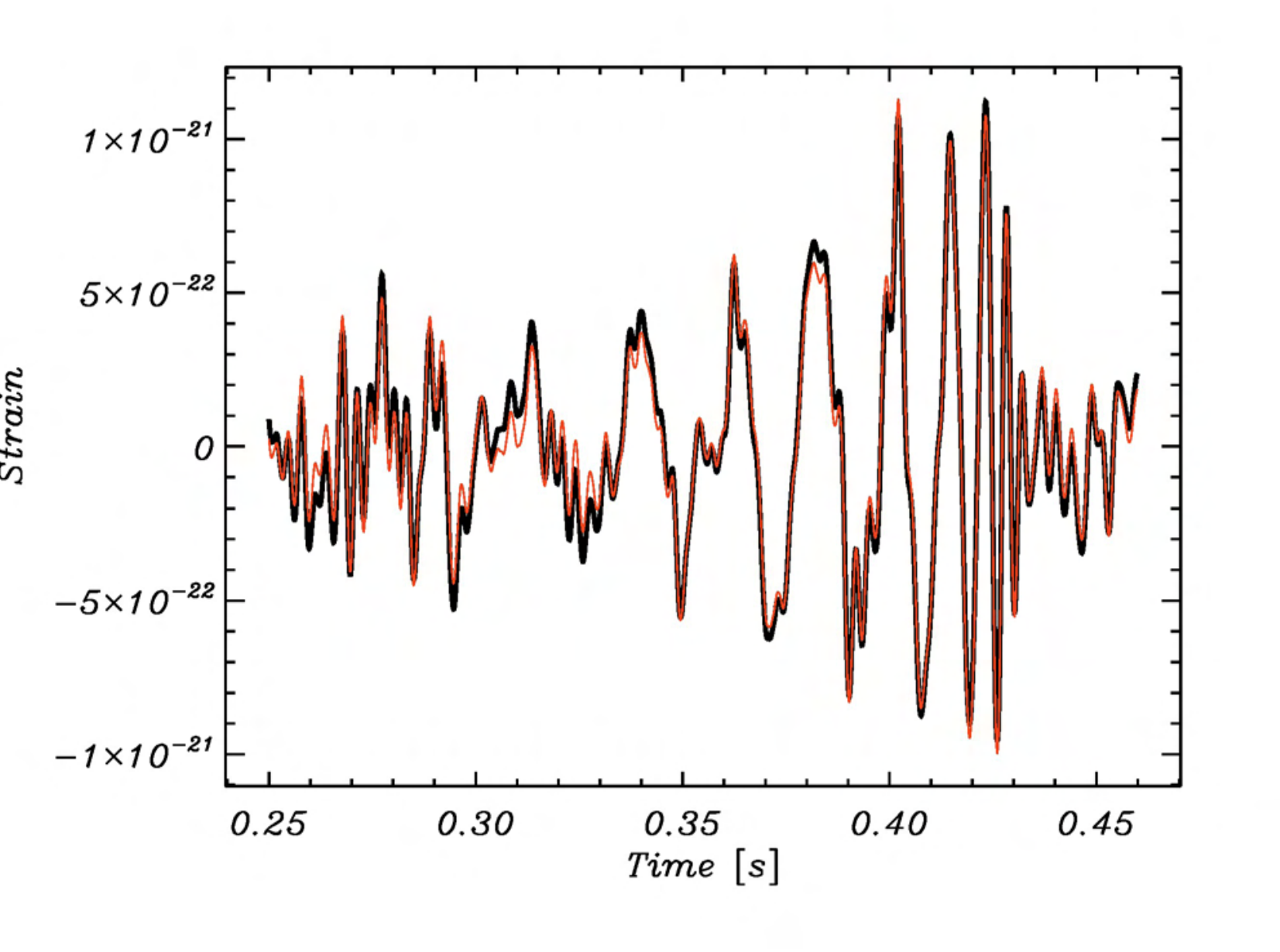}
  \includegraphics[width=0.32\textwidth]{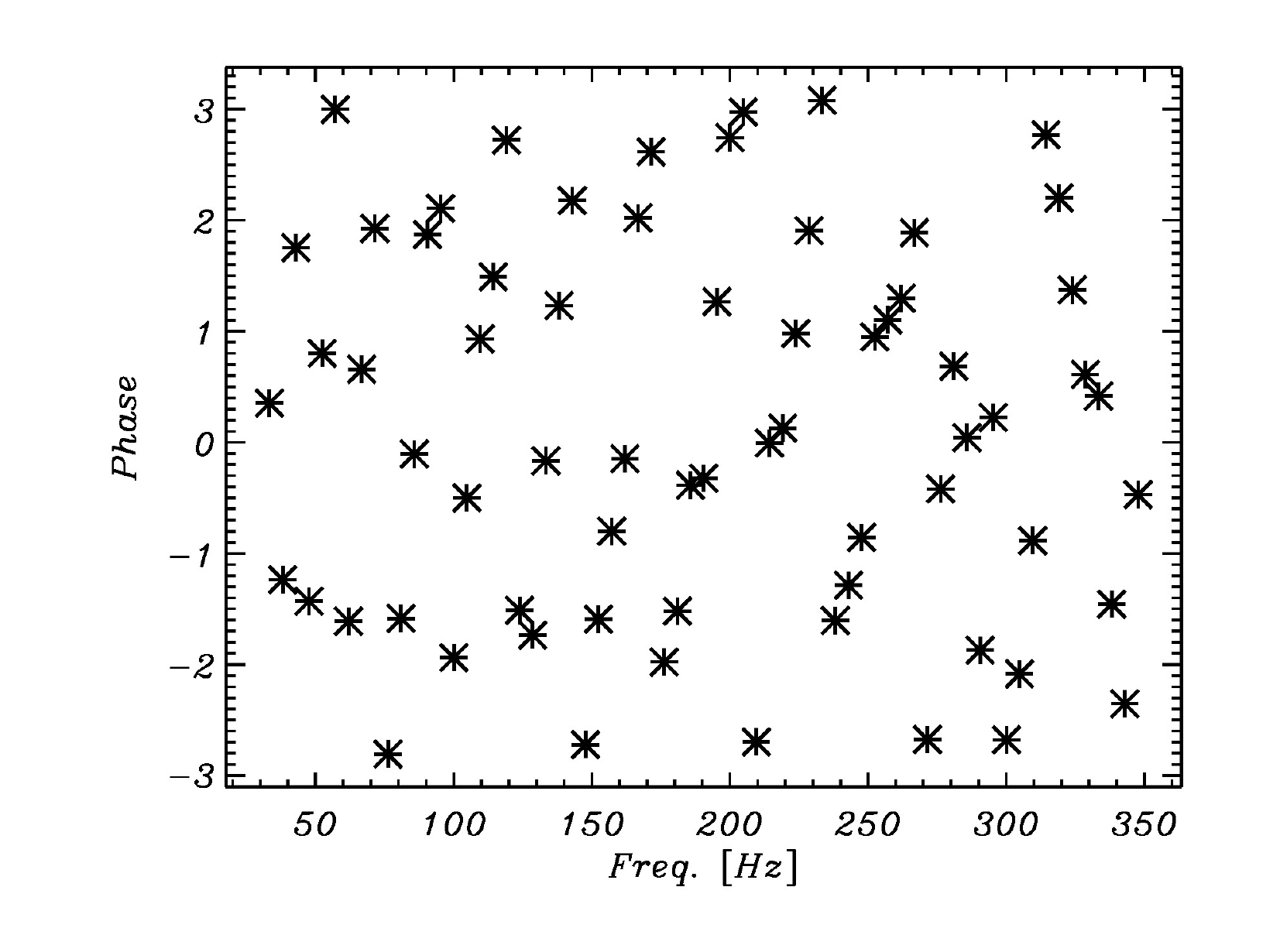}
  \includegraphics[width=0.32\textwidth]{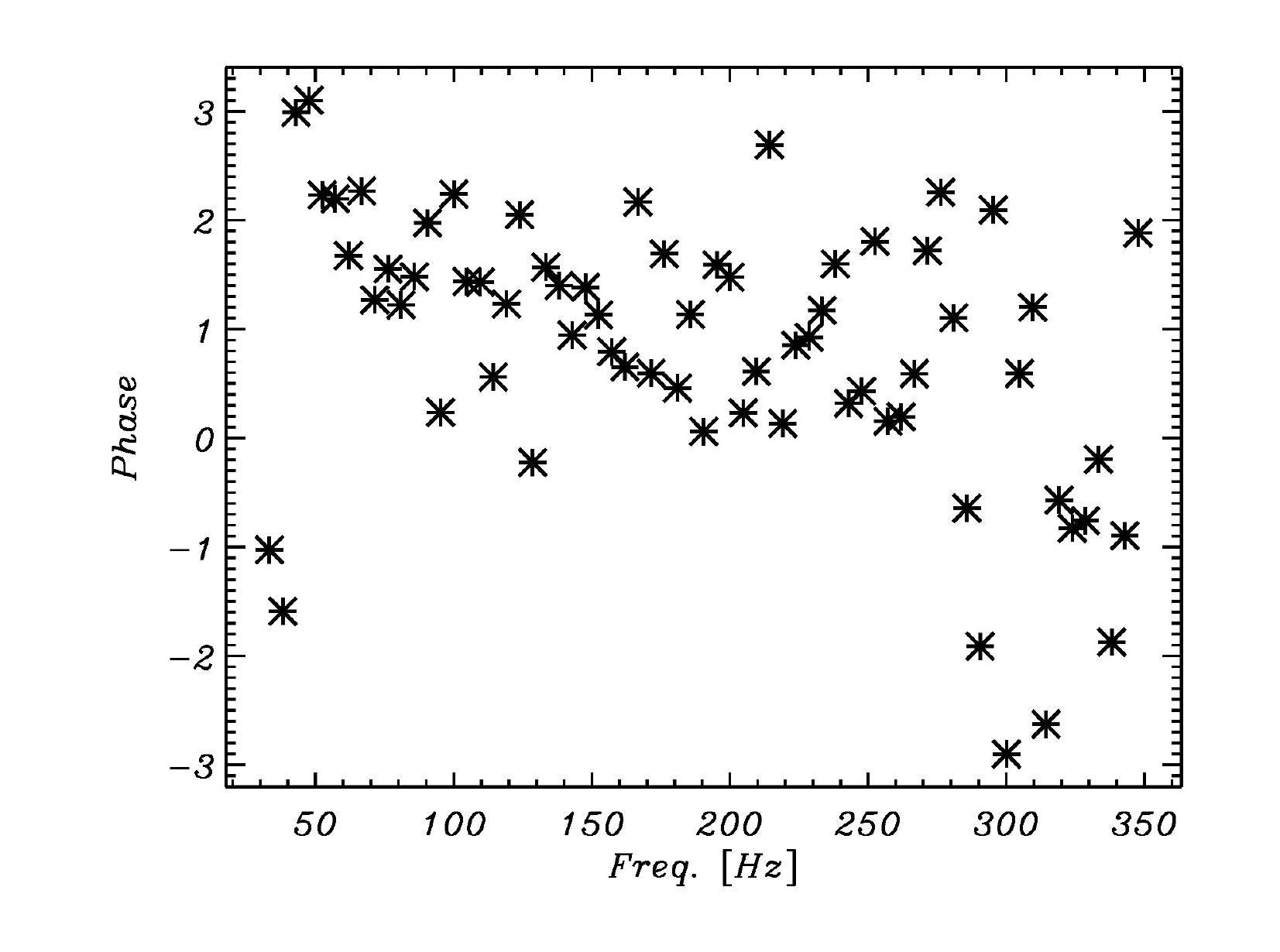}

  \includegraphics[width=0.32\textwidth]{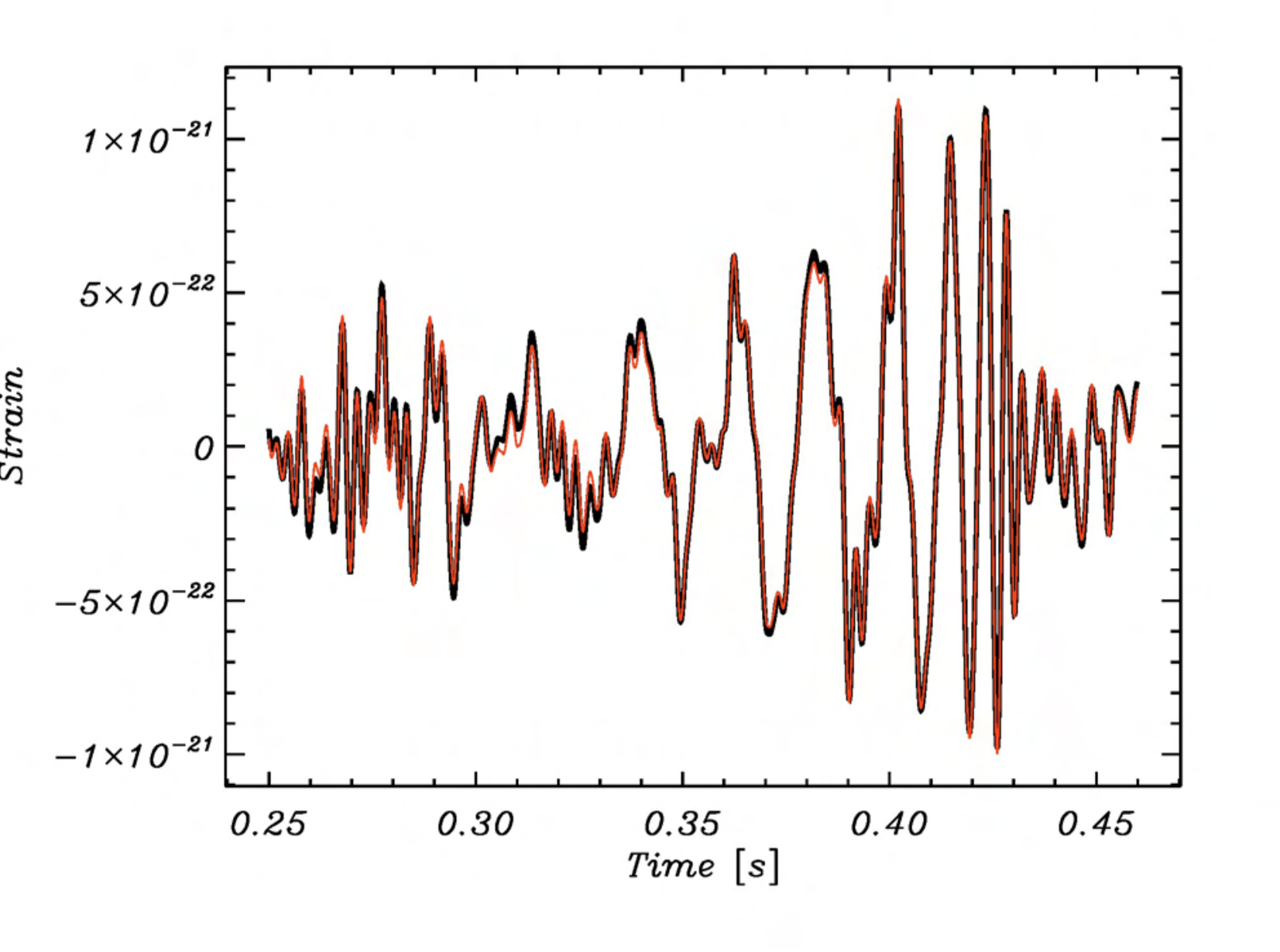}
  \includegraphics[width=0.32\textwidth]{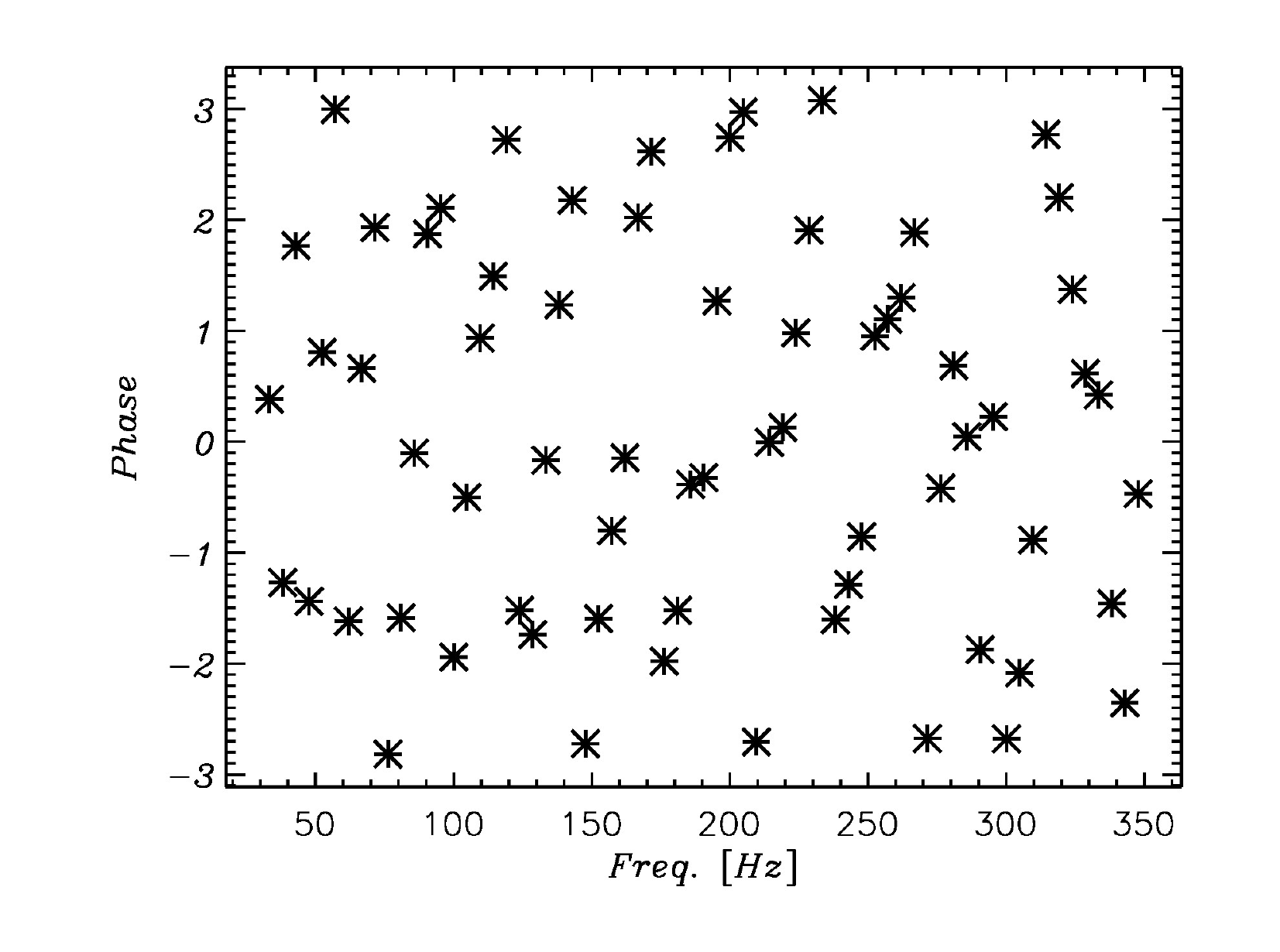}
  \includegraphics[width=0.32\textwidth]{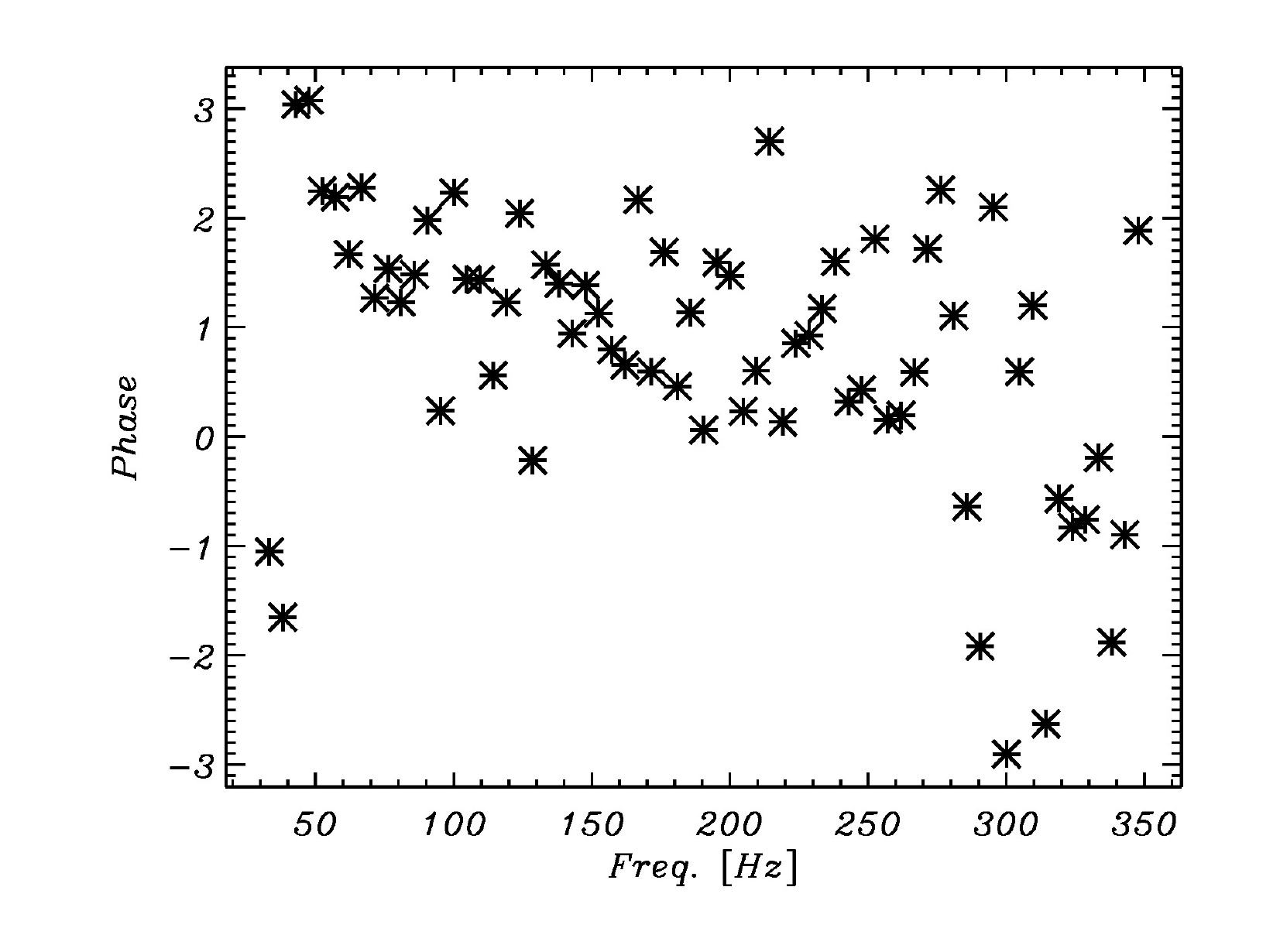}
  \caption{The Hanford 0.2 s data with GW150914 event after cleaning with different choices of the low frequency cutoffs.  The left column is for strain data for 20, 25, 30 and 32 Hz low frequency of BPF (from the top down to the bottom). The middle column show the phases of the cleaned signal and the right column is for the phase difference between two adjacent frequency bins. The black line corresponds to the variation of the lowest frequency. The red line stands for the GW150914 with BPF 35-350 Hz.}
\label{fig4.3}
\end{figure}

We summarize the results of the reconstruction including the phases of the signal and the phase difference in Fig.~\ref{fig4.3}.

Note that the apparent homogeneity and uncorrelated  phases (seen from the middle column) actually correspond to strong correlations (see the right column).

\begin{figure}[tbh!]
  \centering
  \includegraphics[height=0.28\textheight]{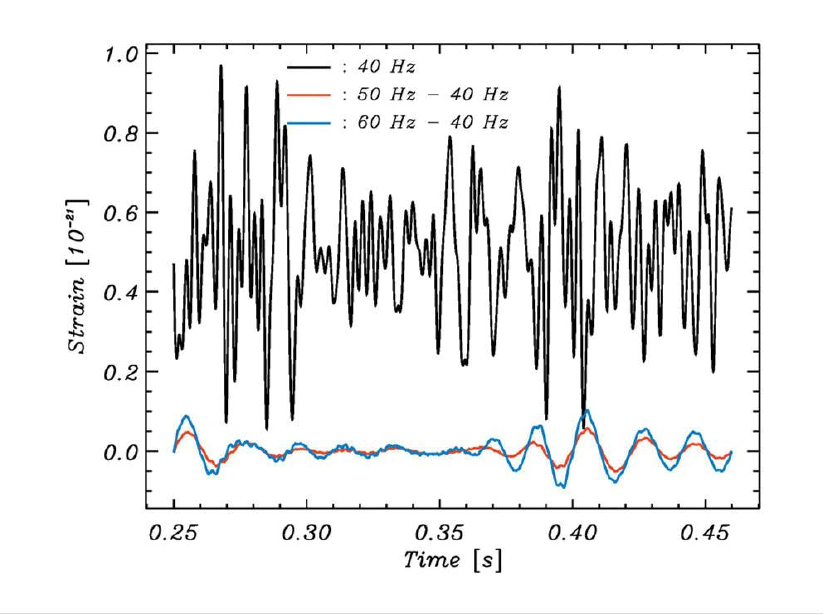}
  \caption{The residuals of the signals minus the templates for the low frequencies 40, 50 and 60 Hz of BPF. For better visual effect, 50 and 60 Hz are shown as difference to 40 Hz.}
  \label{fig4.4}
\end{figure}

There is also an interesting question related to increasing the low frequency of the BPF above 35 Hz up to 100 Hz. In this case we will change the GW-signal in the chirp-effect
zone, which is primarily related to the  60-180 Hz patch of the power spectrum. In Fig.~\ref{fig4.4} we show the residuals after subtraction of the filtered LIGO template from the filtered
data. The highest frequency of BPF is still 350 Hz.

\section{The cross-correlation of the residuals for different BPF pass band}\label{app:chaning the lower limit of bp}

\begin{figure}[tbh!]
  \centering
  \includegraphics[width=0.48\textwidth]{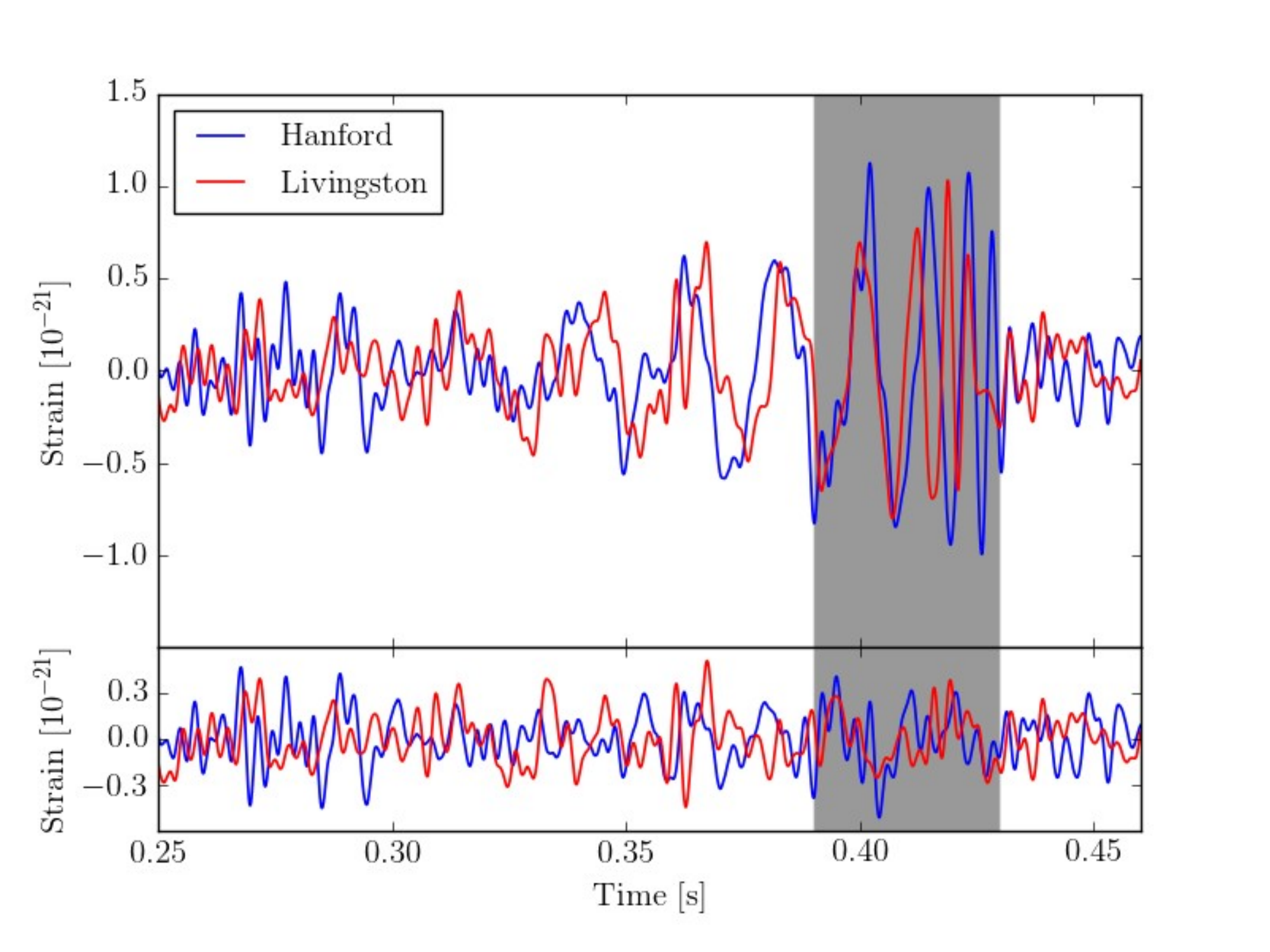}
  \includegraphics[width=0.48\textwidth]{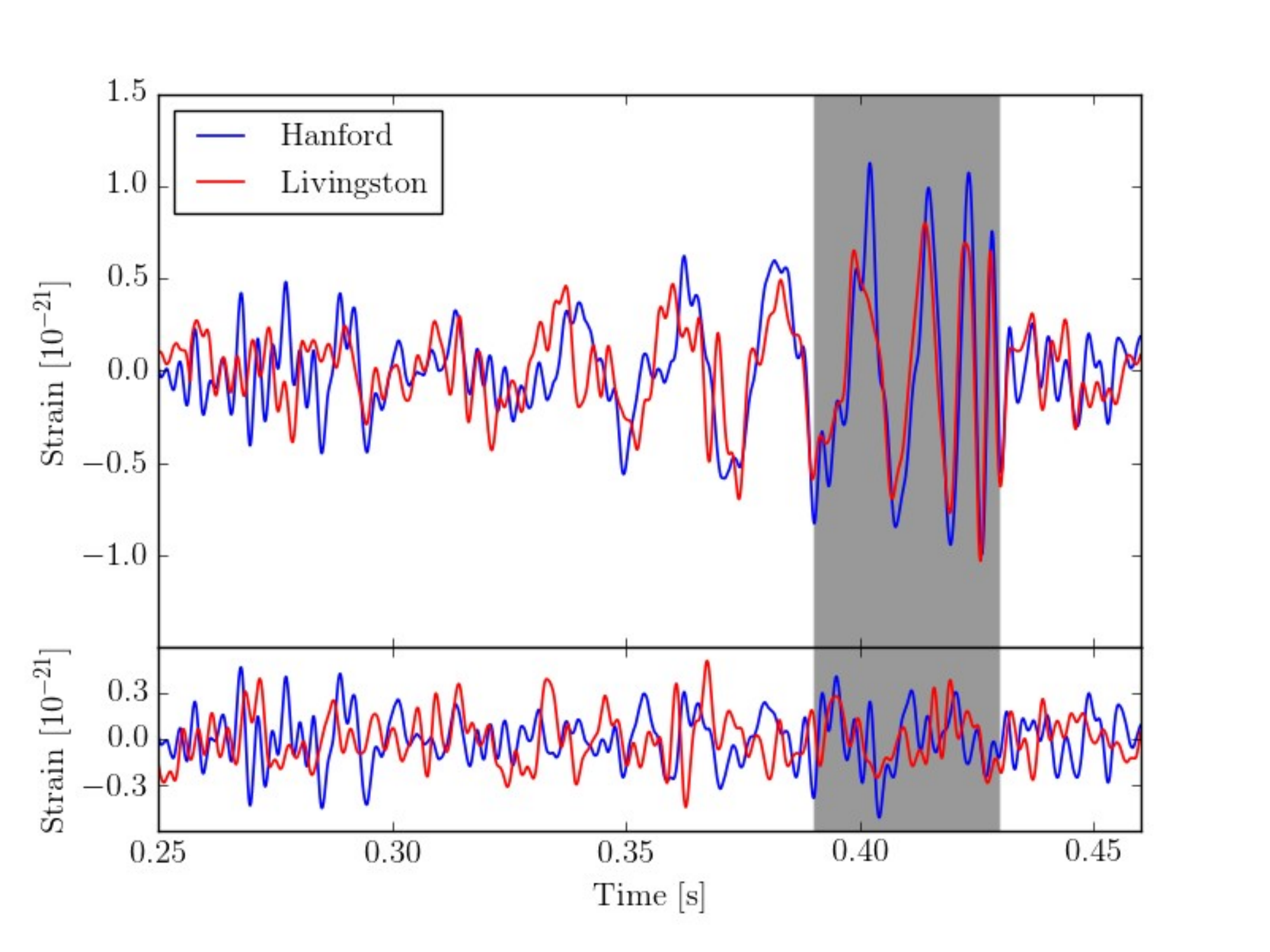}
  \caption{The Hanford (blue) and Livingston (red) records and their residuals after subtraction of the respective templates before (left panel) and after (right panel) shifting the Livingston record by 7 ms and inverting it. The BPF is 35-350 Hz.}
  \label{fig5.1}
\end{figure}

In this section we will illustrate the dependence of the cross-correlation of the residuals (signal minus template) for various lower frequencies of the BPF. We perform our analysis on the 32 s records cleaned as described above with Butterworth filters using various lower frequencies for a fixed higher frequency of 350 Hz. For all the 32 s records we extract the 0.2 s time domain containing GW150914 and focus on the particular zone shown in Fig.~\ref{fig5.1} in grey.

\begin{figure}[tbh!]
  \centering
  \includegraphics[height=0.15\textheight]{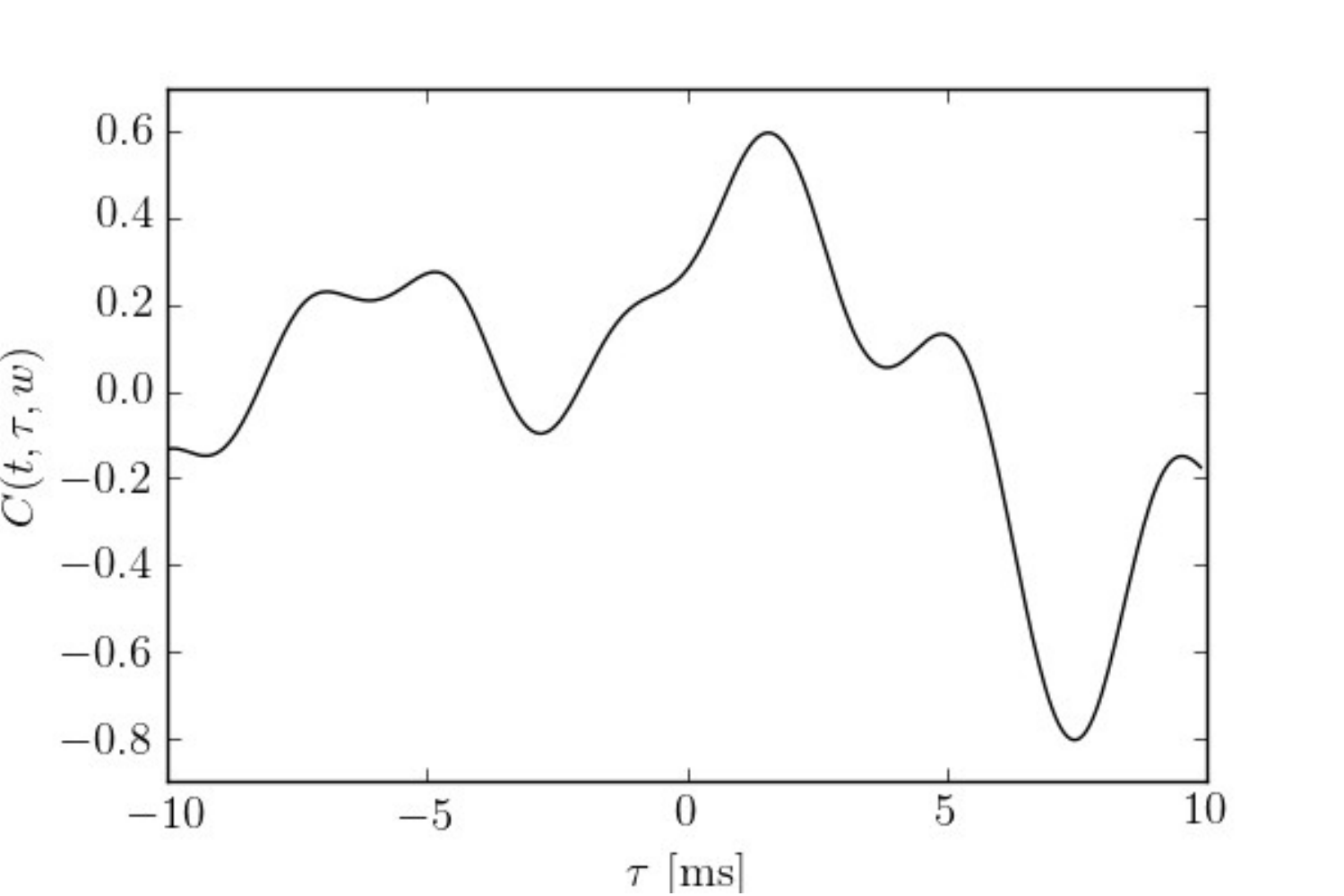}
  \includegraphics[height=0.15\textheight]{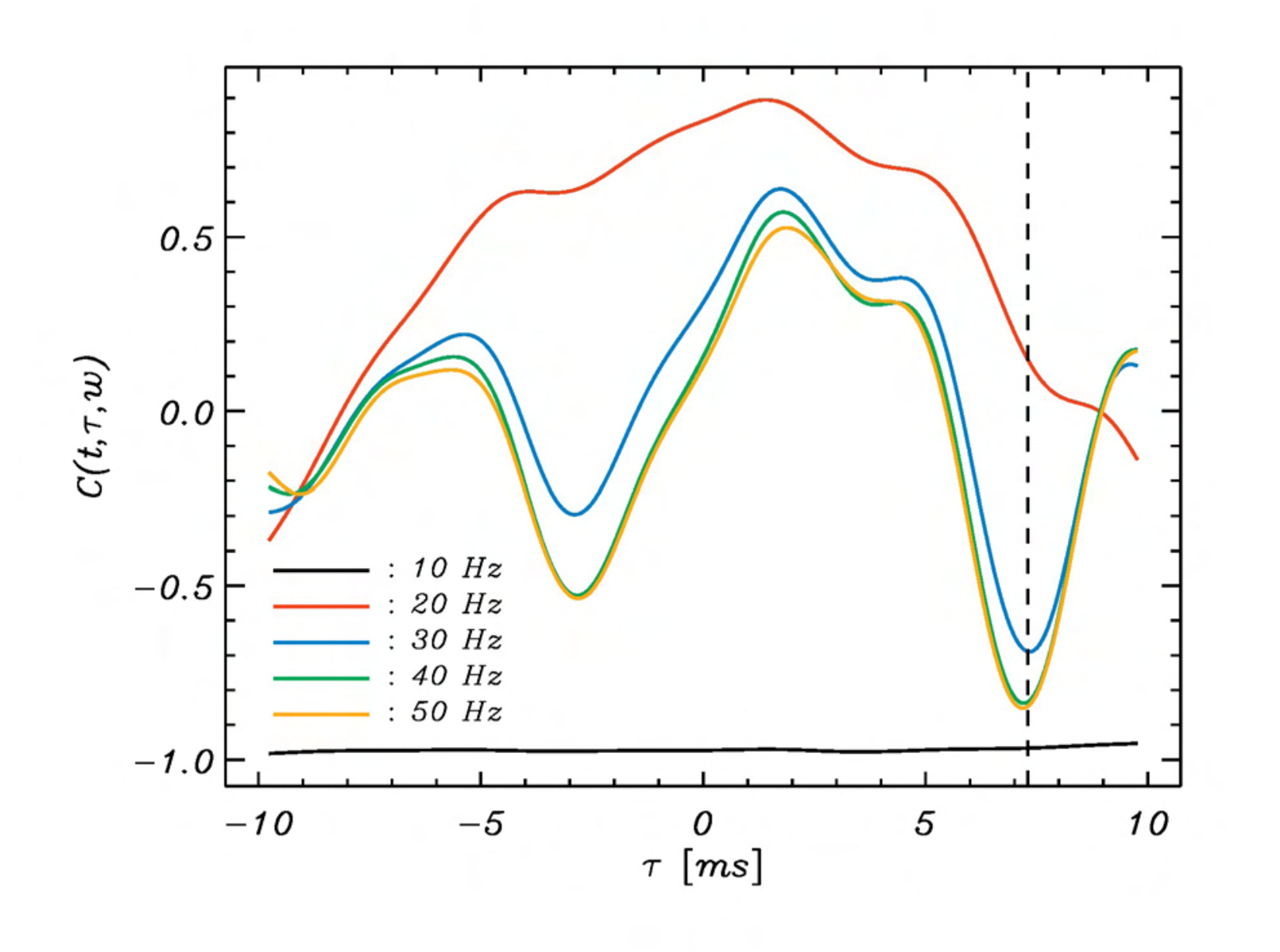}
  \includegraphics[height=0.15\textheight]{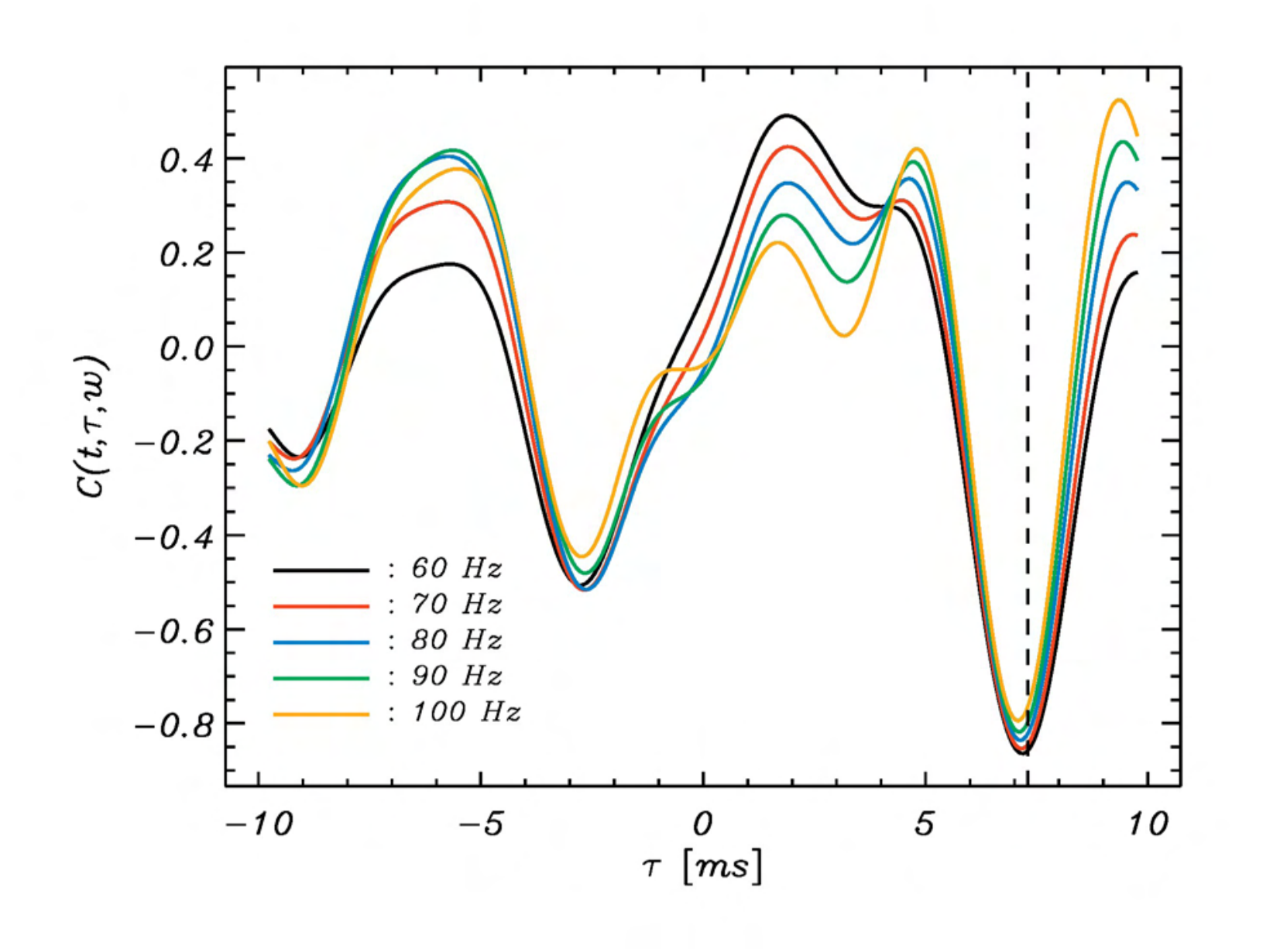}
  \caption{The cross-correlation function for 40 ms time domain from Fig.\ref{fig5.1} and lowest frequency 10-50 Hz(middle) and 60-100 Hz(right) in comparison with 35-350 Hz $C(t,\tau,w)$ from our paper (left).}
\label{fig5.2}
\end{figure}

We denote the strain data, $H(t)$ and $L(t)$, within a given time interval $t_a\leq t\leq t_b$ as $H_{t_a}^{t_b}$ and $L_{t_a}^{t_b}$, respectively, while the arrival time delay between the two sites is $\tau$.  The cross-correlation coefficient between the H and L strain data in a window of width $w$ starting at time $t$ is then given as (also shown in \citep{Liu16})
\begin{equation}
C(t,\tau,w) = {\rm Corr}(H_{t+\tau}^{t+\tau+w},L_{t}^{t+w}),
%\label{eq:runningwindowcorrelation}
\end{equation}
where we allow for a relative shift of the Hanford and Livingston data in the time domain. Here, ${\rm Corr}(x,y)$
is the standard Pearson cross-correlation function between records $x$ and $y$ constrained to the window $w$:
%\blue{
\begin{align}
{\rm Corr}(x_{t+\tau}^{t+\tau+w},y_{t}^{t+w}) = \frac{{\rm Cov}(x_{t+\tau}^{t+\tau+w},y_{t}^{t+w})}{\sqrt{{\rm Cov}(x_{t+\tau}^{t+\tau+w},x_{t+\tau}^{t+\tau+w}) \cdot {\rm Cov}(y_{t}^{t+w},y_{t}^{t+w})}},
\label{corr}
\end{align}
%}
where ${\rm Cov}(x,y)=\langle (x-\langle x \rangle)(y - \langle y \rangle) \rangle$ and $\langle ... \rangle$ denotes the average within the window considered.
In Fig.\ref{fig5.2} we plot the cross-correlation function $C(t,\tau,w)$ for various low frequencies of BPF in comparison with 35-350 Hz domain. For all  variants the LIGO template
is filtered out by the same BPF.

Note the strong dependence of the cross-correlations on the BPF low frequency. For the 35-350 Hz domain, the cross-correlation function for the residuals has a global minimum at $\tau\simeq 7$ ms and a strong minimum at $\tau\simeq 2$ ms.  Since we are interested in the most pronounced Hanford-Livingston anti-correlations, the global minima at $\tau\simeq 7$ ms
is of greatest significance. The second negative peak at $\tau\simeq -3$ ms is less significant for the given BPF domain. However, this tendency depends critically on the
lower frequency of the BPF. From Fig.~\ref{fig5.2} one can see that starting from $f_{min}=40$Hz and for all $f_{min}>40$ Hz the depth at  $\tau\simeq -3$ ms becomes more pronounced but does not reach the amplitude of the $\tau\simeq 7$ ms peak. At the same time, the width of these negative peaks decreases when $f_{min}$ increases.

\section{Precursor and echo of the GW1509014 event}
\begin{figure}[tbh!]
  \centering
  \includegraphics[width=0.78\textwidth]{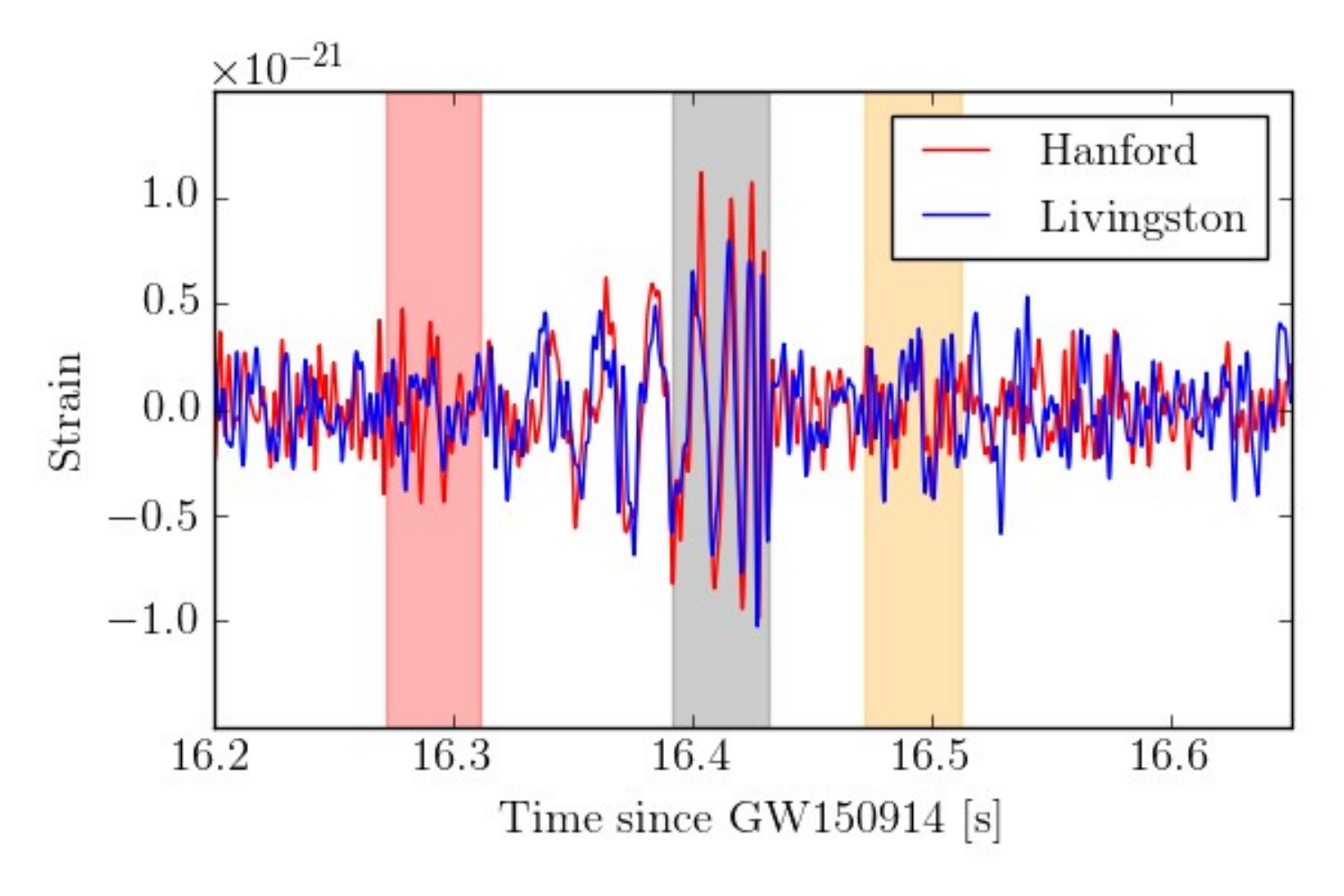}
  \caption{The cleaned records and their residuals for Hanford and Livingston  (The Livingston record is inverted and shifted by 7\,ms for a direct comparison). Here, we focus on the time region from 0.05 to 0.09 s which is shaded in orange.}
  \label{fig5.3}
\end{figure}

In the presentation of the results and their interpretation, it has been necessary to estimate the residuals of the signal in the time domain of the GW-event. The simplest, but not the best, way to do this is merely to subtract from the cleaned data the best fit template or in the case of varying the BPF, its theoretical waveform. As we have pointed out above, the analysis of the properties of the cleaned residuals is a major challenge for LIGO. This analysis should be performed in a manner similar to that employed in CMB science where the primordial CMB signal must be extracted from a combination of  $f^{-1}$ noise, kinematic dipole, Galactic and extragalactic foregrounds, etc. There is, however, an important difference between LIGO data analysis and the CMB approach due to the non-stationarity of the GW signals. The point is that soon after the chirp effect there is no gravitational wave in the detectors, and the signal from the Hanford and Livingston detector should represent pure instrumental noise and potential residuals of the cleaning. This is why we extend the domain of the cross-correlation analysis slightly as shown in Fig.~\ref{fig5.3}. Regarding the GW event, we use the 35-350 Hz BPF.

For a more general investigation of the possible correlation between noise in the Hanford and Livingston in the above context, we invert and shift the Livingston data by 7 ms and apply a running window correlation as
\begin{equation}\label{equ:running CC}
C(t,w) = {\rm Corr}(H_{t}^{t+w},\tilde{L}_{t}^{t+w}),
\end{equation}
where $\tilde{L}$ means shifted and inverted Livingston data. We use $w=20$ ms, which is equivalent to the length adopted in Sec~\ref{sub:residual}-\ref{sub:nulloutput}. The difference in value is because here Livingston is already shifted, which leads to a difference in the edge cut.
\begin{figure}[tbh!]
 \centering
 \includegraphics[width=0.85\textwidth]{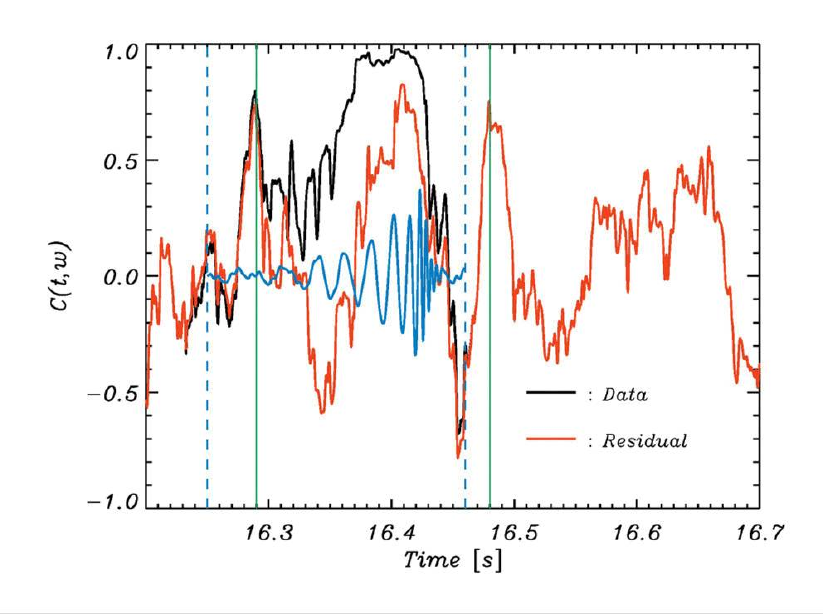}
  \caption{The running window correlation $C(t,w)$ (Eq.~(\ref{equ:running CC})) with Livingston inverted and shifted by 7 ms. Black for the clean strain data, and red for the residual (data minus template). The precursor and echo peaks are marked by the green vertical lines.}
  \label{fig5.5}
\end{figure}

One can clearly see from Fig.~\ref{fig5.5} that, for the precursor and the echo region (marked by green vertical lines), removal of the template will make almost no difference, which means they are nearly unaffected by the template. Simulations with white noise subject to the same 35-350 Hz BPF suggest that the height of both the precursor and the echo peaks have a significance of about $0.5\%$.

%For these shifted and inverted cleaned records we get the cross-correlation coefficient for the same window $w=40$ms, shown in Fig.\ref{fig5.4}. Since we  already inverted and shifted the Livingston record by 7 ms, we should find a positive peak in the cross-correlation in the vicinity of. This is exactly the case.
%
%\begin{figure}[tbh!]
% \centering
% \includegraphics[width=0.5\textwidth]{./figure/Method1_H_Corr.pdf}
%  \caption{The correlation function for the domain shown in the right panel of Fig. 16.}
%  \label{fig5.4}
%\end{figure}

%Thus, right after the end of the GW150914 event we have found the ripples (or echo) in both detectors with the same time lag, as discussed in our paper.

\end{document}